\documentclass[10pt]{book}
\usepackage[english]{babel,varioref}

\usepackage{epsfig}
\usepackage{color}
\usepackage{textcomp}
\usepackage{mathrsfs,amsmath} 
\usepackage{bbm}

\newcommand{\red}{\color{red}}

\def\gtrsim{\mathrel{\raise.4ex\hbox{$>$}\kern-0.8em\lower.7ex\hbox{$\sim$}}}
\def\lesssim{\mathrel{\raise.4ex\hbox{$<$}\kern-0.8em\lower.7ex\hbox{$\sim$}}}

\newcommand{\bone}{\mathbbm{1}}

\newcommand{\sigmab}{\mbox{\boldmath $\sigma $}}
\newcommand{\deltab}{\mbox{\boldmath $\delta $}}
\newcommand{\etab}{\mbox{\boldmath $\eta $}}

\newcommand{\Pib}{\mbox{\boldmath $\Pi $}}
\newcommand{\Pibtilde}{\mbox{\boldmath $\tilde{\Pi} $}}

\newcommand{\bp}{{\bf p}}
\newcommand{\bq}{{\bf q}}
\newcommand{\bk}{{\bf k}}
\newcommand{\br}{{\bf r}}
\newcommand{\bA}{{\bf A}}
\newcommand{\ba}{{\bf a}}

\newcommand{\be}{{\bf e}}

\newcommand{\bj}{{\bf j}}

\newcommand{\bR}{{\bf R}}

\newcommand{\bB}{{\bf B}}
\newcommand{\bE}{{\bf E}}
\newcommand{\bK}{{\bf K}}

\newcommand{\Pitilde}{\tilde{\Pi}}
\newcommand{\Smath}{\mathcal{S}}
\newcommand{\Omath}{\mathcal{O}}
\newcommand{\Hmath}{\mathcal{H}}
\newcommand{\Cmath}{\mathcal{C}}
\newcommand{\Zmath}{\mathcal{Z}}
\newcommand{\Amath}{\mathcal{A}}
\newcommand{\Pmath}{\mathcal{P}}
\newcommand{\Tmath}{\mathcal{T}}
\newcommand{\Emath}{\mathcal{E}}
\newcommand{\Mmath}{\mathcal{M}}

\newcommand{\ua}{\uparrow}
\newcommand{\da}{\downarrow}

\newcommand{\beq}{\begin{equation}}
\newcommand{\beqn}{\begin{eqnarray}}
\newcommand{\eeq}{\end{equation}}
\newcommand{\eeqn}{\end{eqnarray}}
\newcommand{\nn}{\nonumber}

\title{Quantum Hall Effects}
\author{Mark O. Goerbig\\
Laboratoire de Physique des Solides, CNRS UMR 8502\\ Universit\'e Paris-Sud, France} 

\begin{document}

\bibliographystyle{cj} 


\maketitle

\subsection*{Preface}

The present notes cover a series of three lectures on the quantum Hall effect
given at the Singapore session ``Ultracold Gases and Quantum Information''
at {\sl Les Houches Summer School} 2009. Almost 30 years after the discovery of the
quantum Hall effect, the research subject of quantum Hall physics has definitely
acquired a high degree of maturity that is reflected by a certain number of
excellent reviews and books, of which we can cite only a few \cite{PG,yoshioka,ezawa}
for possible further or complementary reading. Also the different sessions
of {\sl Les Houches Summer School} have covered in several aspects quantum Hall
physics, and S. M. Girvin's series of lectures in 1998 \cite{GirvinLH} have certainly become a reference 
in the field.\footnote{These lectures are also available on the preprint server, http://arxiv.org/abs/cond-mat/9907002} 
Girvin's lecture notes were indeed extremely useful for myself when I started
to study the quantum Hall effect at the beginning of my Master and PhD studies.

The present lecture notes are complementary to the existing literature in several aspects.
One should first mention its introductory character to the field, which is in no
way exhaustive. As a consequence, the presentation of one-particle physics and a detailed discussion
of the integer quantum Hall effect occupy the major part of these lecture notes,
whereas the -- certainly more interesting -- fractional quantum Hall effect, with its
relation to strongly-correlated electrons, its fractionally charged quasi-particles
and fractional statistics, is only briefly introduced. 

Furthermore, we have tried
to avoid as much as possible the formal aspects of the fractional quantum Hall effect,
which is discussed only in the framework of trial wave functions {\sl \`a la Laughlin}. 
We have thus omitted, e.g., a presentation of Chern-Simons theories and related
quantum-field theoretical approaches, such as the Hamiltonian theory of the fractional
quantum Hall effect \cite{MS}, as much as the relation between the quantum Hall effect and
conformal field theories. Although these theories are extremely fruitful and still
promising for a deeper understanding of quantum Hall physics, 
a detailed discussion of them would require more space than these lecture notes
with their introductory character can provide. 

Another complementary aspect of the present lecture notes as compared to existing textbooks
consists of an introduction to Landau-level
quantisation that treats in a parallel manner the usual non-relativistic electrons in
semiconductor heterostructures and relativistic electrons in 
graphene (two-dimensional graphite). Indeed, the 2005 discovery of a quantum Hall effect
in this amazing material \cite{graph1,graph2}
has given a novel and unexpected boost to research in quantum Hall
physics. 

As compared to the (oral) lectures, the present notes contain slightly more information. 
An example is Laughlin's plasma analogy, which is described in Sec. \ref{PlasmaLaugh}, although
it was not discussed in the oral lectures. Furthermore, I have decided to add a chapter
on multi-component quantum Hall systems, which, for completeness, needed to be at least briefly
discussed. 

Before the Singapore session of {\sl Les Houches Summer School}, this series of lectures 
had been presented in a similar format at the (French) Summer School of the Research Grouping
``Physique M\'esoscopique'' at the Institute of Scientific Research, Carg\`ese, Corsica, in 2008.
Furthermore, a longer series of lectures on the quantum Hall effect was prepared in collaboration
with my colleague and former PhD advisor Pascal Lederer (Orsay, 2006). Its aim was somewhat different,
with an introduction to the Hamiltonian theories of the fractional quantum Hall effect and
correlation effects in multi-component systems. As already mentioned above, the latter aspect
is only briefly introduced within the present lecture notes and a discussion of Hamiltonian theories is
completely absent. The Orsay series of lectures was repeated by Pascal Lederer
at the {\sl Ecole Polytechnique F\'ed\'erale} in Lausanne Switzerland, in 2006, and
at the University of Recife, Brazil, in 2007. The finalisation of these longer and more detailed 
lecture notes (in French) is currently
in progress. The graphene-related aspects of the quantum Hall effect have furthermore been
presented in a series of lectures on graphene (Orsay, 2008) prepared in collaboration with
Jean-No\"el Fuchs, whom I would like to thank for a careful reading of the present notes.

\tableofcontents



\chapter{Introduction}

\begin{figure}
\begin{center}
\epsfig{figure=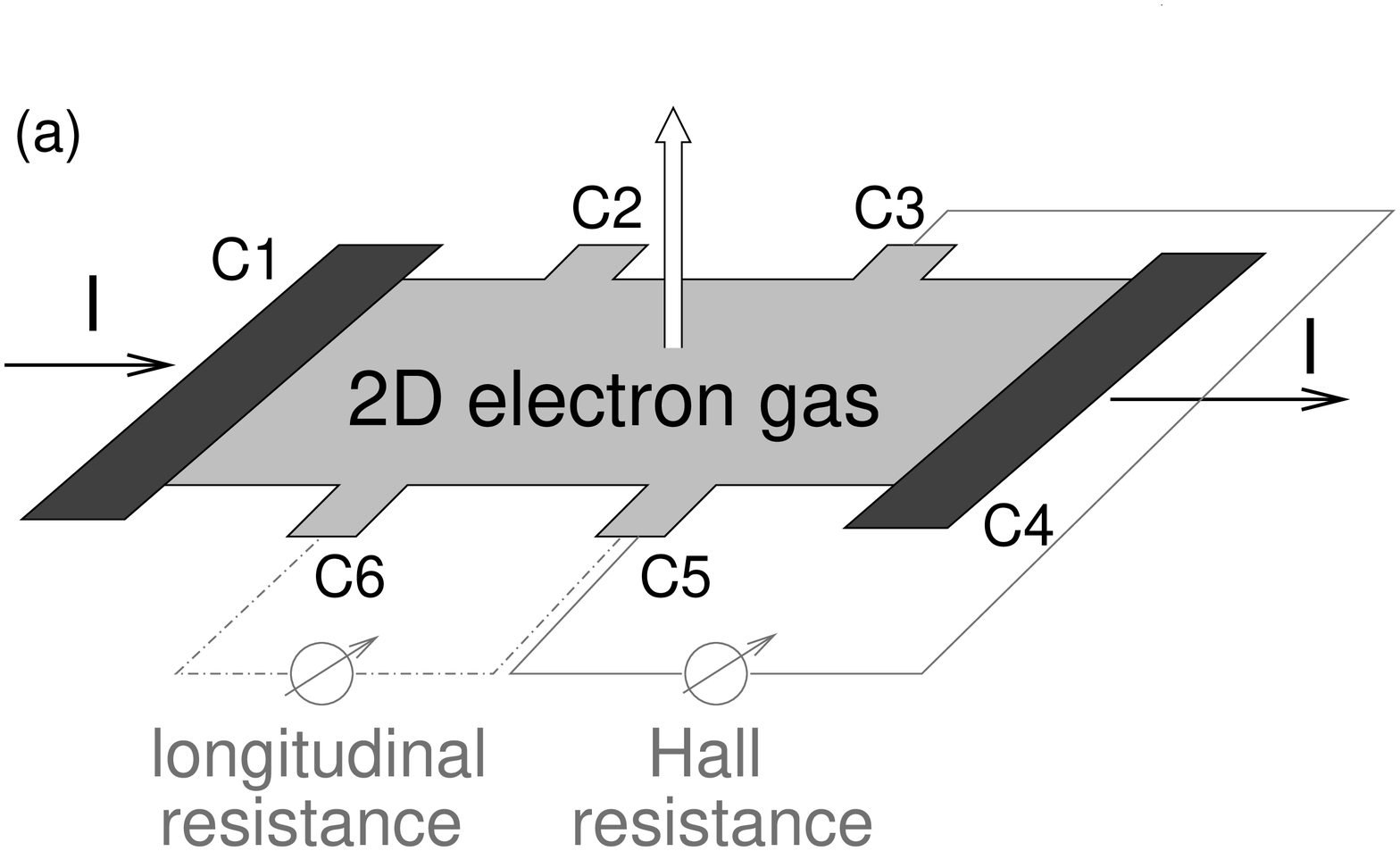,width=6cm,clip}
\epsfig{figure=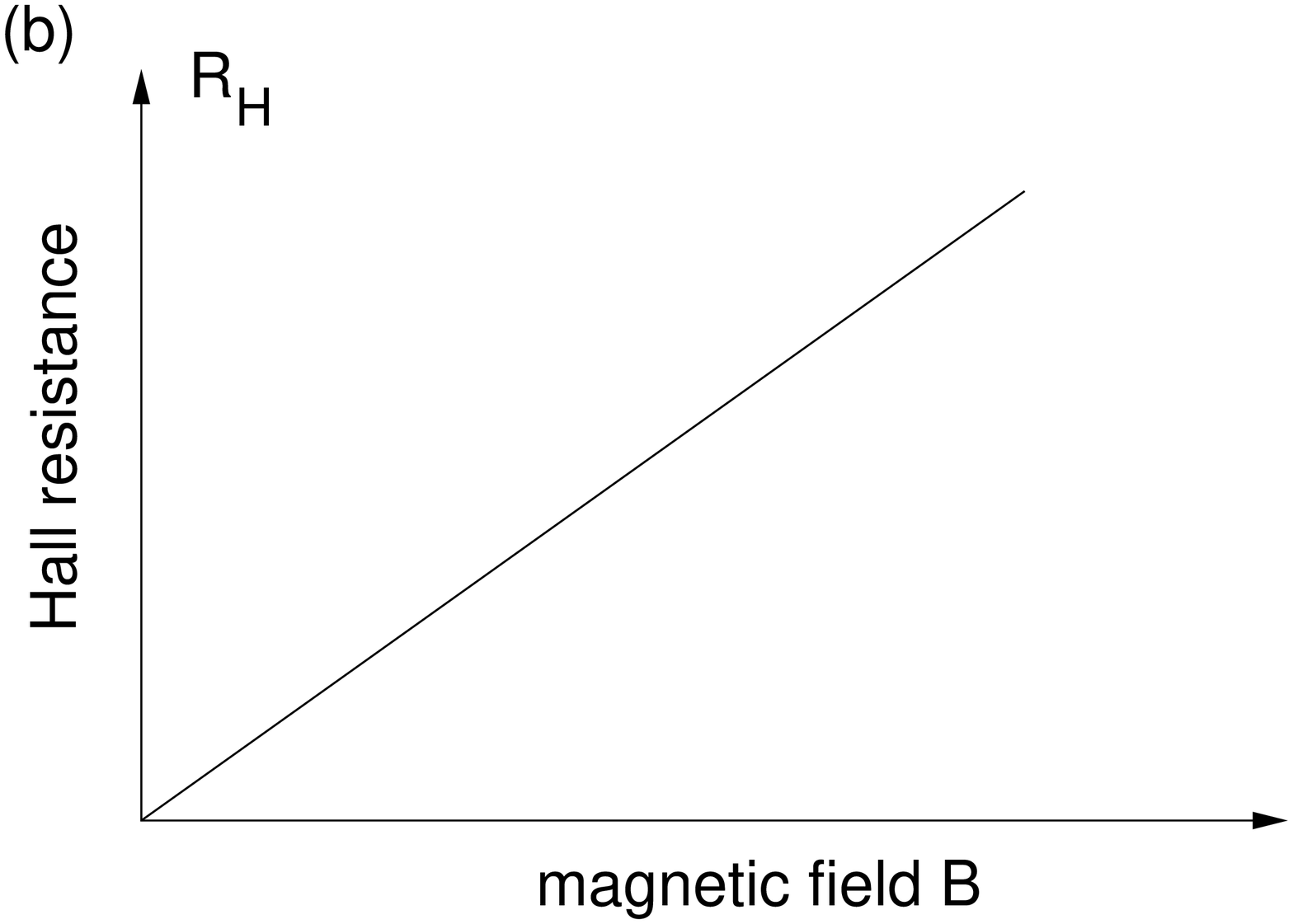,width=5cm,clip}
\end{center}
\caption{ (a) 2D electrons in a perpendicular magnetic field (quantum Hall system). 
In a typical transport measurement, a current $I$ is driven
through the system via the contacts C1 and C4. The longitudinal resistance may be measured between the contacts C5 and C6 (or 
alternatively between C2 and C3). The transverse (or Hall) resistance is measured, e.g., between the contacts C3 and C5. 
(b) Classical Hall resistance as a function of the magnetic field.}
\label{fig01}
\end{figure}

Quantum Hall physics -- the study of two-dimensional (2D) electrons in a strong perpendicular magnetic field [see Fig. \ref{fig01}(a)] --
has become an extremely important research subject during the last two and a half decades. The interest for
quantum Hall physics stems from its position at the borderline between low-dimensional quantum systems and systems 
with strong electronic correlations, probably the major issues of modern condensed-matter physics. From a theoretical point of
view, the study of quantum Hall systems required the elaboration of novel concepts some of which were better known in quantum-field
theories used in high-energy rather than in condensed-matter physics, such e.g. charge fractionalisation, non-commutative geometries 
and topological field theories. 

The motivation of the present lecture notes is to provide in an accessible manner the basic 
knowledge of quantum Hall physics and to enable thus interested graduate students to pursue on her or his own further studies 
in this subject. We have therefore tried, whereever we feel that a more detailed discussion of some aspects in this large 
field of physics would go beyond the introductory character of these notes, to provide references to detailed and pedagogical
references or complementary textbooks.

\section{History of the (Quantum) Hall Effect}
\label{hist}

\markboth{Introduction}{History of the (Quantum) Hall Effect}

\subsection{The physical system}

Our main knowledge of quantum Hall systems, i.e. a system of 2D electrons in a perpendicular magnetic field, stems 
from electronic transport measurements, where one drives a current $I$ through the sample and where one measures both the 
{\sl longitudinal} and the {\sl transverse} resistance (also called Hall resistance). The difference between these two resistances 
is essential and may be defined topologically: consider a current that is driven through the sample via two arbitrary contacts [C1 and C4
in Fig. \ref{fig01}(a)] and draw (in your mind) a line between these two contacts. 
A longitudinal resistance is a resistance measured between
two (other) contacts that may be connected by a line that does not cross the line connecting C1 and C4. In Fig. \ref{fig01}(a), 
we have chosen
the contacts C5 and C6 for a possible longitudinal resistance measurement. The transverse resistance is measured between two contacts that
are connected by an imaginary line that necessarily crosses the line connecting C1 and C4 [e.g. C3 and C5 in Fig. \ref{fig01}(b)].


\subsection{Classical Hall effect}
\label{CHE}

Evidently, if there is a {\sl quantum} Hall effect, it is most natural to expect that there exists also a {\sl classical} Hall effect.
This is indeed the case, and its history goes back to 1879 when Hall showed that the transverse resistance 
$R_H$ of a thin metallic 
plate varies linearly with the strength $B$ of the perpendicular magnetic field [Fig. \ref{fig01}(b)],
\beq\label{eq01}
R_H=\frac{B}{q n_{el}}\ ,
\eeq
where $q$ is the carrier charge ($q=-e$ for electrons in terms of the elementary charge $e$ that we define positive in the remainder 
of these lectures) and $n_{el}$ is the 2D carrier density. Intuitively, one may understand the effect as due to the Lorentz force, 
which bends the trajectory of a charged particle such that a density gradient is built up between the two opposite sample sides that are
separated by the contacts C1 and C4. 
Notice that the classical Hall resistance is still used today to determine, in material science, the carrier charge and density of 
a conducting material.

More quantitatively, the classical Hall effect may be understood within the Drude model for diffusive transport in a metal. Within 
this model, one considers independent charge carriers of momentum $\bp$ described by the equation of motion
$$\frac{d\bp}{dt}=-e\left(\bE+\frac{\bp}{m_b}\times \bB\right)-\frac{\bp}{\tau},
$$
where $\bE$ and $\bB$ are the electric and magnetic fields, respectively. Here, we consider transport of 
negatively charged particles (i.e. electrons with $q=-e$) with band mass $m_b$. The last term takes into account relaxation processes
due to the diffusion of electrons by generic impurities, with a characteristic relaxation time $\tau$. The macroscopic transport 
characteristics, i.e. the resistivity or conductivity of the system, are obtained from the static solution of the equation of motion,
$d\bp/dt = 0$, and one finds for 2D electrons with $\bp=(p_x,p_y)$
\beqn
\nn
eE_x &=& -\frac{eB}{m_b}p_y - \frac{p_x}{\tau},\\
\nn
eE_y &=& \frac{eB}{m_b}p_x -\frac{p_y}{\tau}\ ,
\eeqn
where we have chosen the magnetic field in the $z$-direction. In the above expressions, one notices the appearence of a characteristic
frequency,
\beq\label{cycl}
\omega_C=\frac{eB}{m_b}\ ,
\eeq
which is called {\sl cyclotron frequency} because it characterises the cyclotron motion of a charged particle in a magnetic field. With
the help of the Drude conductivity, 
\beq\label{Drude}
\sigma_0=\frac{n_{el} e^2 \tau}{m_b}\ ,
\eeq
one may rewrite the above equations as
\beqn
\nn
\sigma_0 E_x&=&-e n_{el}\frac{p_x}{m_b}-e n_{el}\frac{p_y}{m_b}(\omega_C\tau),\\
\nn
\sigma_0 E_y&=&e n_{el}\frac{p_x}{m_b}(\omega_C\tau)-e n_{el}\frac{p_y}{m_b},
\eeqn
or, in terms of the current density
\beq\label{curr}
\bj = -e n_{el} \frac{\bp}{m_b}\ ,
\eeq
in matrix form as $\bE=\rho\, \bj$, with the resistivity tensor
\beq\label{restens}
\rho=\sigma^{-1}
=\frac{1}{\sigma_0}\left(\begin{array}{cc} 1 & \omega_C\tau\\
- \omega_C\tau & 1 \end{array}\right)=\frac{1}{\sigma_0}\left(\begin{array}{cc} 1 & \mu B\\
- \mu B & 1 \end{array}\right),
\eeq
where we have introduced, in the last step, the mobility
\beq\label{mobility}
\mu= \frac{e\tau}{m_b}\, .
\eeq
From the above expression, one may immediately read off the Hall resistivity (the off-diagonal terms of the resistivity
tensor $\rho$)
\beq\label{HallRes}
\rho_H= \frac{\omega_C\tau}{\sigma_0}=\frac{eB}{m_b}\tau\, \times\, \frac{m_b}{n_{el}e^2\tau}=\frac{B}{e n_{el}}\ .
\eeq
Furthermore, the conductivity tensor is obtained from the resistivity (\ref{restens}), by matrix inversion,
\beq\label{condtens}
\sigma=\rho^{-1}=\left(\begin{array}{cc} \sigma_L & -\sigma_H\\
\sigma_H& \sigma_L \end{array}\right),
\eeq
with $\sigma_{L}=\sigma_0/(1+\omega_C^2\tau^2)$ and
$\sigma_{H}=\sigma_0\omega_C\tau/(1+\omega_C^2\tau^2)$. It is instructive to discuss, based on these
expressions, the theoretical limit of vanishing impurities, i.e. the limit $\omega_C\tau\rightarrow
\infty$ of very long scattering times. In this case the resistivity and conductivity tensors read
\beq\label{equ01b}
\rho=\left(\begin{array}{cc} 0 & \frac{B}{en_{el}}\\
- \frac{B}{en_{el}}& 0 \end{array}\right)\qquad {\rm and} \qquad
\sigma=\left(\begin{array}{cc} 0 & -\frac{en_{el}}{B}\\
\frac{en_{el}}{B}& 0 \end{array}\right),
\eeq
respectively. Notice that if we had put under the carpet the matrix character of the conductivity and resistivity and if we had only
considered the {\sl longitudinal} components, we would have come to the counter-intuitive  conclusion that the (longitudinal) resistivity
would vanish at the same time as the (longitudinal) conductivity. The transport properties in the clean limit $\omega_C\tau\rightarrow
\infty$ are therefore entirely governed, in the presence of a magnetic field, by the off-diagonal, i.e. transverse, components of the 
conductivity/resistivity. We will come back to this particular feature of quantum Hall systems when discussing the integer quantum Hall
effect below.

\subsubsection{Resistivity and resistance}

The above treatment of electronic transport in the framework of the Drude model allowed us to calculate the conductivity or resistivity
of classical diffusive 2D electrons in a magnetic field. However, an experimentalist does not measure a conductivity or resistivity, i.e.
quantities that are easier to calculate for a theoretician, but a {\sl conductance} or a {\sl resistance}. Usually, 
these quantities are related to one another
but depend on the geometry of the conductor -- the resistance $R$ is thus related to the resistivity 
$\rho$ by $R=(L/A)\rho$, where $L$ is the length of the conductor and $A$ its cross section. From the scaling point of view of 
a $d$-dimensional conductor, the cross section scales as $L^{d-1}$, such that the scaling relation between the resistance and the
resistivity is
\beq\label{res_scale}
R\sim \rho L^{2-d},
\eeq
and one immediately notices that a 2D conductor is a special case. From the dimensional point of view, resistance and resistivity are the
same in 2D, and the resistance is scale-invariant. Naturally, this scaling argument neglects the fact that the length $L$ and the width
$W$ (the 2D cross section) do not necessarily coincide: indeed, the resistance of a 2D conductor depends in general on the so-called
{\sl aspect ratio} $L/W$ via some factor $f(L/W)$ \cite{montam}. However, in the case of the transverse Hall resistance it is the length 
of the conductor itself that plays the role of the cross section, such that the Hall resistivity and the Hall resistance truely 
coincide, i.e. $f=1$. We will see in Chap. \ref{IQHE} that this conclusion also holds in the case of the quantum Hall effect and not only
in the classical regime. Moreover, the quantum Hall effect is highly insensitive to the particular geometric properties of the 
sample used in the transport measurement, such that the quantisation of the Hall resistance is surprisingly precise (on the order 
of $10^{-9}$) and the quantum Hall effect is used nowadays in the definition of the resistance standard.

\subsection{Shubnikov-de Haas effect}

\begin{figure}
\begin{center}
\epsfig{figure=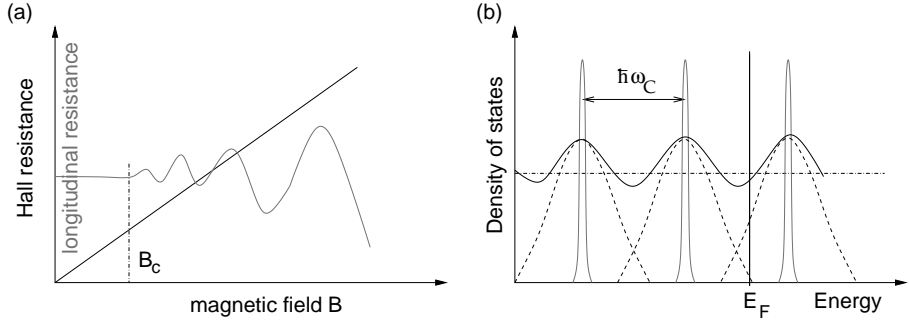,width=12cm,clip}
\end{center}
\caption{ (a) Sketch of the Shubnikov-de Haas effect. 
Above a critical field $B_c$, the longitudinal resistance (grey) starts to oscillate as
a function of the magnetic field. The Hall resistance remains linear in $B$.
(b) Density of states (DOS). In a clean system, the DOS consists of equidistant delta peaks (grey) at the energies 
$\epsilon_n=\hbar\omega_C(n+1/2)$, whereas in a sample with a stronger impurity concentration, the peaks are broadened
(dashed lines). The continuous black line represents the sum of overlapping peaks, and $E_F$ denotes the Fermi energy.}
\label{fig02}
\end{figure}

A first indication for the relevance of quantum phenomena in transport measurements of 2D electrons in a strong magnetic field was
found in 1930 with the discovery of the Shubnikov-de Haas effect \cite{SdH}.
Whereas the classical result (\ref{restens}) for the 
resistivity tensor stipulates that the longitudinal resistivity $\rho_L=1/\sigma_0$ (and thus the longitudinal resistance) 
is independent of the magnetic field, Shubnikov and
de Haas found that above some characteristic magnetic field the longitudinal resistance oscillates as a function of the magnetic field. 
This is schematically depicted in Fig. \ref{fig02}(a). In contrast to this oscillation in the longitudinal resistance, the Hall resistance
remains linear in the $B$ field, in agreement with the classical result from the Drude model (\ref{HallRes}).

The Shubnikov-de Haas effect is a consequence of the energy quantisation of the 2D electron in a strong magnetic field, as it has
been shown by Landau at roughly the same moment. 
This so-called {\sl Landau quantisation} will be presented in great detail in Sec. \ref{Landau}. In a nutshell, Landau quantisation
consists of the quantisation of the cyclotron radius, i.e. the radius of the circular trajectory of an electron in a magnetic field.
As a consequence this leads to the quantisation of its kinetic energy into so-called Landau levels (LLs), 
$\epsilon_n=\hbar \omega_C(n+1/2)$,
where $n$ is an integer. In order for this quantisation to be relevant, the magnetic field must be so strong that the electron
performs at least one complete circular period without any collision, i.e. $\omega_C\tau>1$. This condition defines the 
critical magnetic field $B_c\simeq m_b/e\tau=\mu^{-1}$ above which the longitudinal resistance starts to oscillate, in terms of the 
mobility (\ref{mobility}). Notice that today's samples of highest mobility are characterised by $\mu\sim 10^{7}$ 
cm$^2$/Vs $=10^3$ m$^2$/Vs such that one may obtain Shubnikov-de Haas oscillations at magnetic fields as low as $B_c\sim 1$ mT.

The effect may be understood within a slightly
more accurate theoretical description of electronic transport (e.g. with the help of the Boltzmann
transport equation) than the
Drude model. The resulting Einstein relation relates then the conductivity to a diffusion equation, and 
the longitudinal conductivity 
\beq\label{kubo}
\sigma_L=e^2 D \rho(E_F)
\eeq 
turns out to be proportional to the density of states (DOS) $\rho(E_F)$
at the Fermi energy $E_F$ rather than the electronic density,\footnote{Notice, however, that the Fermi energy and thus the DOS
is a function of the electronic density. Furthermore we mention that in a fully consistent treatment also the diffusion constant
$D$ depends on the density of states and eventually the magnetic field. This affects the precise form of the oscillation but 
not its periodicity.} 
Due to Landau quantisation, the DOS
of a clean system consists of a sequence of delta peaks at the energies $\epsilon_n=\hbar \omega_C(n+1/2)$,
$$\rho(\epsilon)=\sum_n g_n \delta(\epsilon-\epsilon_n),
$$
where $g_n$ is takes into account the degeneracy of the energy levels. These peaks are eventually
impurity-broadened in real samples and may even overlap [see Fig. \ref{fig02}(b)], such that the DOS oscillates in energy with maxima
at the positions of the energy levels $\epsilon_n$. Consider a fixed  number of electrons in the
sample that fixes the zero-field Fermi energy the $B$-field dependence of which we omit in the argument.\footnote{Naturally, this
is a crude assumption because if the density of states $\rho(\epsilon,B)$ depends on the magnetic field, so does the
Fermi energy via the relation $$\int_0^{E_F}d\epsilon\,\rho(\epsilon,B)=n_{el}.
$$
However, the basic features of the Shubnikov-de Haas oscillation may be understood when keeping the Fermi energy constant.} 
When sweeping the magnetic field, one varies the energy distance between the LLs, and the DOS thus becomes maximal when $E_F$
coincides with the energy of a LL and minimal if $E_F$ lies between two adjacent LLs. The resulting oscillation in the DOS as a function
of the magnetic field translates via the relation (\ref{kubo}) into an oscillation of the longitudinal conductivity (or resistivity),
which is the essence of the Shubnikov-de Haas effect. 

\subsection{Integer quantum Hall effect}

An even more striking manifestation of quantum mechanics in the transport properties of 2D electrons in a strong magnetic field
was revealed 50 years later with the discovery of the integer quantum Hall effect (IQHE) by v. Klitzing, Dorda, and Pepper in 1980
\cite{KDP}. The Nobel Prize was attributed in 1985 to v. Klitzing for this extremely important discovery. 

Indeed, the discovery of the IQHE was intimitely related to technological advances in material science, namely in the fabrication
of high-quality field-effect transistors for the realisation of 2D electron gases. These technological aspects will be briefly reviewed
in separate a section (Sec. \ref{2DEG}). 

\begin{figure}
\begin{center}
\epsfig{figure=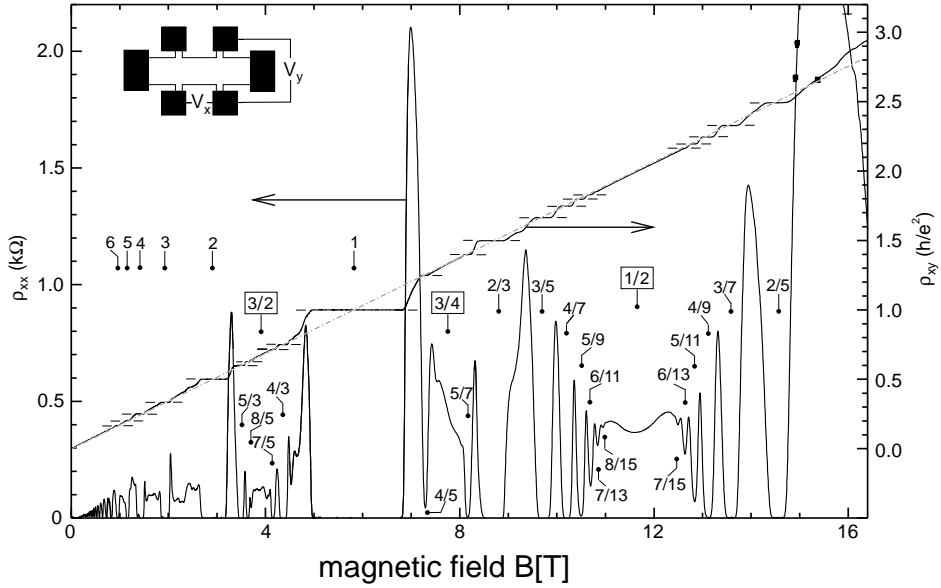,width=13cm,clip}
\end{center}
\caption{ Typical signature of the quantum Hall effect (measured by J. Smet, MPI-Stuttgart). Each plateau in the Hall resistance is
accompanied by a vanishing longitudinal resistance. The classical Hall resistance is indicated by the dashed-dotted line. The numbers
label the plateaus: integral $n$ denote the IQHE and $n=p/q$, with integral $p$ and $q$, indicate the FQHE.}
\label{fig03}
\end{figure}

The IQHE occurs at low temperatures, when the energy scale set by the temperature $k_BT$ is significantly smaller than the LL 
spacing $\hbar\omega_C$. It consists of a quantisation of the Hall resistance, which is no longer linear in $B$, as one would expect
from the classical treatment presented above, but reveals plateaus at particular values of the magnetic field (see Fig. \ref{fig03}). 
In the plateaus,
the Hall resistance is given in terms of universal constants -- it is indeed a fraction of the inverse quantum
of conductance $e^2/h$, and one observes
\beq\label{HallResQ}
R_H = \left(\frac{h}{e^2}\right)\frac{1}{n}\ ,
\eeq
in terms of an integer $n$.
The plateau in the Hall resistance is accompanied by a vanishing longitudinal resistance. This is at first sight reminiscent of the 
Shubnikov-de Haas effect, where the longitudinal resistance also reveals minima although it never vanished. The vanishing of the
longitudinal resistance at the Shubnikov-de Haas minima may indeed be used to determine 
the crossover from the Shubnikov-de Haas regime to the IQHE. 

It is noteworth to mention that the quantisation of the Hall resistance (\ref{HallResQ}) is a {\sl universal} phenomenon, i.e. 
independent of the particular properties 
of the sample, such as its geometry, the host materials  used to fabricate the 2D electron gas and, even more importantly, its 
impurity concentration or distribution. This universality is the 
reason for the enormous precision of the Hall-resistance quantisation (typically $\sim 10^{-9}$), which is nowadays -- since 1990 --
used as the resistance standard,\footnote{The subscript $K$ honours v. Klitzing and $90$ stands for the date since which the 
unit of resistance is defined by the IQHE.}
\beq\label{klitz}
 R_{K-90}=h/e^2 = 25\, 812.807\, \Omega,
\eeq
which is also called the Klitzing constant \cite{metrology1,metrology2}.
Furthermore, as already mentioned in Sec. \ref{CHE}, the vanishing of the longitudinal resistance indicates that 
the scattering time tends to infinity [see Eq. (\ref{equ01b})] in the IQHE. This is another indication of the 
above-mentioned universality of the effect, i.e. that IQHE does not depend on a particular impurity (or scatterer) arrangement.

A detailed presentation of the IQHE, namely the role of impurities, may be found in Chap. \ref{IQHE}.

\subsection{Fractional quantum Hall effect}

Three years after the discovery of the IQHE, an even more unexpected effect was observed in a 2D electron system of 
higher quality, i.e. higher mobility:
the {\sl fractional} quantum Hall effect (FQHE). The effect ows its name to the fact that contrary to the IQHE, where 
the number $n$ in Eq. (\ref{HallResQ}) is an integer, 
a Hall-resistance quantisation was discovered by Tsui, St\"ormer and Gossard with $n=1/3$ \cite{TSG}.
From the phenomenological point of view, the effect is extremely reminiscent of the IQHE: whereas the Hall resistance is quantised
and reveals a plateau, the longitudinal resistance vanishes (see Fig. \ref{fig03}, where different instances of both the IQHE and 
the FQHE are shown).
However, the origins of the two effects are completely different: whereas the IQHE may be understood from Landau quantisation, i.e. 
the kinetic-energy quantisation of independent electrons in a magnetic field, the FQHE is due to strong electronic correlations, when 
a LL is only partially filled and the Coulomb interaction between the electrons becomes relevant.
Indeed, in 1983 Laughlin showed that the origin of the observed FQHE with $n=1/3$, as well as any $n=1/q$ with $q$ being an odd integer, 
is due to the formation
of a {\sl correlated incompressible electron liquid} with extremely exotic properties
\cite{laughlin}, which will be reviewed in Chap. \ref{FQHE}. As for the
IQHE, the discovery and the theory of the FQHE was awarded a Nobel Prize (1998 for Tsui, St\"ormer and Laughlin).

After the discovery of the FQHE with $n=1/3$,\footnote{The quantity $n$ determines the filling of the LLs, usually described by the
Greek letter $\nu$, as we will discuss in Sec. \ref{Landau}.}
a plethora of other types of FQHE has been dicovered and theoretically described. One should first mention the $2/5$ and $3/7$ states
(i.e. with $n=2/5$ and $n=3/7$), 
which are part of the series $p/(2sp\pm 1)$, with the integers $s$ and $p$. This series has found a compelling interpretation within
the so-called {\sl composite-fermion} (CF) theory according to which the {\sl F}QHE may be viewed as an {\sl I}QHE of a novel
quasi-particle that consists of an electron that ``captures'' an even number of flux quanta \cite{Jain1,Jain2}. The basis of
this theory is presented in Sec. \ref{FQHE3}. Another intriguing FQHE was discovered in 1987 by Willet {\sl et al.},
with $n=5/2$ and $7/2$ \cite{willett} -- it is in so far intriguing as up to this moment only states $n=p/q$ with {\sl odd} denominators
had been observed in monolayer systems. From a theoretical point of view, it was shown in 1991 by Moore and Read \cite{MR} and by
Greiter, Wilczek and Wen \cite{GWW} that this FQHE may be described in terms of a very particular, so-called {\sl Pfaffian}, wave function,
which involves particle pairing and the excitations of which are anyons with non-Abelian statistics. These particles are 
intensively studied in today's research because they may play a relevant role in quantum computation. 
The physics of anyons will be introduced briefly in Sec. \ref{FQHE2}. Finally, we would mention in this brief (and naturally incomplete) 
historical overview a FQHE with $n=4/11$ discovered in 
2003 by Pan {\sl et al.} \cite{Pan}: it does not fit into the above-mentioned CF series, but it would correspond to a FQHE of CFs rather
than an IQHE of CFs.

\subsection{Relativistic quantum Hall effect in graphene}

Recently, quantum Hall physics experienced another unexpected boost with the discovery of a ``relativistic'' quantum Hall effect 
in graphene, a one-atom-thick layer of graphite \cite{graph1,graph2}. Electrons in graphene behave as if they were relativistic massless 
particles. Formally, their quantum-mechanical behaviour is no longer described in terms of a (non-relativistic) Schr\"odinger equation, 
but rather by a relativistic 2D Dirac equation \cite{antonioRev}. As a consequence, Landau quantisation of the electrons' kinetic energy
turns out to be different in graphene than in conventional (non-relativistic) 2D electron systems, as we will discuss in Sec. 
\ref{Landau}. This yields a ``relativistic'' quantum Hall effect with an unusual series for the Hall plateaus. Indeed rather than
having plateaus with a quantised resistance according to $R_H=h/e^2 n$, with integer values of $n$, one finds plateaus with
$n=\pm 2(2n'+1)$, in terms of an integer $n'$, i.e. with $n=\pm 2,\pm 6, \pm 10, ...$. The different signs in the series ($\pm$) 
indicate that there are two different carriers, electrons in the conduction band and holes in the valence band, 
involved in the quantum Hall effect in graphene. As we will briefly discuss in Sec. \ref{2DEG}, one may easily change the character
of the carriers in graphene with the help of the electric field effect.

Interaction effects may be relevant in the formation of other integer Hall plateaus,
such as $n=0$ and $n=\pm 1$ \cite{zhang}, which do not occur naturally in the series $n=\pm 2(2n'+1)$ characteristic of the relativistic
quantum Hall effect. Furthermore, a FQHE with $n=1/3$ has very recently been observed, although in a simpler geometric 
(two-terminal) configuration than the standard one depicted in Fig. \ref{fig01}(a) \cite{grapheneFQHE1,grapheneFQHE2}.

\section{Two-Dimensional Electron Systems}
\label{2DEG}

\markboth{Introduction}{Two-Dimensional Electron Systems}

As already mentioned above, the history of the quantum Hall effect is intimitely related to technological advances in the
fabrication of 2D electron systems with high electronic mobilities. The increasing mobility allows one to probe the fine 
structure of the Hall curve and thus to observe those quantum Hall states which are more fragile, such as some exotic FQHE 
states (e.g. the 5/2, 7/2 or the 4/11 states). This 
may be compared to the quest for high resolutions in optics: the higher the optical resolution, the better the chance of
observing tinier objects. In this sense, electronic mobility means resolution and the tiny object is
the quantum Hall state. As an order of magnitude, today's best 2D 
electron gases (in GaAs/AlGaAs heterostructures) are characterised by mobilities $\mu \sim 10^{7}$ cm$^2$/Vs.

\subsection{Field-effect transistors}

\begin{figure}
\begin{center}
\epsfig{figure=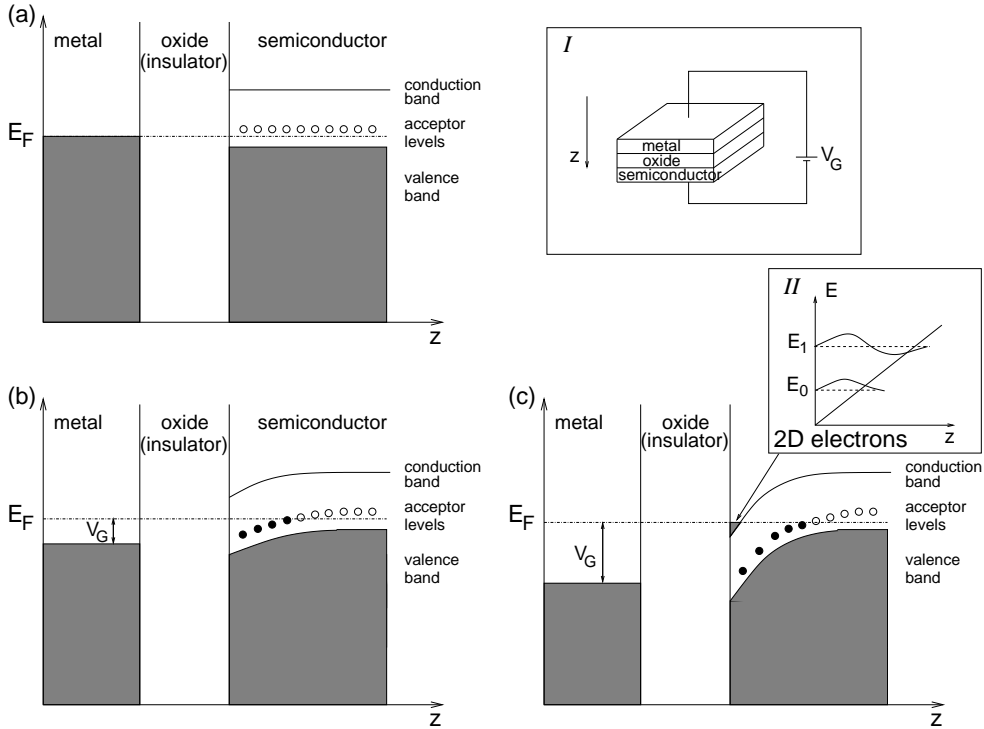,width=13cm,clip}
\end{center}
\caption{ MOSFET. The inset {\sl I} shows a sketch of a MOSFET. {\sl (a)} Level structure at $V_G=0$. In the metallic part, the band
is filled up to the Fermi energy $E_F$ whereas the oxide is insulating. In the semiconductor, the Fermi energy lies in the band gap
(energy gap between the valence and the conduction bands). Close to the valence band, albeit above $E_F$, are the acceptor levels. 
{\sl (b)} The chemical potential in the metallic part may be controled by the gate voltage $V_G$ via the electric field effect. 
As a consequence of the introduction of holes the semiconductor bands are bent downwards, and above a threshold voltage {\sl (c)}, 
the conduction band is filled in the vicinity of the interface with the insulator. One thus obtains a 2D electron gas. Its confinement
potential of which is of triangular shape, the levels (electronic subbands) of which are represented in the inset {\sl II}.
}
\label{fig04}
\end{figure}

The samples used in the discovery and in the first studies of the IQHE were so-calles {\sl metal-oxide-semiconductor field-effect 
transistors} (MOSFET). A metallic layer is seperated from a semiconductor (typically doped silicon) by an insulating oxide
(e.g. SiO$_2$) layer (see inset {\sl I} in Fig. \ref{fig04}). The chemical potential in the metallic layer may be varied with the
help of a gate voltage $V_G$. At $V_G=0$, the Fermi energy in the semiconductor lies in the band gap below the acceptor levels of
the dopants [Fig. \ref{fig04}(a)]. When lowering the chemical potential in the metal with the help of a positive gate voltage $V_G>0$,
one introduces holes in the metal that attract, via the electric field effect, electrons from the semiconductor to the 
semiconductor-insulator interface. These electrons populate the acceptor levels, and as a consequence, the semiconductor bands are bent 
downwards when they approach the interface, such that the filled acceptor levels lie now below the Fermi energy [Fig. \ref{fig04}(b)].

Above a certain threshold of the gate voltage, the bending of the semiconductor bands becomes so strong 
that not only the acceptor levels are below the Fermi energy, but also the conduction band in the vicinity of the interface 
which consequently gets filled with electrons [Fig. \ref{fig04}(c)].
One thus obtains a confinement potential of triangular shape for the electrons in the conduction band, the dynamics of which is
quantised into discrete electronic subbands in the perpendicular $z$-direction (see inset {\sl II} in Fig. \ref{fig04}).
Naturally, the electronic wave functions are then extended in the $z$-direction, but in typical MOSFETs only the lowest electronic 
subband $E_0$ is filled, such that the electrons are purely 2D from a dynamical point of view, i.e. there is no electronic motion in
the $z$-direction. 

The typical 2D electronic densities in these systems are on the order of $n_{el}\sim 10^{11}$ cm$^{-2}$, i.e. much lower than
in usual metals. This turns out to be important in the study of the IQHE and FQHE, because the 
effects occur, as we will show below, when the 2D electronic density is on the order of the density of magnetic flux $n_B=B/(h/e)$
threading the system,
in units of the flux quantum $h/e$. This needs to be compared to metals where 
the surface density is on the order of $10^{14}$ cm$^{-2}$, which would require
inaccessibly high magnetic fields (on the order of 1000 T) in order to probe the regime $n_{el}\sim n_B$.

\subsection{Semiconductor heterostructures}

\begin{figure}
\begin{center}
\epsfig{figure=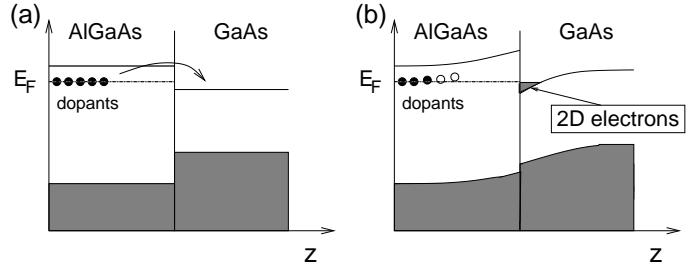,width=9cm,clip}
\end{center}
\caption{ Semiconductor heterostructure (GaAs/AlGaAs). {\sl (a)} Dopants are introduced in the AlGaAs layer at a certain distance 
from the interface. The Fermi energy lies below the in the band gap and is pinned by the dopant levels. The GaAs conduction band
has an energy that is lower than that of the dopant levels, such that it is energetically favourable for the electrons in 
the dopant layer to populate the GaAs conduction band in the vicinity of the interface. {\sl (b)} This polarisation bends the
bands in the vicinity of the interface between the two semiconductors, and thus a 2D electron gas is formed there on the GaAs side.
}
\label{fig05}
\end{figure}

The mobility in MOSFETs, which is typically on the order of $\mu \sim 10^{6}$ cm$^2$/Vs, 
is limited by the quality of
the oxide-semiconductor interface (surface roughness). This technical difficulty is circumvented in semiconductor heterostructures --
most popular are GaAs/AlGaAs heterostructures -- which are grown by molecular-beam epitaxy (MBE), where high-quality interfaces 
with almost atomic precision may be achieved, with mobilities on the order of $\mu \sim 10^{7}$ cm$^2$/Vs. These mobilities
were necessary to observe the FQHE, which was indeed first observed in a GaAs/AlGaAs sample \cite{TSG}.

In the (generic) case of GaAs/AlGaAs, the two semiconductors do not possess the same band gap -- indeed that of GaAs is smaller than
that of AlGaAs, which is chemically doped by donor ions at a certain distance from the interface between GaAs and AlGaAs
[Fig \ref{fig05}(a)].
The Fermi energy is pinned by these donor levels in AlGaAs, which may have a higher energy than the originally unoccupied conduction band
in the GaAs part, such that it becomes energetically favourable for the electrons in the donor levels to occupy the GaAs conduction
band in the vicinity of the interface. As a consequence, the energy bands of AlGaAs are bent upwards, whereas those of GaAs are bent
downwards. Similarly to the above-mentioned MOSFET, one thus obtains a 2D electron gas at the interface on the GaAs side, with
a triangular confinement potential.

\subsection{Graphene}
\label{SecGraph}

\begin{figure}
\begin{center}
\epsfig{figure=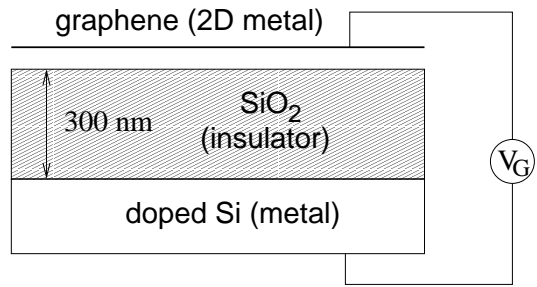,width=7cm,clip}
\end{center}
\caption{ Schematic view of graphene on a SiO$_2$ substrate with 
a doped Si (metallic) backgate. The system graphene-SiO$_2$-backgate may be viewed
as a capacitor the charge density of which is controled by a gate voltage $V_G$.}
\label{fig06}
\end{figure}

Graphene, a one-atom thick layer of graphite, presents a novel 2D electron system, which, from the electronic point of view, is
either a zero-overlap semi-metal or a zero-gap semiconductor, where the conduction and the valence bands are no longer separated
by an energy gap. Indeed, in the absence of doping, the Fermi energy lies exactly at the points where the valence band touches 
the conduction band and where the density of states vanishes linearly. 

In order to vary the Fermi energy in graphene, one usually places a graphene flake 
on a 300 nm thick insulating 
SiO$_2$ layer which is itself placed on top of a positively doped metallic silicon substrate (see Fig. \ref{fig06}).
This sandwich structure, with the metallic silicon layer that serves as a backgate, may thus be
viewed as a capacitor (Fig. \ref{fig06}) the capacitance of which is 
\beq\label{condenser}
C=\frac{Q}{V_G}=\frac{\epsilon_0\epsilon A}{d},
\eeq
where $Q=en_{2D}A$ is the capacitor charge, in terms of the total surface
$A$, $V_G$ is the gate voltage, and $d=300$ nm is the thickness of the SiO$_2$ layer
with the dielectric constant $\epsilon=3.7$. 
The field-effect induced 2D carrier density is thus given by
\beq\label{InducedDens}
n_{2D}=\alpha V_G \qquad {\rm with} \qquad \alpha\equiv\frac{\epsilon_0\epsilon}{ed}
\simeq 7.2\times 10^{10} \frac{{\rm cm}^{-2}}{\rm V}.
\eeq
The gate voltage may vary roughly between $-100$ and $100$ V, such that one may
induce maximal carrier densities on the order of $10^{12}$ cm$^{-2}$, on top 
of the intrinsic carrier density which turns out to be zero in graphene, as 
will be discussed in the next chapter. At gate voltages above $\pm 100$ V, the
capacitor breaks down (electrical breakdown).

In contrast to 2D electron gases in semiconductor heterostructures, the
mobilities achieved in graphene are rather low: they are typically on the order of
$\mu\sim 10^{4}-10^{5}$ cm$^2$/Vs. Notice, however, that these graphene samples are fabricated 
in the so-called exfoliation technique, where one ``peals'' thin graphite crystals,
under ambiant condictions, whereas the highest-mobility GaAs/AlGaAs laboratory samples are
fabricated with a very high technological effort. The mobilities of graphene samples
are comparable to those of commercial silicon-based electronic elements.

\chapter{Landau Quantisation}
\label{Landau}

\markboth{Landau Quantisation}{Landau Quantisation}

The basic ingredient for the understanding of both the IQHE and the FQHE is Landau quantisation,
i.e. the kinetic-energy quantisation of a (free) charged 2D particle in a perpendicular magnetic
field. In this chapter, we give a detailed introduction to the different aspects of Landau
quantisation. We have chosen a very general presentation of this quantisation in order to account 
for both a non-relativistic and a relativistic 2D particle some properties of which, such as
the level degeneracy, are identical. In Sec. \ref{zeroB}, we introduce the basic Hamiltonians for 
2D particles in the absence of a magnetic field and discuss both Schr\"odinger- and Dirac-type
particles, and discuss the case of a non-zero $B$-field in Sec. \ref{B}. Sec. \ref{LL} is devoted
to the discussion of the LL structure of non-relativistic and relativistic particles.

\section{Basic One-Particle Hamiltonians for $B=0$}
\label{zeroB}

\markboth{Landau Quantisation}{Basic One-Particle Hamiltonians for $B=0$}

In this section, we introduce the basic Hamiltonians which we treat in a quantum-mechanical manner in the following
parts. Quite generally, we consider a Hamiltonian for a 2D particle\footnote{All vector quantities (also in the quantum-mechanical
case of operators) ${\bf v}=(v_x,v_y)$ are hence 2D, unless stated explicitly.}
that is translation invariant, i.e. the momentum $\bp=(p_x,p_y)$ is a constant of motion, in the absence of a magnetic field.
In quantum mechanics, this means that the momentum operator commutes with the Hamiltonian, $[\bp,H]=0$, and that the eigenvalue of 
the momentum operator is a good quantum number.

\subsection{Hamiltonian of a free particle}

In the case of a free particle, this is a very natural assumption, and one has for the non-relativistic case,
\beq\label{free}
H = \frac{\bp^2}{2m},
\eeq
in terms of the particle mass $m$.\footnote{The statement that $\bp$ is a constant of motion 
remains valid also in the case of a relativistic particle. However,
the Hamiltonian description depends on the frame of reference because the energy is not Lorentz-invariant, i.e. invariant under
a transformation into another frame of reference that moves at constant velocity with respect to the first one.
For this reason a Lagrangian rather than a Hamiltonian formalism is often prefered in relativistic quantum mechanics.}
However, we are interested, here, in the motion of electrons in some material (in a metal or at the interface of to semiconductors).
It seems, at first sight, to be a very crude assumption to describe the motion of an electron in a crystalline environment in 
the same manner as a particle in free space. Indeed, a particle in a lattice in not described by the Hamiltonian (\ref{free}) but rather
by the Hamiltonian
\beq\label{LatticeHam}
H=\frac{\bp^2}{2m} + \sum_i^N V(\br-\br_i),
\eeq
where the last term represents the electrostatic potential caused by the ions situated at the lattice sites $\br_i$. Evidently, 
the Hamiltonian now depends on the position $\br$ of the particle with respect to that of the ions, and the momentum $\bp$ is 
therefore no longer a constant of motion or a good quantum number. 

This problem is solved with the help of Bloch's theorem: although an arbitrary spatial translation is not an allowed symmetry
operation as it is the case for a free particle (\ref{free}), the system is invariant under a translation by an arbitrary
lattice vector if the lattice is of infinite extension -- an assumption we make here.\footnote{Although this may seem to be a typical
``theoretician's assumption'', it is a very good approximation when the lattice size is much larger than all other relevant
length scales, such as the lattice spacing or the Fermi wave length.}
In the same manner as for the free particle, where one defines the momentum as the generator of a spatial translation, one may then
define a generator of a lattice translation. This generator is called the {\sl lattice momentum} or also the {\sl quasi-momentum}. 
As a consequence of the discreteness of the lattice translations, not all values of this lattice momentum are physical, but only those
within the first Brillouin zone (BZ) -- any vibrational mode, be it a lattice vibration or an electronic wave, with a wave vector outside
the first BZ can be described by a mode with a wave vector within the first BZ. Since these lecture notes cannot include a full
course on basic solid-state physics, we refer the reader to standard textbooks on solid-state physics \cite{AM,kittel}.

The bottom line is that also in a (perfect) crystal, the electrons may be described in terms of a Hamiltonian 
$H(p_x,p_y)$ if one keeps in mind that the momentum $\bp$ in this expression is a lattice momentum restricted to the 
first BZ. Notice, however, that although the resulting Hamiltonian
may often be written in the form (\ref{free}), the mass is generally not the free electron mass but a {\sl band mass} $m_b$ 
that takes into account the particular features of the energy bands\footnote{In GaAs, e.g., the band mass is 
$m_b=0.068 m_0$, in terms of the free electron
mass $m_0$. } -- indeed, the mass may even depend on the direction of propagation, such
that one should write the Hamiltonian more generally as 
$$
H=\frac{p_x^2}{2m_x} + \frac{p_y^2}{2m_y}\ .
$$

\subsection{Dirac Hamiltonian in graphene}

\begin{figure}
\begin{center}
\epsfig{figure=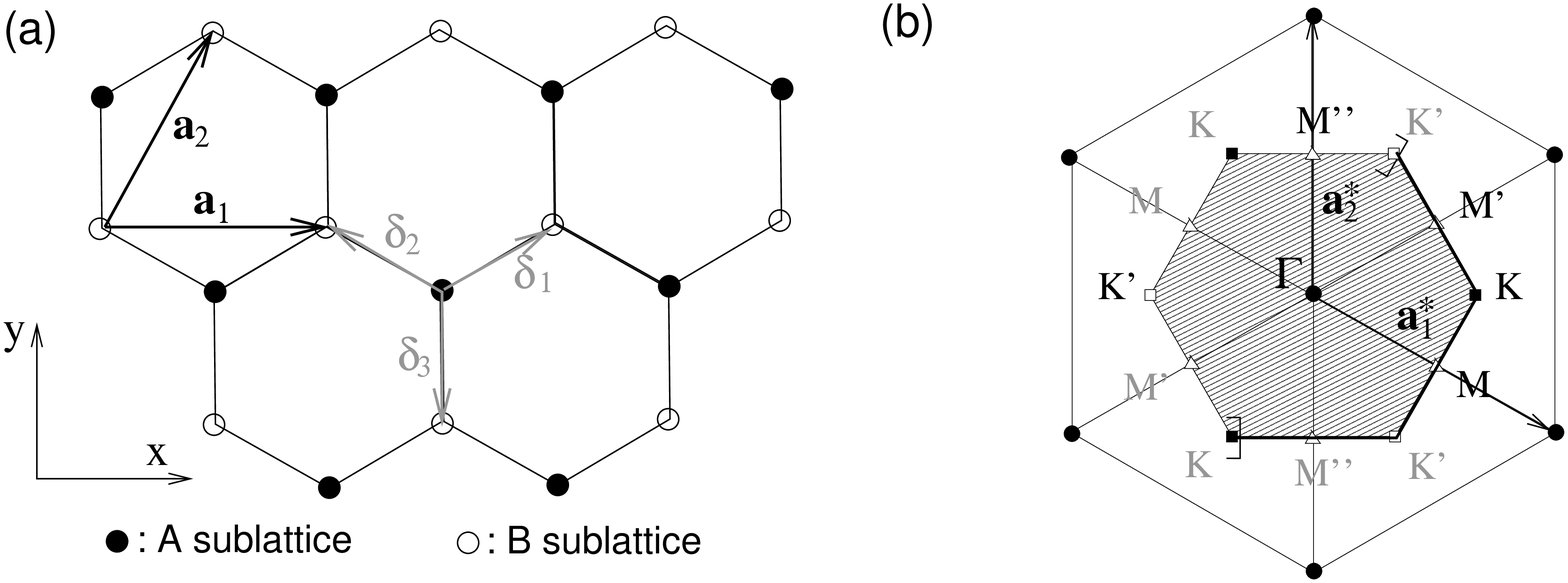,width=12cm,clip}
\end{center}
\caption{ {\sl (a)} Honeycomb lattice. 
The vectors $\deltab_1$, $\deltab_2$,
and $\deltab_3$ connect {\sl nn} carbon atoms, separated by a distance $a=0.142$ nm.
The vectors $\ba_1$ and $\ba_2$ are basis vectors of the triangular Bravais
lattice. 
{\sl (b)} Reciprocal lattice of the triangular lattice. Its primitive lattice
vectors are $\ba_1^*$ and $\ba_2^*$. The shaded region represents the first Brillouin
zone (BZ), with its centre $\Gamma$ and 
the two inequivalent corners $K$ (black squares)
and $K'$ (white squares). The thick part of the border of the first BZ represents
those points which are counted in the definition such that no points are
doubly counted. The first BZ, defined in a strict manner, is, thus, the shaded 
region plus the thick part of the border. For completeness, we have also
shown the three inequivalent cristallographic points $M$, $M'$, and $M''$
(white triangles).}
\label{fig07}
\end{figure}

The above considerations for electrons in a 2D lattice are only valid in the case of a {\sl Bravais} lattice, i.e. a lattice 
in which all lattice sites are equivalent from a crystallographic point of view. However, 
some lattices, such as the honeycomb lattice that 
describes the arrangement of carbon atoms in graphene due to the sp$^2$ hybridisation of the valence electrons, are not 
Bravais lattices. In this case, one may describe the lattice as a Bravais lattice plus a particular pattern of $N_s$ sites,
called the {\sl basis}. This is illustrated in Fig. \ref{fig07}(a) for the case of the honeycomb lattice. When one compares 
a site A (full circle) with a site B (empty circle), one notices that the environment of these two sites is different:
whereas a site A has nearest neighbours in the directions north-east, north-west and south, a site B has nearest neighbours
in the directions north, south-west and south-east. This precisely means that the two sites are not equivalent from a crystallographic
point of view -- although they may be equivalent from a chemical point of view, i.e. occupied by the same atom or ion type
(carbon in the case of graphene). However, all sites A form a triangular Bravais lattice as well as all sites B. Both subsets
of lattice sites form the two {\sl sublattices}, and the honeycomb lattice may thus be viewed as a triangular Bravais lattice 
with a two-atom basis, e.g. the pattern of two A and B sites connected by the vector $\deltab_3$. 

\begin{figure}
\begin{center}
\epsfig{figure=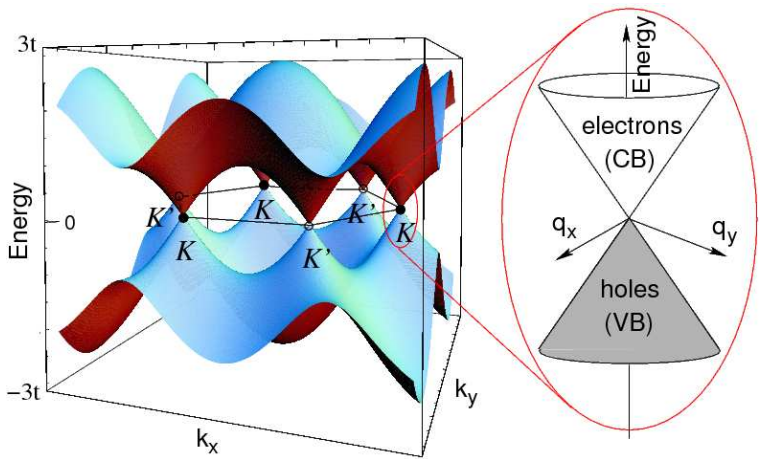,width=10cm,clip}
\end{center}
\caption{ Energy bands of graphene. The valence band touches the conduction band in the two inequivalent BZ corners $K$ and
$K'$. For undoped graphene, the Fermi energy lies precisely in the contact points, and the band dispersion in the vicinity of these
points is of conical shape.}
\label{fig08}
\end{figure}

In order to calculate the electronic bands in a lattice with $N_s$ Bravais sublattices, i.e. a basis with $N_s$ sites, one needs to
describe the general electronic wave function as a superposition of $N_s$ different wave functions, which satisfy each Bloch's 
theorem for all sublattices \cite{AM,kittel}. 
Formally, this may be described in terms of a $N_s\times N_s$ matrix, the eigenvalues of which yield
$N_s$ different energy bands. In a lattice with $N_s$ 
different sublattices, one therefore obtains one energy band per sublattice, and for 
graphene, one obtains two different bands for the conducting electrons, the valence band and the conduction band. 

The Hamiltonian for low-energy electrons in reciprocal space reads
\beq\label{raw}
H(\bk)=t\left(\begin{array}{cc}
0 & \gamma_{\bk}^* \\ \gamma_{\bk} & 0
        \end{array}\right),
\eeq
which is obtained within a tight-binding model, where one considers electronic hopping between nearest-neighbouring sites
with a hopping amplitude $t$. Because
the nearest neighbour of a site A is a site B and {\sl vice versa} [see Fig. \ref{fig07}(a)], the Hamiltonian is off-diagonal, 
and the off-diagonal elements are related by complex conjugation due to time-reversal symmetry [$H(-\bk)^*=H(\bk)$]. As already 
mentioned above, the lattice momentum $\bk$ is restricted to the first BZ, which is of hexagonal shape and which we have depicted in 
Fig. \ref{fig07}(b) for completeness. The precise form of the functions $\gamma_{\bk}$ is derived in Appendix \ref{TBgraphene}
[Eq. (\ref{eq2:18})]. The 
band structure is obtained by diagonalising the Hamiltonian, and one finds the two bands, labelled by $\lambda=\pm$, 
$\epsilon_{\lambda}(\bk)=\lambda t|\gamma_{\bk}|$, which are plotted in Fig. \ref{fig08}. 
The valence band ($\lambda=-$) touches the conduction band ($\lambda=+$)
in the two inequivalent corners $K$ and $K'$ of the first BZ. Because there are as many electrons in the $\pi$-orbitals, that determine
the low-energy conduction properties of graphene, as lattice sites, the overall energy band structure is half-filled. This is due to
the two spin orientations of the electrons, which allow for a quantum-mechanical double occupancy of each $\pi$-orbital. As a consequence,
the Fermi energy lies exactly in the contact points $K$ and $K'$ of the two bands unless the graphene sheet is doped, e.g. with the
help of the electric field effect, as described in Sec. \ref{2DEG} of the previous chapter.

The inset in Fig. \ref{fig08} shows the band dispersion in the vicinity of the contact points $K$ and $K'$, the linearity of 
which is sufficient to
describe the low-energy electronic properties in graphene, i.e. when all relevant energy scales are much smaller than the full
band width.\footnote{Indeed, in graphene, the relevant low-energy scales are in the $10-100$ meV regime, whereas non-linear corrections
of the band dispersion become relevant in the eV regime.}
The conical form of the two bands is reminiscent of that of relativistic particles, the general dispersion of which
is $E=\pm\sqrt{m^2c^4 + \bp^2 c^2}$, in terms of the light velocity $c$ and the particle mass $m$. If the latter is zero, one obtains
precisely $E=\pm c|\bp|$, as in the case of low-energy electrons in graphene (inset of Fig. \ref{fig08}),
which may thus be treated as {\sl massless Dirac fermions}. Notice that in the continuum description of electrons in graphene,
we have two electron types -- one for the $K$ point and another one for the $K'$ point. This doubling is called 
{\sl valley degeneracy} which is two-fold here.

The analogy between electrons in graphene and 
massless relativistic particles is corroborated by a low-energy expansion of the Hamiltonian (\ref{raw}) around
the contact points $K$ and $K'$, at the momenta $\bK$ and $\bK'=-\bK$ [see Fig. \ref{fig07}(a)], $\bk=\pm \bK +\bp/\hbar$, 
where $|\bp/\hbar|\ll |\bK|$. One may then expand the function $\gamma_{\pm \bK+\bp/\hbar}$ to first order, and one 
obtains formally\footnote{Notice that $\gamma_{\pm\bK}=0$ by symmetry.}
$$
H=t\left(\begin{array}{cc}
0 & \nabla \gamma_{\bK}^*\cdot\bp \\ \nabla \gamma_{\bK}\cdot\bp & 0
        \end{array}\right) = 
v \left(\begin{array}{cc}
0 & p_x- i p_y \\ p_x+ i p_y & 0
        \end{array}\right) = v \bp\cdot \sigmab
$$
where $\sigmab=(\sigma^x,\sigma^y)$ in terms of the Pauli matrices
$$\sigma^x=\left(\begin{array}{cc} 0 & 1 \\ 1 & 0 \end{array}\right) , \qquad 
\sigma^y=\left(\begin{array}{cc} 0 & -i \\ i & 0 \end{array}\right) \qquad {\rm and} \qquad
\sigma^z=\left(\begin{array}{cc} 1 & 0\\ 0 & -1 \end{array}\right)
$$
and where we have chosen to expand the 
Hamiltonian (\ref{raw}) around the $K$ point.\footnote{One obtains a similar result at the $K'$ point, see Eq. (\ref{DirHamKp}) 
in Appendix \ref{TBgraphene}.} Here, the Fermi velocity
$v$ plays the role of the velocity of light $c$, which is though roughly 300 times larger, $c\simeq 300 v$. 
The details of the above derivation
may be found in Appendix \ref{TBgraphene}. The above Hamiltonian is indeed formally that of massless 2D particles, and it is
sometimes called Weyl or Dirac Hamiltonian.

We will discuss, in the remainder of this chapter, how the two Hamiltonians
\beq\label{0BHams}
H_S=\frac{\bp^2}{2m_b} \qquad {\rm and}\qquad H_D = v\bp\cdot\sigmab\ ,
\eeq
for non-relativistic and relativistic particles, respectively,
need to be modified in order to account for a non-zero magnetic field.

\section{Hamiltonians for Non-Zero $B$ Fields}
\label{B}

\markboth{Landau Quantisation}{Hamiltonians for Non-Zero $B$ Fields}

\subsection{Minimal coupling and Peierls substitution}

In order to describe free electrons in a magnetic field, one needs to replace the momentum by its gauge-invariant form \cite{jackson}
\beq\label{mom}
\bp \rightarrow \Pib = \bp + e\bA(\br), 
\eeq
where $\bA(\br)$ is the vector potential that generates the magnetic field $\bB=\nabla\times \bA(\br)$. This 
gauge-invariant momentum is proportional the electron velocity ${\bf v}$, which must naturally be 
gauge-invariant because it is a physical quantity.
Notice that because $\bA(\br)$ is not gauge invariant,
neither is the momentum $\bp$. Remember that adding the gradiant of an arbitrary 
derivable function $\lambda(\br)$,
$\bA(\br)\rightarrow \bA(\br) + \nabla \lambda(\br)$, does not change the magnetic field because the rotational of a gradient is zero.
Indeed, the momentum transforms as $\bp\rightarrow \bp - e\nabla \lambda(\br)$ 
under a gauge transformation in order to compensate the transformed vector potential, such that $\Pib$ is gauge-invariant. 
The substitution (\ref{mom}) is also called {\sl minimal substitution}.

In the case of electrons on a lattice, this substitution is more tricky because of the presence of several bands. Furthermore,
the vector potential is unbound, even for a finite magnetic field; this becomes clear if one chooses a particular gauge, such
as e.g. the Landau gauge 
$\bA_L(\br)=B(-y,0,0)$, in which case the value of the vector potential may become as large as $B\times L_y$, where 
$L_y$ is the macroscopic extension of the system in the $y$-direction. However, it may be shown that the substitution (\ref{mom}), which
is called {\sl Peierls substitution} in the context of electrons on a lattice, remains correct as long as the lattice spacing 
$a$ is much smaller than the {\sl magnetic length}
\beq\label{lB}
l_B = \sqrt{\frac{\hbar}{eB}}\ ,
\eeq
which is the fundamental length scale in the presence of a magnetic field. Because $a$ is typically an atomic scale ($\sim 0.1$ to 10 nm)
and $l_B\simeq 26\, {\rm nm}/\sqrt{B{\rm [T]}}$, this condition is fulfilled in all atomic lattices for the magnetic fields, which may
be achieved in today's high-field laboratories ($\sim 45$ T in the continuous regime and $\sim 80$ T in the pulsed 
regime).\footnote{Higher magnetic fields may be obtained only in {\sl semi-destructive experiments}, in which the sample survives
the experiment but not the coil that is used to produce the magnetic field.}

With the help of the (Peierls) substitution (\ref{mom}), one may thus immediately write down the Hamiltonian for charged particles in 
a magnetic field if one knows the Hamiltonian in the absence of a magnetic field,
$$
H(\bp) \rightarrow H(\Pib)  = H(\bp + e\bA)= H^B(\bp,\br). 
$$
Notice that because of the spatial dependence of the vector potential, the resulting Hamiltonian is no longer translation invariant, 
and the (gauge-dependent) momentum $\bp$ is no longer a conserved quantity. We will limit the discussion to the $B$-field Hamiltonians
corresponding to the Hamiltonians (\ref{0BHams})
\beq\label{BHamS}
H_S^B = \frac{[\bp + e\bA(\br)]^2}{2m_b}
\eeq
for non-relativistic and
\beq\label{BHamD}
H_D^B = v[\bp +e \bA(\br)]\cdot \sigmab
\eeq
for relativistic 2D charged particles, respectively.

\subsection{Quantum mechanical treatment}

In order to analyse the one-particle Hamiltonians (\ref{BHamS}) and (\ref{BHamD}) in a quantum-mechanical treatment, we use the 
standard method, the {\sl canonical quantisation}
\cite{CT}, where one interprets the physical quantities as
operators that act on state vectors in a Hilbert space. These operators do in general not commute with each other, i.e. the 
order matters in which they act on the state vector that describe the physical system. Formally one introduces the {\sl commutator}
$[\Omath_1,\Omath_2]\equiv \Omath_1\Omath_2 - \Omath_2\Omath_1$ between the two operators $\Omath_1$ and $\Omath_2$, which are said to 
commute when $[\Omath_1,\Omath_2]=0$ or else not to commute.
The basic physical quantities in the argument of the Hamiltonian are the 2D position $\br=(x,y)$ and 
its canonical momenta $\bp=(p_x,p_y)$, which satisfy the commutation relations
\beq\label{canQ}
[x,p_x]=i\hbar, \qquad [y,p_y]=i\hbar \qquad {\rm and} \qquad [x,y]=[p_x,p_y]=[x,p_y]=[y,p_x]=0,
\eeq
i.e. each component of the position operator does not commute with the momentum in the corresponding direction.
This non-commutativity between the position and its associated momentum is the origin of the Heisenberg inequality according
to which one cannot know precisely both the position of a quantum-mechanical particle and, at the same moment, its momentum,
$\Delta x\Delta p_x\gtrsim h$ and $\Delta y\Delta p_y\gtrsim h$.

As a consequence of the commutation relations (\ref{canQ}), the components of the gauge-invariant momentum no longer commute
themselves,
\beqn 
\nn
\left[\Pi_x,\Pi_y\right] &=& \left[p_x + eA_x(\br), p_y + eA_y(\br)\right] = e\left(\left[p_x,A_y\right] - \left[p_y,A_x\right]\right)\\
\nn
&=& e\left(\frac{\partial A_y}{\partial x}[p_x,x] + \frac{\partial A_y}{\partial y}[p_x,y] -
\frac{\partial A_x}{\partial x}[p_y,x] - \frac{\partial A_x}{\partial y}[p_y,y]\right),
\eeqn
where we have used the relation\footnote{More precisely we have used a 
gradient generalisation of this relation to operator functions that
depend on several different operators,
$$[\Omath_0,f(\Omath_1,...,\Omath_J)] = \sum_{j=1}^J\frac{\partial f}{\partial \Omath_j} [\Omath_0,\Omath_j]
$$
which is valid if $[[\Omath_0,\Omath_j],\Omath_0]=[[\Omath_0,\Omath_j],\Omath_j]=0$ for all $j=1,...,N$.} 
\beq\label{Haus}
[\Omath_1,f(\Omath_2)] = \frac{d f}{d \Omath_2} [\Omath_1,\Omath_2]
\eeq
between two arbitrary operators, the commutator of which is a c-number or an operator that commutes itself with both 
$\Omath_1$ and $\Omath_2$
\cite{CT}. With the help of the commutation relations (\ref{canQ}), one finds that
$$
\left[\Pi_x,\Pi_y\right] = -ie\hbar \left(\frac{\partial A_y}{\partial x} - \frac{\partial A_x}{\partial y}\right) = 
-ie\hbar \left(\nabla\times \bA\right)_z = -i e \hbar B,
$$
and, in terms of the magnetic length (\ref{lB}),
\beq\label{ComMom}
\left[\Pi_x,\Pi_y\right] = -i \frac{\hbar^2}{l_B^2}\ .
\eeq
This equation is the basic result of this section and merits some further discussion.
\begin{itemize}
\item As one would have expected for gauge-invariant quantities (the two components of $\Pib$), their commutator is itself
gauge-invariant. Indeed, it only depends on universal constants and the (gauge-invariant) magnetic field $B$, and not
on the vector potential $\bA$.
\item The components of the gauge-invariant momentum $\Pib$ are mutually {\sl conjugate} in the same manner as $x$ and $p_x$ or
$y$ and $p_y$. Remember that $p_x$ generates the translations in the $x$-direction (and $p_y$ those in the $y$-direction). This is
similar here: $\Pi_x$ generates a ``boost'' of the gauge-invariant momentum in the $y$-direction, and similarly $\Pi_y$ one in
the $x$-direction. 
\item As a consequence, one may not diagonalise at the same time $\Pi_x$ {\sl and} $\Pi_y$, in contrast to the zero-field case, where
the arguments of the Hamiltonian, $p_x$ and $p_y$, commute. 
\end{itemize}

For solving the Hamiltonians (\ref{BHamS}) and (\ref{BHamD}), it is convenient to use the pair of conjugate operators $\Pi_x$ 
and $\Pi_y$ to introduce {\sl ladder operators} in the same manner as in the quantum-mechanical treatment of the one-dimensional
harmonic oscillator. Remember from your basic quantum-mechanics class that the ladder operators may be viewed as the complex 
position of the one-dimensional oscillator in the phase space, which is spanned by the position ($x$-axis) and the momentum ($y$-axis),
$$\tilde{a}=\frac{1}{\sqrt{2}}\left(\frac{x}{x_0} - i \frac{p}{p_0}\right) \qquad {\rm and} \qquad
\tilde{a}^{\dagger}=\frac{1}{\sqrt{2}}\left(\frac{x}{x_0} + i \frac{p}{p_0}\right),$$
where $x_0=\sqrt{\hbar/m_b\omega}$ and $p_0=\sqrt{\hbar m_b\omega}$ are normalisation constants in terms of the oscillator frequency
$\omega$ \cite{CT}. The fact that the position $x$ and the momentum $p$ are conjugate variables and the particular choice of the
normalisation constants yields the commutation relation $[\tilde{a},\tilde{a}^{\dagger}]=1$ for the ladder operators.

In the case of the 2D electron in a magnetic field, the ladder operators play the role of a {\sl complex} gauge-invariant 
momentum (or velocity), and they read
\beq\label{ladder}
a = \frac{l_B}{\sqrt{2}\hbar}\left(\Pi_x - i\Pi_y\right) \qquad {\rm and} \qquad
a^{\dagger} = \frac{l_B}{\sqrt{2}\hbar}\left(\Pi_x + i\Pi_y\right),
\eeq
where we have chosen the appropriate normalisation such as to obtain the usual commutation relation
\beq\label{ComLad}
[a,a^{\dagger}]=1.
\eeq
It turns out to be helpful for future calculations to invert the expression for the ladder operators (\ref{ladder}),
\beq\label{ladder1}
\Pi_x = \frac{\hbar}{\sqrt{2}l_B}\left(a^{\dagger}+a\right) \qquad {\rm and} \qquad
\Pi_y = \frac{\hbar}{i\sqrt{2}l_B}\left(a^{\dagger}-a\right).
\eeq

\section{Landau Levels}
\label{LL}

\markboth{Landau Quantisation}{Landau Levels}

The considerations of the preceding section are extremely useful in the calculation of the level spectrum associated with 
the Hamiltonians (\ref{BHamS}) and (\ref{BHamD}) of both the non-relativistic and the relativistic particles, respectively.
The understanding of this level spectrum is the issue of the present section. Because electrons do not only possess a charge
but also a spin, each level is split into two spin branches separated by the energy difference $\Delta_Z\epsilon = g \mu_B B$,
where $g$ is the $g$-factor of the host material and $\mu_B = e\hbar/2m_0$ the Bohr magneton. In order to simplify the following
presentation of the quantum-mechanical treatment and the level structure, we neglect this effect associated with the spin
degree of freedom. Formally, this amounts to considering {\sl spinless fermions}. Notice, however, that there exist
interesting physical properties related to the spin degree of freedom, which will be treated separately in Chap. \ref{MultiC}.

\subsection{Non-relativistic Landau levels}

In terms of the gauge-invariant momentum, the Hamiltonian (\ref{BHamS}) for non-relativistic electrons reads
$$H_S^B=\frac{1}{2m_b}\left(\Pi_x^2 + \Pi_y^2\right).$$
The analogy with the one-dimensional harmonic oscillator is apparent if one notices that both conjugate operators $\Pi_x$ and
$\Pi_y$ occur in this expression in a quadratic form. If one replaces these operators with the ladder operators (\ref{ladder1}),
one obtains, with the help of the commutation relation (\ref{ComLad}),
\beqn \label{HamLadS}
\nn
H_S^B &=& \frac{\hbar^2}{4ml_B^2}\left[a^{\dagger 2} + a^{\dagger}a + aa^{\dagger} + a^2 -\left(a^{\dagger 2} - a^{\dagger}a - 
aa^{\dagger} + a^2\right)\right] \\
\nn
&=& \frac{\hbar^2}{2m l_B^2} \left(a^{\dagger}a + aa^{\dagger}\right) = \frac{\hbar^2}{ml_B^2}\left(a^{\dagger}a + \frac{1}{2}\right)\\
&=& \hbar\omega_C\left(a^{\dagger}a + \frac{1}{2}\right),
\eeqn
where we have used the relation $\omega_c=\hbar/m_bl_B^2$ between the cyclotron frequency (\ref{cycl})
and the magnetic length (\ref{lB}) in the last step.

As in the case of the one-dimensional harmonic oscillator, the eigenvalues and eigenstates of the Hamiltonian (\ref{HamLadS})
are therefore those of the {\sl number operator} $a^{\dagger}a$, with
$a^{\dagger}a |n\rangle = n |n\rangle$. The ladder operators act on these states in the usual manner \cite{CT}
\beq\label{nlad}
a^{\dagger}|n\rangle = \sqrt{n+1}|n+1\rangle \qquad {\rm and} \qquad
a|n\rangle = \sqrt{n} |n-1\rangle,
\eeq
where the last equation is valid only for $n>0$ -- the action of $a$ on the ground state $|0\rangle$ gives zero, 
\beq\label{0lad}
a|0\rangle = 0.
\eeq
This last equation turns out to be helpful in the calculation of the eigenstates associated with the level of lowest energy, 
as well as the construction of states in higher levels $n$ (see Sec. \ref{WFsym})
\beq\label{constrN}
|n\rangle = \frac{\left(a^{\dagger}\right)^n}{\sqrt{n!}} |0\rangle.
\eeq

\begin{figure}
\begin{center}
\epsfig{figure=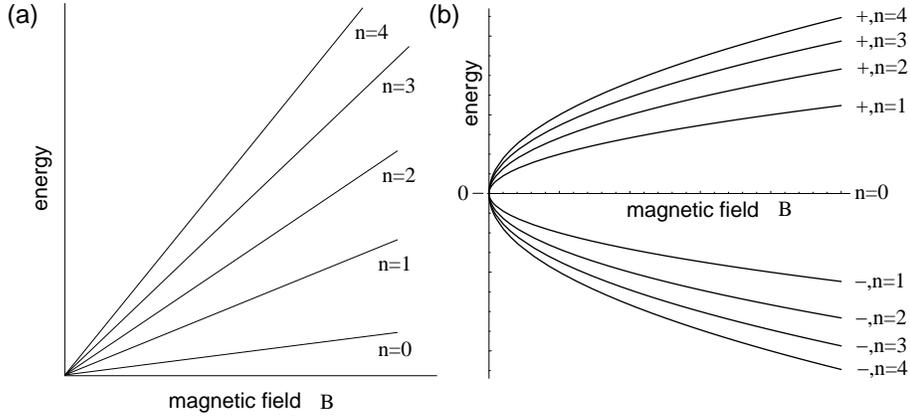,width=12cm,clip}
\end{center}
\caption{ Landau levels as a function of the magnetic field. {\sl (a)} Non-relativistic case with $\epsilon_n=\hbar\omega_C(n+1/2)\propto 
B(n+1/2)$. {\sl (b)} Relativistic case with $\epsilon_{\lambda,n}=\lambda(\hbar v/l_B)\sqrt{2n}\propto \lambda \sqrt{Bn}$.}
\label{fig09}
\end{figure}

The energy levels of the 2D charged non-relativistic particle are therefore discrete and labelled by the integer $n$,
\beq\label{Llevels}
\epsilon_n = \hbar\omega_C\left(n + \frac{1}{2}\right).
\eeq
These levels, which are also called {\sl Landau levels} (LL), are depicted in Fig. \ref{fig09}(a) as a function of the magnetic 
field. Because of the linear field-dependence of the cyclotron frequency, the LLs disperse linearly themselves with the magnetic field.

\subsection{Relativistic Landau levels}
\label{RelLLsec}

The relativistic case (\ref{BHamD}) for electrons in graphene
may be treated exactly in the same manner as the non-relativistic one. In terms of the 
ladder operators (\ref{ladder}), the Hamiltonian reads
\beq\label{HamLadD}
H_D^B= v \left(\begin{array}{cc}
0 & \Pi_x- i \Pi_y \\ \Pi_x+ i \Pi_y & 0
        \end{array}\right) = \sqrt{2}\frac{\hbar v}{l_B}\left(\begin{array}{cc}
0 & a \\ a^{\dagger} & 0
        \end{array}\right) .
\eeq
One notices the occurence of a characteristic frequency $\omega'=\sqrt{2}v/l_B$, which plays the role of the cyclotron frequency
in the relativistic case. Notice, however, that this frequency may not be written in the form $eB/m_b$ because the band mass 
is strictly zero in graphene, such that the frequency would diverge.\footnote{Sometimes, a {\sl cyclotron mass} $m_C$ 
is formally introduced via the equality $\omega'\equiv eB/m_C$.
However, this mass is a somewhat artificial quantity, which turns out to depend on the 
carrier density. We will therefore not use this quantity in the present lecture notes.}

In order to obtain the eigenvalues and the eigenstates of the Hamiltonian (\ref{HamLadD}), one needs to solve the eigenvalue equation
$H_D^B\psi_n = \epsilon_n \psi_n$. Because the Hamiltonian is a $2\times 2$ matrix, the eigenstates are 2-spinors, 
$$\psi_n=\left(\begin{array}{c} u_n \\ v_n \end{array} \right),$$
and we thus need to solve the system of equations
\beq\label{eigen}
\hbar\omega'a\, v_n = \epsilon_n \, u_n \qquad {\rm and} \qquad \hbar\omega' a^{\dagger}\, u_n= \epsilon_n\, v_n\ ,
\eeq
which yields the equation 
\beq\label{eigen2}
a^{\dagger}a\, v_n = \left(\frac{\epsilon_n}{\hbar\omega'}\right)^2 v_n
\eeq
for the second spinor component. One notices that this component is an eigenstate of the number operator
$n=a^{\dagger}a$, which we have already encountered in the preceding subsection. We may therefore identify, up to a 
numerical factor, the second spinor component $v_n$ with the eigenstate 
$|n\rangle$ of the {\sl non-relativistic} Hamiltonian (\ref{HamLadS}),
$v_n\sim |n\rangle$. Furthermore, one observes that the square of the energy is proportional to this quantum number,
$\epsilon_n^2 = (\hbar\omega')^2 n$. This equation has two solutions, a positive and a negative one, and one needs to
introduce another quantum number $\lambda=\pm$, which labels the states of positive and negative energy, respectively.
This quantum number plays the same role as the band index ($\lambda=+$ for the conduction and $\lambda=-$ for the
valence band) in the zero-$B$-field case discussed in Sec. \ref{zeroB}. One thus obtains the level spectrum \cite{mcclure}
\beq\label{RelLLs}
\epsilon_{\lambda,n} = \lambda \frac{\hbar v}{l_B}\sqrt{2n}
\eeq
the energy levels of which are depicted in Fig. \ref{fig09}(b). These {\sl relativistic Landau
levels} disperse as $\lambda\sqrt{Bn}$ as a function of the magnetic field.

Once we know the second spinor component, the first spinor component is obtained from Eq. (\ref{eigen}), which
reads $u_n\propto a\,v_n\sim a|n\rangle \sim |n-1\rangle$. One then needs to distinguish the zero-energy LL
($n=0$) from all other levels. Indeed, for $n=0$, the first component is zero as one may see from Eq. (\ref{0lad}). In this
case one obtains the spinor 
\beq\label{spinN0}
\psi_{n=0} = \left(\begin{array}{c} 0 \\ |n=0\rangle  \end{array}\right).
\eeq
In all other cases ($n\neq 0$), one has positive and negative energy solutions, which differ among each other by a relative sign
in one of the components. A convenient representation of the associated spinors is given by
\beq\label{spinN}
\psi_{\lambda,n\neq 0} = \frac{1}{\sqrt{2}}\left(\begin{array}{c} |n-1\rangle \\ \lambda |n\rangle  \end{array}\right).
\eeq

\subsubsection{Experimental observation of relativistic Landau levels}

\begin{figure}
\begin{center}
\epsfig{figure=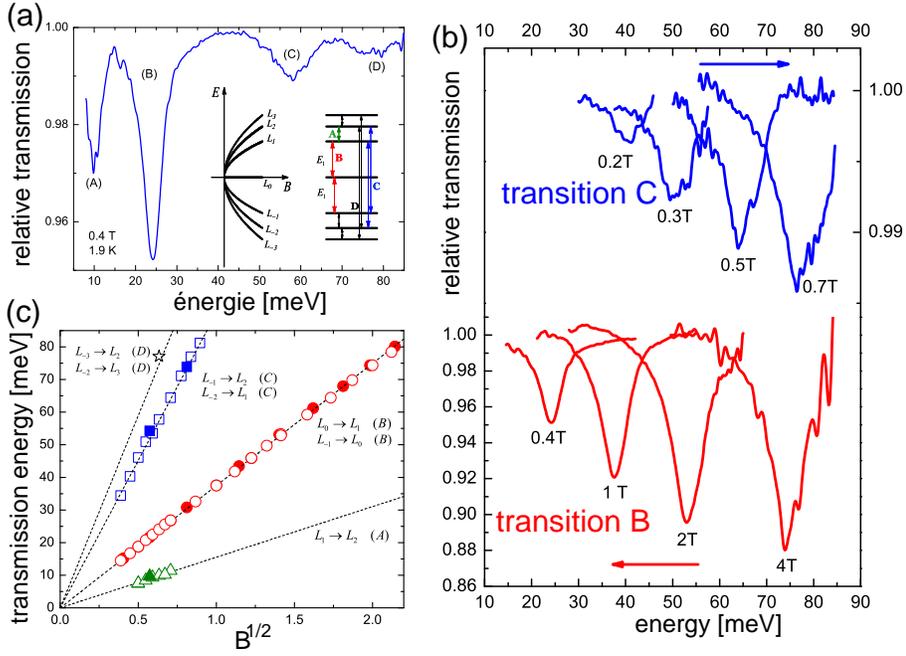,width=12cm,clip}
\end{center}
\caption{ LL spectroscopy in graphene (from Sadowski {\sl et al.}, 2006). {\sl (a)} For a fixed magnetic field (0.4 T), one
observes resonances in the transmission spectrum as a function of the irradiation energy. The resonances are associated
with allowed dipolar transitions between relativistic LLs. {\sl (b)} These resonances are shifted as a function of the
magnetic field. {\sl (c)} If one plots the resonance energies as a function of the square root of the magnetic field, $\sqrt{B}$,
a linear dependence is observed as one would expect for relativistic LLs. }
\label{fig09bis}
\end{figure}

Relativistic LLs have been observed experimentally in transmission spectroscopy, where one shines light on the sample and measures
the intensity of the transmitted light. Such experiments have been performed on so-called epitaxial graphene\footnote{Epitaxial
graphene is obtained from a thermal graphitisation process of an epitaxially grown SiC crystal \cite{berger}}
\cite{sadowski} and later on exfoliated graphene \cite{jiang}.
When the monochromatic light is in resonance with a dipole-allowed transition from the (partially) filled LL $(\lambda,n)$ 
to the (partially) unoccupied LL $(\lambda',n\pm 1)$,
the light is absorbed due to an electronic excitation between the two levels
[see Fig. \ref{fig09bis}(a)]. Notice that, in a non-relativistic 2D electron gas,
the only allowed dipolar transition is that from the last occupied LL $n$ to the first unoccupied one $n+1$. The transition
energy is $\hbar\omega_C$, independently of $n$, and one therefore observes a single absorbtion line (cyclotron resonance).
In graphene, however, there are many more allowed transitions due to the presence 
of two electronic bands, the conduction and the valence band, and the transitions have the energies
$$\Delta_{n,\xi}=\frac{\hbar v}{l_B}\left[\sqrt{2(n+1)}-\xi \sqrt{2n}\right],$$
where $\xi=+$ denotes an intraband and $\xi=-$ an interband transition. One therefore obtains families of resonances
the energy of which disperses as $\Delta_{n,\xi}\propto \sqrt{B}$, as it has been observed in the experiments  
[see Fig. \ref{fig09bis}(c), where we show the results from Sadowski {\sl et al.} \cite{sadowski}]. 
Notice that the dashed lines in Fig. \ref{fig09bis}(c) are fits with a single fitting parameter 
(the Fermi velocity $v$), which matches well all experimental points for different values of $n$.

\subsection{Level degeneracy}

In the preceding subsection, we have learnt that the energy of 2D (non-)relativistic charged particles is characterised by
a quantum number $n$, which denotes the LLs (in addition to the band index $\lambda$ in for relativistic particles). However, 
the quantum system is yet underdetermined, as may be seen from the following dimensional argument. The original Hamiltonians
(\ref{BHamS}) and (\ref{BHamD}) are functions that depend on {\sl two} pairs of conjugate operators, $x$ and $p_x$, and $y$ and
$p_y$, respectively, whereas when they are expressed in terms of the gauge-invariant momentum $\Pib$ or else the ladder
operators $a$ and $a^{\dagger}$ the Hamiltonians (\ref{HamLadS}) and (\ref{HamLadD}) depend only on a {\sl single} pair
of conjugate operators. From the original models, one would therefore expect the quantum states to be described by {\sl two}
quantum numbers (one for each spatial dimension). This is indeed the case in the zero-field models (\ref{0BHams}),
where the quantum states are characterised by the two quantum numbers $p_x$ and $p_y$, i.e. 
the components of the 2D momentum. For a complete description of the quantum states,
we must therefore search for a second pair of conjugate operators, which necessarily
commutes with the Hamiltonian and which therefore gives rise to the {\sl level degeneracy} of the LLs -- in addition to the 
degeneracy due to internal degrees of freedom such as the spin\footnote{The quantum states are naturally only degenerate if
one neglects the Zeeman effect.} or, in the case of graphene, the two-fold valley degeneracy.

In analogy with the gauge-invariant momentum, $\Pib=\bp + e\bA(\br)$, we consider the same combination with the 
{\sl opposite} relative sign,
\beq\label{Pitild}
\Pibtilde= \bp - e\bA(\br),
\eeq
which we call {\sl pseudo-momentum} to give a name to this operator.
One may then express the momentum operator $\bp$ and the vector potential $\bA(\br)$ in terms of $\Pib$ and $\Pibtilde$,
\beq\label{inverse}
\bp = \frac{1}{2}(\Pib+\Pibtilde) \qquad {\rm and} \qquad \bA(\br)=\frac{1}{2e}(\Pib - \Pibtilde).
\eeq
Notice that, in contrast to the gauge-invariant momentum, the pseudo-momentum {\sl depends} on the gauge and, therefore, does
not represent a physical quantity.\footnote{We will nevertheless try to give a physical interpretation to this operator below,
within a semi-classical picture.} However, the commutator between the two components of the pseudo-momentum turn out to be 
gauge-invariant,
\beq\label{ComPM}
\left[\Pitilde_x,\Pitilde_y\right] = i\frac{\hbar^2}{l_B^2}\ .
\eeq
This expression is calculated in the same manner as the commutator (\ref{ComMom}) between $\Pi_x$ and $\Pi_y$, as well as the
mixed commutators between the gauge-invariant momentum and the pseudo-momentum,
\beqn\label{MixedCom}
\nn
\left[\Pi_x,\Pitilde_x\right] &=& 2ie\hbar\frac{\partial A_x}{\partial x}\ ,\\
\left[\Pi_y,\Pitilde_y\right] &=& 2ie\hbar\frac{\partial A_y}{\partial y}\ ,\\
\nn
\left[\Pi_x,\Pitilde_y\right] &=&  ie\hbar\left(\frac{\partial A_x}{\partial y} + \frac{\partial A_y}{\partial x}\right) = 
- \left[\Pitilde_x,\Pi_y\right] .
\eeqn
These mixed commutators are unwanted quantities because they would induce unphysical dynamics due to the fact that the
components of the pseudo-momentum would not commute with the Hamiltonian, $[\Pitilde_{x/y},H]\neq 0$. However, this 
embarrassing situation may
be avoided by choosing the appropriate gauge, which turns out to be the {\sl symmetric} gauge
\beq\label{symG}
\bA_S(\br) = \frac{B}{2}(-y,x,0),
\eeq
with the help of which all mixed commutators (\ref{MixedCom}) vanish such that the components of the pseudo-momentum
also commute with the Hamiltonian.

Notice that there exists a second popular choice for the vector potential, namely the {\sl Landau} gauge, which we have
already mentioned above,
\beq\label{LG}
\bA_L(\br) = B(-y,0,0),
\eeq
for which the last of the mixed commutators (\ref{MixedCom}) would not vanish. This gauge choice may even occur simpler:
because the vector potential only depends on the $y$-component of the position, the system remains then translation invariant 
in the $x$-direction. Therefore, the associated momentum $p_x$ is a good quantum number, which may be used to label the
quantum states in addition to the LL quantum number $n$. For the Landau gauge, which is useful in the description of
geometries with translation invariance in the $y$-direction, the wave functions are calculated in Sec. (\ref{WFLandau}).
However, the symmetric gauge, the wave functions of which are presented in Sec. (\ref{WFsym}), plays an 
important role in two different aspects; first, it allows for a semi-classical interpretation more easily than the Landau
gauge, and second, the wave functions obtained from the symmetric gauge happen to be the basic ingredient in the 
construction of trial wave functions {\sl \`a la Laughlin} for the description of the FQHE, as we will see in Chap. \ref{FQHE}.

The pseudo-momentum, with its mutually conjugate components $\Pitilde_x$ and $\Pitilde_y$, allows us to introduce, in the same
manner as for the gauge-invariant momentum $\Pib$, ladder operators,
\beq\label{ladderb}
b=\frac{l_B}{\sqrt{2}\hbar}\left(\Pitilde_x + i\Pitilde\right) \qquad {\rm and} \qquad
b^{\dagger}=\frac{l_B}{\sqrt{2}\hbar}\left(\Pitilde_x - i\Pitilde\right),
\eeq
which again satisfy the usual commutation relations $[b,b^{\dagger}]=1$ and which, in the symmetric gauge, commute with
the ladder operators $a$ and $a^{\dagger}$, $[b,a^{(\dagger)}]=0$, and thus with the Hamiltonian, $[b^{(\dagger)},H_B]=0$.
One may then introduce a number operator $b^{\dagger}b$ associated with these ladder operators, the eigenstates of 
which satisfy the eigenvalue equation
$$b^{\dagger}b|m\rangle = m|m\rangle.$$ 
One thus obtains a second quantum number, an integer $m\geq 0$, 
which is necessary to describe, as expected from the above dimensional argument,
the full quantum states in addition to the LL quantum number $n$. The quantum states therefore become tensor products of the 
two Hilbert vectors
\beq\label{QstateNR}
|n,m\rangle = |n\rangle\otimes |m\rangle
\eeq
for non-relativistic particles. In the relativistic case, one has
\beq\label{QstateR}
\psi_{\lambda n,m} = \psi_{\lambda n,m}\otimes |m\rangle = \frac{1}{\sqrt{2}}\left(\begin{array}{c} |n-1,m\rangle \\ \lambda |n,m\rangle  \end{array}\right)
\eeq
for $n\neq 0$  and
\beq\label{QstateN0}
\psi_{n=0,m} = \psi_{n=0}\otimes |m\rangle = \left(\begin{array}{c} 0 \\ |n=0,m\rangle  \end{array}\right)
\eeq
for the zero-energy  LL.

\subsection{Semi-classical interpretation of the level degeneracy}

\begin{figure}
\begin{center}
\epsfig{figure=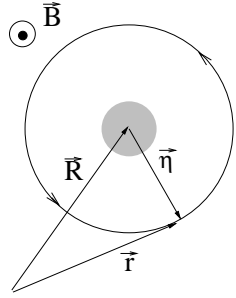,width=3cm,clip}
\end{center}
\caption{ Cyclotron motion of an electron in a magnetic field around the guiding centre $\bR$. The grey region indicates the
quantum-mechanical uncertainty of the guiding-centre position due to the non-commutativity (\ref{ComGC}) of its components.}
\label{fig10}
\end{figure}

How can we illustrate this somewhat mysterious pseudo-momentum introduced formally above? Remember that, because the pseudo-momentum
is a gauge-dependent quantity, any physical interpretation needs to be handled with care. However, within a semiclassical treatment, 
the symmetric gauge allows us
to make a connection with a classical constant of motion that one obtains from solving 
the classical equations of motion for a massive electron in a magnetic field,
\beq\label{EqM}
m_b\ddot{\br} = -e(\dot{\br}\times \bB) \qquad \Leftrightarrow\qquad  \left\{
\begin{array}{ccc}
 \ddot{x} &=& -\omega_C \dot{y}\\ \vspace*{-0.1cm}
\\
 \ddot{y} &=&  \omega_C \dot{x}
\end{array}
\right.
\eeq
which is nothing other than the electron's acceleration due to the Lorentz force. These equations may be integrated, 
and one then finds
\beq\label{constM}
\left.
\begin{array}{ccc}
 \dot{x} = \frac{\Pi_x}{m_b} &=& -\omega_C (y-Y)\\ \vspace*{-0.1cm}
\\
 \dot{y} = \frac{\Pi_y}{m_b} &=&  \omega_C (x-X)
\end{array}
\right\} \qquad \Leftrightarrow \qquad \left\{
\begin{array}{ccc}
 y &=& Y - \frac{\Pi_x}{eB}\\ \vspace*{-0.1cm}
\\
 x &=& X + \frac{\Pi_y}{eB}
\end{array}
\right. 
\eeq
where $\bR=(X,Y)$ is an integration constant, which physically describes a constant of motion. 
This quantity may easily be interpreted: it represents the centre of the electronic cyclotron motion (see Fig. \ref{fig10}).
Indeed, further integration of the equations (\ref{constM}) yields the classical cyclotron motion
$$
x(t)= X-r\sin(\omega_Ct+\phi)\qquad {\rm and} \qquad y(t)=Y+r\cos(\omega_Ct+\phi),
$$
where the phase $\phi$ is another constant of motion. The cyclotron motion itself is described by the velocities (or else
the gauge-invariant momenta) $\Pi_x/m$ and $\Pi_y/m$. Whereas the energy depends on these velocities that determine the radius $r$
of the cyclotron motion, it is completely independent of the position of its centre $\bR$, which we call {\sl guiding centre} from 
now on, as one would expect from the translational invariance of the equations of motion (\ref{EqM}).

In order to relate the guiding centre $\bR$ to the pseudo-momentum $\Pibtilde$, we use Eq. (\ref{inverse}) for the vector
potential in the symmetric gauge,
$$
e\bA(\br)=\frac{eB}{2}\left(\begin{array}{c} -y \\ x\end{array}\right)=\frac{1}{2}(\Pib-\Pibtilde).
$$
The postions $x$ and $y$ may then be expressed in terms of the momenta $\Pib$ and $\Pibtilde$,
\beqn
\nn
y &=& \frac{\Pitilde_x}{eB} - \frac{\Pi_x}{eB} \\
\nn
x &=& - \frac{\Pitilde_y}{eB} + \frac{\Pi_y}{eB}\ .
\eeqn
A comparison of these expresssions with Eq. (\ref{constM}) allows us to identify 
\beq\label{PM:GC}
X=-\frac{\Pitilde_y}{eB} \qquad {\rm and} \qquad Y=\frac{\Pitilde_x}{eB}\ .
\eeq
This means that, in the symmetric gauge, the components of the pseudo-momentum are nothing other, apart from a factor 
to translate a momentum into a position, than the the components of the guiding centre, which are naturally constants of 
motion. In a quantum-mechanical treatment, these operators therefore necessarily commute with the Hamiltonian, as we have already
seen above. Furthermore, the commutation relation (\ref{ComPM}) between the components of the pseudo-momentum, 
$[\Pitilde_x,\Pitilde_y]=i\hbar^2/l_B^2$ induces the commutation relation
\beq\label{ComGC}
[X,Y]=i l_B^2
\eeq
between the components of the guiding-centre operator. This means that there is a Heisenberg uncertainty associated with the
guiding-centre position of a quantum-mechanical state -- one cannot know its $x$- and $y$-components simultaneously,
and the guiding centre is, therefore, smeared out over a surface 
\beq\label{minsurf}
\Delta X\Delta Y=2\pi l_B^2
\eeq 
(see grey region in Fig. \ref{fig10}).\footnote{Mathematicians speak of a {\sl non-commutative geometry} in this context, and
the charged 2D particle may be viewed as a pardigm of this concept.}
This minimal surface plays the same role as the surface (action) $h$ in phase space and therefore allows us to count the number of 
possible quantum states of a given (macroscopic) surface $\Amath$,
$$N_B=\frac{\Amath}{\Delta X\Delta Y} =  \frac{\Amath}{2\pi l_B^2}= n_B \times \Amath,$$
where we have introduced the flux density
\beq\label{fluxdens}
n_B= \frac{1}{2\pi l_B^2} = \frac{B}{h/e},
\eeq 
which is nothing other than the magnetic field measured in units of the flux quantum $h/e$. Therefore, {\sl the number of quantum states
in a LL equals the number of flux quanta threading the sample surface $\Amath$, and each LL is macroscopically degenerate}. 
We will show in a more quantitative manner than in the above argument based on the Heisenberg inequality that the number of
states per LL is indeed given by $N_B$ when discussing, in the next section, 
the electronic wave functions in the symmetric and the Landau gauges.

Similarly to the guiding-centre operator, we may introduce the {\sl cyclotron variable} $\etab=(\eta_x,\eta_y)$, which determines 
the cyclotron motion and which fully describes the dynamic properties. The cyclotron variable is perpendicular to the electron's
velocity and may be expressed in terms of the gauge-invariant momentum $\Pib$,
\beq\label{CyclVar}
\eta_x= \frac{\Pi_y}{eB} \qquad {\rm and} \qquad \eta_y = - \frac{\Pi_x}{eB}\ ,
\eeq
as one sees from Eq. (\ref{constM}).
The position of the electron is thus decomposed into its guiding centre and its cyclotron variable, $\br=\bR + \etab$.
Also the components of the cyclotron variable do not commute, and one finds with the help of Eq. (\ref{ComMom})
\beq\label{ComCV}
[\eta_x,\eta_y] = \frac{[\Pi_x,\Pi_y]}{(eB)^2} = -il_B^2
=-[X,Y].
\eeq

Until now, we have only discussed a single particle and its possible quantum states. Consider now $N$ 
independent quantum-mechanical electrons at zero-temperature. In
the absence of a magnetic field, electrons in a metal, due to their fermionic nature and the Pauli principle which prohibits double
occupancy of a single quantum state, fill all quantum states up to the Fermi energy, which depends thus on the number of electrons
itself. The situation is similar in the presence of a magnetic field: the electrons preferentially occupy the lowest LLs, i.e. those
of lowest energy. But once a LL is filled, the remaining electrons are forced to populate higher LLs. In order to describe the
LL filling it is therefore useful to introduce the dimensionless ratio between the number of electrons $N_{el}=n_{el}\times \Amath$ 
and that of the flux quanta,
\beq\label{filling}
\nu = \frac{N_{el}}{N_B}=\frac{n_{el}}{n_B} = \frac{hn_{el}}{eB},
\eeq
which is called {\sl filling factor}. Indeed the integer part, $[\nu]$, of the filling factor counts the number of completely
filled LLs. Notice that one may vary the filling factor either by changing the particle number or by changing the magnetic field. 
At fixed particle number, lowering the magnetic field corresponds to an increase of the filling factor.

\section{Eigenstates}

\markboth{Landau Quantisation}{Eigenstates}

\subsection{Wave functions in the symmetric gauge}
\label{WFsym}

The algebraic tools established above may be used calculate the electronic wave functions, which
are the space representations of the quantum states $|n,m\rangle$, $\phi_{n,m}(x,y)=\langle x,y|n,m\rangle$.\footnote{
We limit the discussion to the non-relativistic case. The spinor wave functions for relativistic electrons are then easily
obtained with the help of Eqs. (\ref{QstateR}) and (\ref{QstateN0}).}
Notice first that one may obtain all quantum state $|n,m\rangle$ from a single state $|n=0,m=0\rangle$, with the help
of 
\beq\label{constrNM}
|n,m\rangle = \frac{\left(a^{\dagger}\right)^n}{\sqrt{n!}} \frac{\left(b^{\dagger}\right)^m}{\sqrt{m!}}|n=0,m=0\rangle,
\eeq
which is a generalisation of Eq. (\ref{constrN}). Naturally, this equation translates into a differential equation for the
wave functions $\phi_{n,m}(x,y)$.

A state in the lowest LL ($n=0$) is characterised by the condition (\ref{0lad})
\beq\label{eq02}
a |n=0,m\rangle = 0,
\eeq
which needs to be translated into a differential equation. Remember from Eq. (\ref{ladder}) that
$a = (l_B/\sqrt{2}\hbar)(\Pi_x - i\Pi_y)$ and, by definition, $\Pib= -i\hbar\nabla + e\bA(\br)$ where we have 
already represented the momentum as a differential operator in position representation, $\bp=-i\hbar\nabla$.
One then finds 
$$a=-i\sqrt{2}\left[\frac{l_B}{2}\left(\partial_x - i\partial_y\right)+\frac{x-iy}{4l_B}\right],$$
where $\partial_x$ and $\partial_y$ are the components of the gradient $\nabla=(\partial_x,\partial_y)$,
and one sees from this expression that it is convenient to introduce {\sl complex coordinates} to describe the 2D plane. We define
$z=x-iy$, $z^*=x+iy$, $\partial = (\partial_x + i\partial_y)/2$ and $\bar{\partial} = (\partial_x - i\partial_y)/2$.
The lowest LL condition (\ref{eq02}) then becomes a differential equation,
\beq\label{eq03}
\left(\frac{z}{4 l_B} + l_B\bar{\partial}\right) \phi_{n=0}(z,z^*) = 0,
\eeq
which may easily be solved by the complex function
\beq\label{eq04}
\phi_{n=0}(z,z^*) = f(z) e^{-|z|^2/4l_B^2},
\eeq
where $f(z)$ is an {\sl analytic} function, i.e. $\bar{\partial} f(z) = 0$, and $|z|^2=zz^*$. 
This means that there is an additional degree of freedom because 
$f(z)$ may be {\sl any} analytic function. It is not unexpected that this degree of freedom is associated with the second 
quantum number $m$, as we will now discuss.

The ladder operators $b$ and $b^{\dagger}$ may be expressed in position representation in a similar manner as $a$, and one
obtains the space representation of the different ladder operators,
\beqn\label{diffLadd}
\nn
a=-i\sqrt{2}\left(\frac{z}{4l_B}+l_B\bar{\partial} \right), \qquad 
a^{\dagger}=i\sqrt{2}\left(\frac{z^*}{4l_B}-l_B\partial \right) \\
b=-i\sqrt{2}\left(\frac{z^*}{4l_B}+l_B\partial \right),  \qquad 
b^{\dagger}=i\sqrt{2}\left(\frac{z}{4l_B}-l_B\bar{\partial} \right).
\eeqn
In the same manner as for a state in the lowest LL, the condition for the reference state with $m=0$ is $b|n,m=0\rangle = 0$,
which yields the differential equation
$$
\left(z^*+4l_B^2\partial\right)\phi_{m=0}'(z,z^*)=0
$$
with the solution
$$
\phi_{m=0}'(z,z^*)=g(z^*)e^{-|z|^2/4l_B^2},
$$
in terms of an {\sl anti-analytic} function $g(z^*)$ with $\partial g(z^*)=0$. The wave function $\phi_{n=0,m=0}(z,z^*)$ must
therefore be the Gaussian with a prefactor that is both analytic and anti-analytic, i.e. a constant that is fixed by the 
normalisation. One finds
\beq\label{N0M0}
\phi_{n=0,m=0}(z,z^*)=\langle z,z^*|n=0,m=0\rangle=\frac{1}{\sqrt{2\pi l_B^2}}
e^{-|z|^2/4l_B^2},
\eeq
and a lowest-LL state with arbitrary $m$ may then be obtained with the help of Eq. (\ref{constrNM}),
\beqn\label{N0M}
\nn
\phi_{n=0,m}(z,z^*)&=&\frac{i^m\sqrt{2^m}}{\sqrt{2\pi l_B^2 m!}}\left(
\frac{z}{4l_B}-l_B\bar{\partial} \right)^m e^{-|z|^2/4l_B^2}  \\
&=&\frac{i^m}{\sqrt{2\pi l_B^2 m!}}\left(
\frac{z}{\sqrt{2}l_B}\right)^m e^{-|z|^2/4l_B^2}.
\eeqn
The states within the lowest LL are therefore, apart from the Gaussian, given by the usual polynomial basis states $z^m$
of analytic functions. In an arbitrary LL, the states may be obtained in a similar manner, but they happen to be more complicated
because the differential operators (\ref{diffLadd}) no longer act on the Gaussian only but also on the polynomial functions. 
They may be expressed in terms on Laguerre polynomials.

To conclude the discussion about the wave functions in the symmetric gauge, we calculate the average value of the guiding-centre
operator in the state $|n=0,m\rangle$. With the help of Eqs. (\ref{ladderb}) and (\ref{PM:GC}), one may express the components 
of the guiding-centre operator in terms of the ladder operators $b$ and $b^{\dagger}$,
\beq\label{GC:ladd}
X=\frac{l_B}{i\sqrt{2}}(b^{\dagger} - b) \qquad {\rm and} \qquad Y=\frac{l_B}{\sqrt{2}}(b^{\dagger} + b),
\eeq
and the ladder operators act, in analogy with Eq. (\ref{nlad}), on the states $|n,m\rangle$ as
$$b^{\dagger}|n,m\rangle = \sqrt{m+1}|n,m+1\rangle \qquad {\rm and} \qquad
b |n,m\rangle = \sqrt{m}|n,m-1\rangle.
$$
The average value of the guiding-centre operator is therefore zero in the states $|n,m\rangle$,
$$\langle \bR\rangle\equiv \langle n=0,m|\bR|n=0,m\rangle=0,$$
but we have 
\beq\label{Rav}
\langle |\bR|\rangle=\left\langle \sqrt{X^2+Y^2}\right\rangle=l_B\left\langle 
\sqrt{2b^{\dagger}b+1}\right\rangle=l_B\sqrt{2m+1}.
\eeq
This means that the guiding centre is situated, in a quantum state $|n,m\rangle$, somewhere on a circle of radius 
$l_B\sqrt{2m+1}$ whereas its angle (or phase) is completely undetermined. 

The symmetric gauge is the natural gauge to describe 
a sample in the form of a disc. Consider the disc to have a radius $R_{max}$
(and a surface $\Amath=\pi R_{max}^2$). How many quantum states may be accomodated within the circle? The quantum state with 
maximal $m$ quantum number, which we call $M$, has a radius $l_B\sqrt{2M+1}$, which must naturally coincide with the 
radius $R_{max}$ of the disc. One therefore obtains $\Amath = \pi l_B^2(2M+1)$, and the number of states within the disc is then,
in the thermodynamic limit $M\gg 1$,
\beq\label{NBsym}
M=\frac{\Amath}{2\pi l_B^2}= n_B\times \Amath=N_B, \eeq
in agreement with the result (\ref{fluxdens}) obtained from the argument based on the Heisenberg uncertainty relation.

\subsection{Wave functions in the Landau gauge}
\label{WFLandau}

If the sample geometry is rectangular, the Landau gauge (\ref{LG}), $\bA_L(\br)=B(-y,0,0)$, is more appropriate than
the symmetric gauge to describe the physical system. As already mentioned above, the momentum $p_x=\hbar k$ 
is a good quantum number due to translational invariance in the $x$-direction. One may therefore use a plane-wave
ansatz 
$$\psi_{n,k}(x,y) = \frac{e^{ikx}}{\sqrt{L}}\chi_{n,k}(y),
$$
for the wave functions. In this case, the Hamiltonian (\ref{BHamS}) becomes
\beq\label{HamLG}
H_S^B= \frac{(p_x- eBy)^2}{2m} + \frac{p_y^2}{2m} = \frac{p_y^2}{2m} + \frac{1}{2} m\omega_C (y-y_0)^2,
\eeq
where we have defined 
\beq\label{GCk}
y_0 = kl_B^2.
\eeq
The Hamiltonian (\ref{HamLG}) is nothing other than the Hamiltonian of a one-dimensional oscillator centred around the 
position $y_0$, and the eigenstates are 
$$\chi_{n,k}(y)=H_n\left(\frac{y-y_0}{l_B}\right)e^{-(y-y_0)^2/4l_B^2},$$
in terms of Hermite polynomials $H_n(x)$ \cite{CT}. The coordinate $y_0$ plays the role of the guiding centre component $Y$, 
the component $X$ being smeared over the whole sample
length $L$, as it is dictated by the Heisenberg uncertainty relation resulting from the commutation relation (\ref{ComGC})
$[X,Y]=il_B^2$.

Using periodic boundary conditions $k=m\times 2\pi/L$ for the wave vector in the $x$-direction, one may count the number of states
in a rectangular surface of length $L$ and width $W$ (in the $y$-direction), similarly to the above arguments in the symmetric
gauge. Consider the sample to range from $y_{min}=0$ to 
$y_{max}=W$, the first corresponding via the above-mentioned condition (\ref{GCk}) to the wave vector $k=0$ and the latter to
a wave vector $k_{max}=M \times 2\pi/L$. Two neighbouring quantum states are separated by the distance 
$\Delta y=\Delta k l_B^2=\Delta m (2\pi/L) l_B^2 = 2\pi l_B^2/L$, 
and each state therefore occupies a surface $\sigma=\Delta y\times L=2\pi l_B^2$, which agrees with the result (\ref{minsurf}) obtained 
above with the help of the consideration based on the Heisenberg uncertainty relation. 
The total number of states is, as in the symmetric gauge and the general argument leading to Eq. (\ref{fluxdens}),
$$M=N_B= n_B\times LW = n_B \times \Amath,$$
i.e. the number of flux quanta threading the (rectangular) surface $\Amath=LW$.

\chapter{Integer Quantum Hall Effect}
\label{IQHE}

\markboth{Integer Quantum Hall Effect}{Integer Quantum Hall Effect}

The quantum-mechanical treatment of the 2D electron in a perpendicular magnetic field is the backbone for the understanding of
the basic properties of the quantum Hall effect. However, we need to relate the kinetic-energy quantisation to the resistance 
quantisation, which is the essential feature of the IQHE. In the present chapter, we discuss the transport properties of 
electrons in the IQHE, namely the somewhat mysterious role that disorder plays in this type of transport. Remember from the
introduction that the Hall resistance is quantised with an astonishingly high precision ($10^{-9}$), such that it
is now used as the standard of resistance [see Eq. (\ref{klitz})]. The resistance quantisation in the IQHE therefore does reflect 
neither a particular disorder distribution nor a particular sample geometry. Nevertheless, disorder turns out to play an
essential role in the occurence of the IQHE, as we will see in this chapter. 

We will first consider, in Sec. \ref{ExtPot}, the motion of a 2D electron in a perpendicular magnetic field when also 
an external electrostatic potential is present, such as the one generated by disorder or the confinement potential that
defines the sample boundaries. In Sec. \ref{LLCond}, we then calculate the conductance of a single LL within a mesoscopic 
picture and discuss the difference between a two-terminal and a six-terminal transport measurement in Sec. \ref{4term}. 
Furthermore, we discuss, in Sec. \ref{PercIQHE}, the IQHE within a percolation picture and present some scaling properties 
that characterise the plateau transitions. We terminate this chapter with a short discussion of the pecularities of the
relativistic quantum Hall effect in graphene the understanding of which requires essentially the same ingredients as the IQHE in
non-relativistic quantum Hall systems.

\section{Electronic Motion in an External Electrostatic Potential}
\label{ExtPot}

\markboth{Integer Quantum Hall Effect}{Electronic Motion in an External Electrostatic Potential}

\begin{figure}
\begin{center}
\epsfig{figure=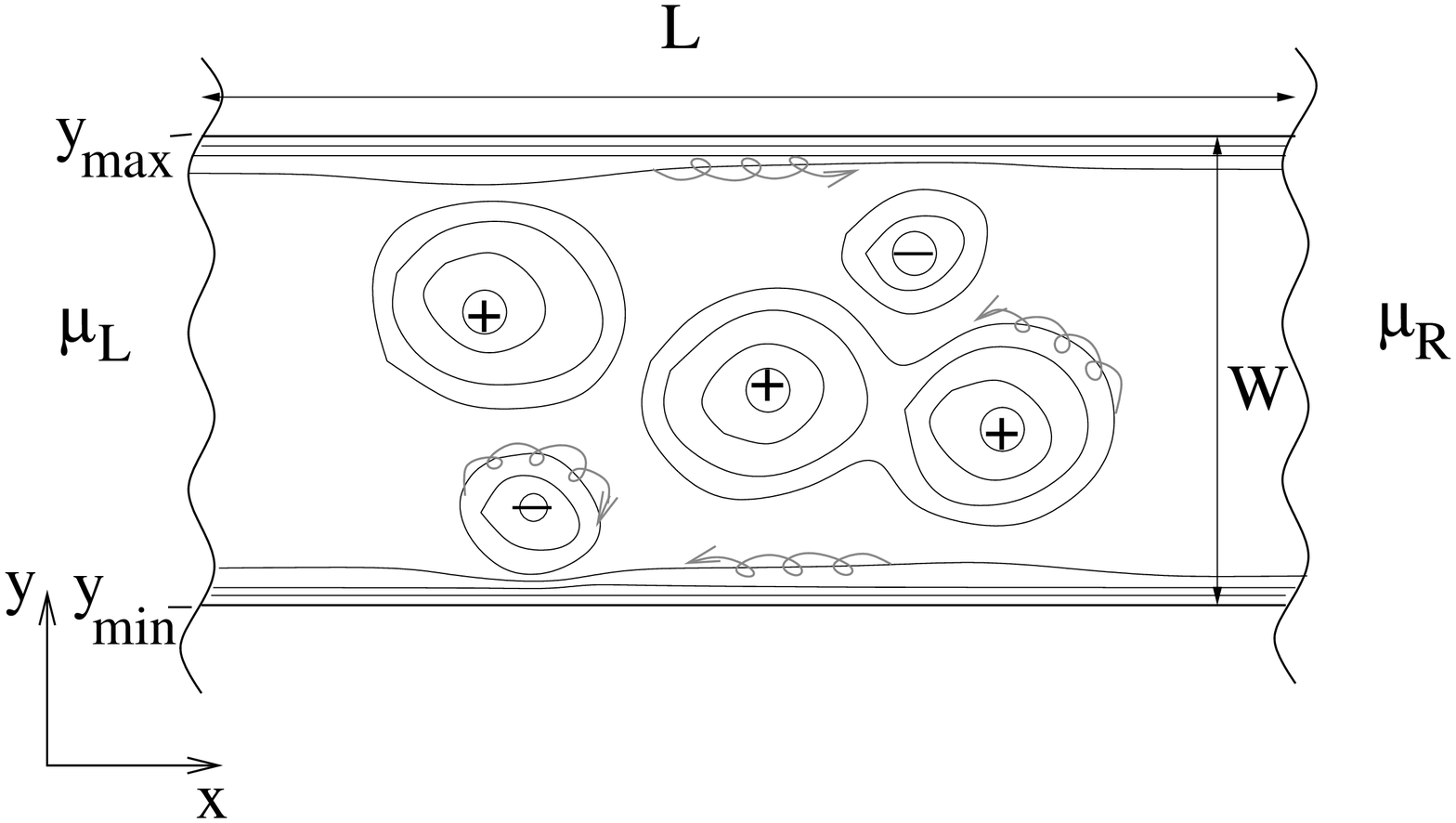,width=10cm,clip}
\end{center}
\caption{ Potential landscape of an electrostatic potential in a sample. The metallic contacts are  described by
the chemical potentials $\mu_L$ and $\mu_R$ for the left and right contacts, respectively. We consider $L\gg W \gg \xi \gg l_B$, where 
$\xi$ is the typical length scale for the variation of the electrostatic potential. The sample is confined in the $y$-direction
between $y_{max}$ and $y_{min}$. The thin lines indicate the equipotential lines. When approaching
one of the sample edges, they become parallel to the edge. 
The grey lines indicate the electronic motion with the guiding centre moving along the equipotential lines.
The electron turns clockwise around a summit of the potential landscape, which is caused e.g. by a negatively charged impurity ($-$),
and counter-clockwise around a valley ($+$). At the sample edges, the equipotential lines due to the confinement potential connect
the two contacts on the left and on the right hand side.}
\label{fig11}
\end{figure}

We consider a system the length $L$ of which is much larger than the width $W$ (see Fig. \ref{fig11}). This may be modeled by 
a confinement potential $V_{\rm conf}(y)$ that only depends on the $y$-direction, i.e. the system remains translation-invariant
in the $x$-direction with respect to this potential.\footnote{Naturally, the system is also confined in the $x$-direction, but
since we consider a sample with $L\gg W$, the system appears as translation-invariant in the $x$-direction when one considers
intermediate length scales. The latter may be taken into account with the help of periodic boundary conditions that
discretise the wave vector in the $x$-direction, as we have seen in the preceding chapter within the quantum-mechanical treatment
of the 2D electron in the Landau gauge (see Sec. \ref{WFLandau}).} 
In addition to the confinement, we consider a smoothly varying electrostatic
potential $V_{\rm imp}(x,y)$ that is caused by the impurities in the sample. This impurity potential breaks the translation invariance
in the $x$-direction as well as that in the $y$-direction, which is already broken by the confinement potential. The Hamiltonian
of a 2D particle in a perpendicular magnetic field then needs to be completed by a potential term
\beq\label{extPot}
V(\br) = V_{\rm conf}(y) + V_{\rm imp}(x,y),
\eeq
which creates a potential landscape that is schematically depicted in Fig. \ref{fig11}.

\subsection{Semi-classical treatment}

In a first step, we consider a potential $V(\br)$ that varies smoothly on the scale set by the magnetic length, i.e. 
$\xi\gg l_B$, where $\xi$ describes the characteristic length scale for the variation of $V(\br)$.
Notice first that the external
electrostatic potential {\sl lifts the LL degeneracy} because the Hamiltonian $H=H_B+V(\br=\bR +\etab)$ no longer commutes 
with the guiding-centre operator $\bR$, in contrast to the ``free'' Hamiltonian $H_B$, $[H,\bR]=[V,\bR]\neq 0$.
Physically, this is not unexpected: the guiding centre is a constant of motion due to translation invariance, i.e. it does
not matter whether the electron performs its cyclotron motion around a point $\bR_1$ or $\bR_2$ in the 2D plane as long as 
the cyclotron radius is the same. However, the electrostatic potential $V(\br)$ breaks this translation invariance
and thus lifts the degeneracy associated with the guiding centre.

In the case where the electrostatic potential varies smoothly on a length scale set by the magnetic length
and does not generate LL mixing, i.e. when
$|\nabla V| \ll \hbar \omega_C/l_B$, we may approximate the argument $\br$ in the potential (\ref{extPot}) 
by the guiding-centre variable $\bR$,\footnote{This approximation may be viewed as the first term of an expansion of the
electrostatic potential in the coherent (or vortex) state basis, where the states are maximally localised around the 
guiding-centre position $\bR$ \cite{champel}.} 
\beq\label{GCpotential}
V(\br)\simeq V(\bR). 
\eeq
Notice that this approximation may seem unappropriate
when we consider the confinement potential in the $y$-direction which may vary abruptly when approaching the sample edges.
The confinement potential with translation invariance in the $x$-direction will be discussed separately in the following subsection.

As a consequence of the non-commutativity of the potential term $V(\bR)$ with the guiding-centre operator, the latter 
quantity acquires dynamics, as may be seen from the Heisenberg equations of motion
\beqn\label{HeisenbergEM}
\nn
i\hbar \dot{X} &=& [X,H] = [X,V(\bR)] = \frac{\partial V}{\partial Y}[X,Y] = il_B^2 \frac{\partial V}{\partial Y}\\
i\hbar \dot{Y} &=& [Y,V(\bR)] = - il_B^2\frac{\partial V}{\partial X},
\eeqn
i.e. $\dot{\bR} \perp \nabla V$. Here, we have used the commutation relation (\ref{ComGC}) for the guiding-centre components
and Eq. (\ref{Haus}). The Heisenberg equations of motion are particularly useful in the discussion of the semi-classical limit
because the averaged equations satisfy the classical equations of motion,
\beq\label{SemCl}
\langle \dot{\bR}\rangle \perp \nabla V,
\eeq
which means that, within the semi-classical picture, {\sl the guiding centres move along the equipotential lines of the 
smoothly varying external electrostatic potential}. This feature, which is also called the {\sl Hall drift}, 
\beq\label{HallDrift}
{\bf v}_D=\frac{\bE\times\bB}{B^2} = \langle \dot{\bR}\rangle = \frac{-\nabla V \times \bB}{eB^2},
\eeq
in terms of the (local) electric field $\bE=-\nabla V/e$,
is depicted in Fig. \ref{fig11} by the grey lines. 

In the bulk,
the potential landscape is created by the charged impurities in the sample, and the electrons turn clockwise on an equipotential
line around a summit that is caused by a negatively charged impurity and counter-clockwise around a valley created by a 
positively charged impurity. If the equipotential lines are closed, as it is the case for most of the equipotential lines in
a potential landscape,\footnote{In order to illustrate this point, consider a hiking tour in the mountains, e.g. around Les Houches
in the French Alps. To 
go from one point to another one at the same height, one usual needs go downhill as well as uphill. It is very rare to be able 
to stay on the same height unless one wants to turn in circles that are just the closed countour lines
which correspond to closed equipotential lines in our potential landscape. For those who participated 
at the Les Houches session which was outsourced to Singapore, where there are no mountains and where the whether is anyway too hot
for hiking, just look at a hiking map of some mountainous region. Then search for countour lines that connect one border of the map
to the opposite border. It turns out to be very hard to find such lines as compared to a large number of closed countour lines.
}
an electron cannot move from one point to another one over a macroscopic distance, e.g. from one contact to the other one.
An electron moving on a closed equipotential line
can therefore not contribute to the electronic transport, and the electron is thus {\sl localised}. Notice that
this type of localisation it different from other popular types. Anderson localisation in 2D, e.g., is due to quantum
interferences of the electronic wave functions \cite{AALR}. Here, however, the localisation is a purely {\sl classical} effect.
The high-field localisation is also different from the interaction-driven Mott insulator, where the electrons freeze out
in order to minimise the mutual Coulomb repulsion between the electrons. 

At the edge, the equipotential lines reflect the confinement potential, which is zero in the bulk but rapidly increases when
approaching the sample edge at $y_{min}$ and $y_{max}$ (see Fig. \ref{fig11}). In this case, the equipotential lines are open
and therefore connect the two different electronic contacts. The electrons occupying quantum states 
at these equipotential lines then contribute
to the electronic transport, in contrast to those on closed equipotential lines in the bulk. These quantum states are called
{\sl extended states},\footnote{In the semi-classical picture the extended states are also called {\sl skipping orbits}.}
as opposed to the {\sl localised states} discussed above. The difference between localised and
extended states turns out to be essential in the understanding of the IQHE, as we will see below (Sec. \ref{PercIQHE}).

\subsection{Electrostatic potential with translation invariance in the $x$-direction}
\label{secInv}

Although the above semi-classical considerations yield the correct physical picture of localised and extended states, 
it is based on the assumption that the electrostatic potential varies smoothly on the scale set by the magnetic length,
such that we may replace the electron's position by that of its guiding centre [Eq. (\ref{GCpotential})].
This assumption is, however, problematic in view of the confinement potential which varies strongly at the sample
edges, i.e. in the vicinity of $y_{min}$ and $y_{max}$. We will therefore treat the $y$-dependent confinement potential
in a quantum treatment. Naturally, the appropriate gauge for the quantum-mechanical treatment is the Landau gauge (\ref{LG}),
which respects the translation invariance in the $x$-direction, and the Hamiltonian (\ref{HamLG}) becomes 
\beq\label{HamLGconf}
H =  \frac{p_y^2}{2m} + \frac{1}{2} m\omega_C (y-y_0)^2 + V_{\rm conf}(y).
\eeq
Remember that for a fixed wave vector $k$ in the $x$-direction, the position around which the one-dimensional harmonic
oscillator is centred is fixed by Eq. (\ref{GCk}), $y_0=kl_B^2$. We may therefore expand the confinement potential, even in the case 
of a strong variation, around this position,
$$V(y)\simeq V(y_0=kl_B^2) - eE(y_0)(y-y_0) + \Omath\left(\frac{\partial^2 V}{\partial y^2}\right),$$
where the local electric field is given in terms of the first derivative of the potential at $y_0$,
$$eE(y_0) = -\left. \frac{\partial V_{\rm conf}}{\partial y}\right|_{y_0} .$$
This expansion yields the Hamiltonian
$$
H =  \frac{p_y^2}{2m} + \frac{1}{2} m\omega_C (y-y_0')^2 + V_{\rm conf}(y_0) - \frac{1}{2}m v_D^2(y_0),
$$
where the local drift velocity reads $v_D=E(y_0)/B$ and the position of the harmonic oscillator is shifted,
$y_0\rightarrow y_0'=y_0 + eE(y_0)/m\omega_C^2$. Notice that the last term is quadratic in the 
electric field $E(y_0)$ and therefore a second-order term in the expansion of the confinement potential. We
neglect this term in the following calculations. The final Hamiltonian then reads
\beq\label{HamLGconf2}
H =  \frac{p_y^2}{2m} + \frac{1}{2} m\omega_C (y-y_0')^2 + V_{\rm conf}(y_0'),
\eeq
where we have replaced the argument $y_0$ by the shifted harmonic-oscillator position $y_0'$, which is valid
at first order in the expansion of the confinement potential. One therefore obtains the energy spectrum 
\beq\label{energyConf}
\epsilon_{n,y_0}=\hbar\omega_C\left(n + \frac{1}{2}\right) + V(y_0),
\eeq
where we have omitted the prime at the shifted harmonic-oscillator position to simplify the notation. One therefore
obtains the same LL spectrum as in the absence of a confinement potential, apart from an energy shift that is 
determined by the value of the confinement potential at the harmonic-oscillator position, which may indeed vary
strongly. This position $y_0$ plays the role of the guiding-centre position, as we have already mentioned in the last chapter,
where we have calculated the electronic wave functions in the Landau gauge (\ref{WFLandau}). One thus obtains
a result that is consistent with the semi-classical treatment presented above.

\section{Conductance of a Single Landau Level}
\label{LLCond}

\markboth{Integer Quantum Hall Effect}{Conductance of a Single Landau Level}

We now calculate the conductance of a completely filled LL for the geometry depicted in Fig. \ref{fig11}, i.e. when all
quantum states (described within the Landau gauge) of the $n$-th LL are occupied. In a first step, we calculate the current of 
the $n$-th LL, which flows from the left to the right contact, with the help of the formula \cite{butt}
\beq\label{currMes}
I_n^x=-\frac{e}{L}\sum_k \langle n,k|v_x|n,k\rangle,
\eeq
i.e. as the sum over all $N_B$ quantum channels labelled by the wave vector $k = 2\pi m/L$, with the velocity 
$$\langle n,k|v_x|n,k\rangle = \frac{1}{\hbar}\frac{\partial \epsilon_{n,k}}{\partial k}=
\frac{L}{2\pi \hbar}\frac{\Delta \epsilon_{n,m}}{\Delta m},$$
in terms of the dispersion relation (\ref{energyConf}).\footnote{This relation may be obtained from the Heisenberg equations
of motion, $i\hbar\dot{x}=[x,H]=(\partial H/\partial p_x)[x,p_x]=i \partial H/ \partial k$, where we have used 
Eq. (\ref{Haus}) and $p_x=\hbar k$. One therefore obtains the operator equation
$$\dot{x}=\frac{1}{\hbar}\frac{\partial H}{\partial k},$$
which we evaluate in the state $|n,k\rangle$. In the last step we have used the periodic boundary conditions.} 
Notice that the velocity in the $y$-direction is zero because 
the energy does not disperse as a function of the $y$-component of the wave vector. 
The above expression is readily evaluated with $\Delta m=1$, and one obtains 
$$\langle n,k|v_x|n,k\rangle = \frac{L}{h}\left(\epsilon_{n,m+1} - \epsilon_{n,m}\right).
$$
With the help of this expression, the current (\ref{currMes}) of the $n$-th LL becomes
$$I_n= -\frac{e}{L}\sum_{m}\frac{L}{h}\left(\epsilon_{n,m+1} - \epsilon_{n,m}\right),
$$
and one notices that all terms in the sum cancel apart from the boundary terms $\epsilon_{n,m_{min}}$ and
$\epsilon_{n,m_{min}}$, which correspond to the chemical potentials $\mu_{min}$ and $\mu_{max}$, respectively.
The difference between these two chemical potentials may be described in terms of the (Hall) voltage $V$ between the 
upper and the lower edge,
$\mu_{max}-\mu_{min} = -eV$. One thus obtains the final result
\beq\label{currLL}
I_n = -\frac{e}{h}\left(\mu_{max} - \mu_{min}\right)=\frac{e^2}{h} V.
\eeq

\begin{figure}
\begin{center}
\epsfig{figure=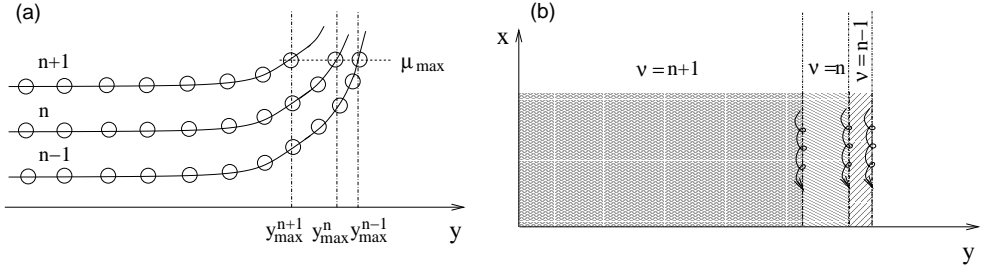,width=13cm,clip}
\end{center}
\caption{ Edge states. {\sl (a)} The LLs are bent upwards when approaching the sample edge, which may be modeled 
by an increasing confinement potential. One may associate with each LL $n$ a maximal value $y_{max}^n$ 
of the $y$-component where the LL crosses the chemical potential $\mu_{max}$.
{\sl (b)} At each position $y_{max}^n$, the filling factor decreases by a jump of 1. The $n$-th edge state is 
associated with the jump at $y_{max}^n$ and the gradient of the confinement potential imposes a direction 
to the Hall drift of this state {\sl (chirality)}. This chirality is the same for all edge states at the same edge.}
\label{fig12}
\end{figure}

This means that each LL contributes one {\sl quantum of conductance} $G_n=e^2/h$ to the electronic transport
and $n$ completely filled LLs contribute a conductance\footnote{Notice that, because the lowest LL is labelled by
$n'=0$, the last one has the index $n-1$ in the case of $n$ completely filled levels.}
\beq\label{LLcond}
G=\sum_{n'=0}^{n-1} G_{n'} = n\frac{e^2}{h}\, .
\eeq 
Notice furthermore that this is a particular example of the Landauer-B\"uttiker formula of quantum transport
$$G_n=\frac{e^2}{h} T_n$$
through a conduction channel $n$, where $T_n$ is the transmission coefficient of the channel
\cite{BILP,butt,datta}. Because $T_n + R_n=1$,
in terms of the reflexion coefficient, the above result (\ref{currLL}) indicates that each filled LL may be viewed
as a conduction channel with perfect transmission $T_n=1$, i.e. where an injected electron is not reflected or backscattered.

\subsection{Edge states}

The astonishing feature of perfect transmission, 
which is independent of the length $L$ (or more precisely of the aspect ratio $L/W$, see the
discussion in Sec. \ref{CHE} of the introduction) or the particular geometry of the sample, may be understood from the 
edge-state picture which we have introduced above (see Fig. \ref{fig12}). Consider the upper edge, without loss of 
generality. The current-transporting edge state
of the $n$-th LL is the one situated at $y_{max}^n$, where the $n$-th LL crosses the Fermi energy and where the 
filling factor jumps from $\nu= n+1$ to $n$.\footnote{Strictly speaking the filling factor does not jump not abruptly when one
takes interactions between the electrons into account. In this case, two incompressible strips, of $\nu=n+1$ and $\nu=n$
are separated by a {\sl compressible} strip of finite width. 
The picture of chiral electron transport remains, however, essentially the same when considering such compressible regions.} 
Due to the upward bent of the confinement potential a particular direction is imposed on
the electronic motion, which is nothing other than the Hall drift (see Fig. \ref{fig11}). This uni-directional motion
is also called 
{\sl chirality} of the edge state. Notice that this is the same chirality which one expects from the semi-classical 
expression (\ref{HallDrift}) for the drift velocity.
The chirality is the same for all edge states $n$ at the same sample edge where
the gradient of the confinement potential does not change its direction. Therefore, even if an electron is scattered
from one channel $n$ to another one $n'$ at the same edge it does not change its direction of motion, and the electron
cannot be backscattered unless it is scattered to the opposite edge with inverse chirality. However, in a usual quantum
Hall system, the opposite edges are separated by a macroscopic distance $\sim W$, and backscattering processes are therefore
strongly (exponentially) suppressed in the ratio $l_B/W$ between the magnetic length, which determines the 
spatial extension of quantum-mechanical state, and the macroscopic sample width $W$. Notice that the quantum Hall system at 
integer filling factors $\nu=n$ is therefore a very unusual electron liquid: it is indeed a bulk insulator with perfectly
conducting (non-dissipative) edges.

\section{Two-terminal versus Six-Terminal Measurement}
\label{4term}

\markboth{Integer Quantum Hall Effect}{Two-terminal versus Six-Terminal Measurement}

\subsection{Two-terminal measurement}

\begin{figure}
\begin{center}
\epsfig{figure=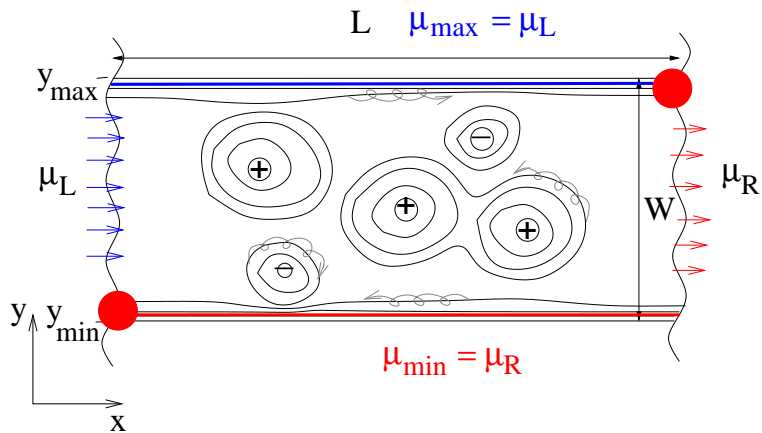,width=10cm,clip}
\end{center}
\caption{ Two-terminal measurement. The current is driven through the sample via the left and the right contacts, where one
also measures the voltage drop and thus a resistance. The upper edge is in thermodynamic equilibrium with the left contact (blue), whereas
the lower one is in equilibrium with the right contact (red). The chemical potential drops abruptly
when the upper edge reaches the right contact, and when the lower edge reaches the left contact. Dissipation occurs in these hot
spots (red dots). The measured resistance between the two contacts thus equals the Hall resistance.}
\label{fig13}
\end{figure}

In the preceding section Sec. (\ref{LLCond}), we have calculated the conductance of a single LL (and $n$ filled LLs)
within a so-called two-terminal measurement, where we inject a current in the left contact with chemical potential 
$\mu_L$ and collect the outcoming current at the right contact with $\mu_R$. As a consequence of Eq. (\ref{currLL}),
this current builds up a voltage $V$ between the upper and the lower sample edge. This voltage drop
is therefore associated with a Hall resistance, which is the inverse of the conductance $G=ne^2/h$,
\beq\label{Hall2Term}
R_H = G^{-1} = \frac{h}{e^2}\frac{1}{n}\ ,
\eeq
and which coincides with the contact (or interface) resistance of a mesoscopic system \cite{datta}.
However, the voltage drop $V_L$ between the left and the right contact is given by the difference of the chemical potentials
in the contacts, $\mu_R - \mu_L = -eV_L$, and the associated longitudinal resistance $V_L/I$ is non-zero,
in contrast to what we have seen in the introduction. This is due to the fact that 
the difference between the longitudinal and the Hall resistance is not clearly defined in such a two-terminal measurement.

This feature may be understood from
Fig. \ref{fig13}. Indeed, due to the above-mentioned absence of backscattering, the chemical potential is constant along a sample
edge, but there is a potential difference between the two edges. This means that the chemical potential must change somewhere
along the edge. Consider
the upper edge that is fed with electrons by the left contact, i.e. the upper edge is in thermodynamic equilibrium with the left 
contact and the chemical potentials therefore coincide, $\mu_L=\mu_{max}$ (see Fig.\ref{fig13}). 
Now, when the upper edge touches the right contact which is at a different chemical potential $\mu_R$, 
the chemical potential of the upper edge must rapidly relax to be in equilibrium with the right contact. In the same manner, the
lower edge is in equilibrium with the right contact, $\mu_{\min}=\mu_R$, and abruptly changes when touching the left contact.
The rapid change in the chemical potential is associated with a dissipation of energy (at so-called {\sl hot spots}) 
that has been observed experimentally \cite{klass}. In this experiment, the sample was put in liquid helium and the heating
at the hot spots caused a local vaporisation of the helium observable in form of a fountain of gas bubbles. 

Due to the equivalence of the chemical potentials $\mu_L=\mu_{max}$ and $\mu_{\min}=\mu_R$, the voltage drops $V$, between the upper
and the lower edge, and $V_L$ between the current contacts are equal, $V=V_L$. An unexpected consequence of this equation is that 
in a resistance measurement between the two contacts, in the two-terminal configuration, the two-terminal resistance equals
the Hall resistance,
\beq\label{Res2Term}
R_{R-L} = R_H = \frac{h}{e^2}\frac{1}{n}\ ,
\eeq
and {\sl not} the (vanishing) longitudinal resistance, when the bulk is insulating (at $\nu=n$).

\subsection{Six-terminal measurement}
\label{6term}

\begin{figure}
\begin{center}
\epsfig{figure=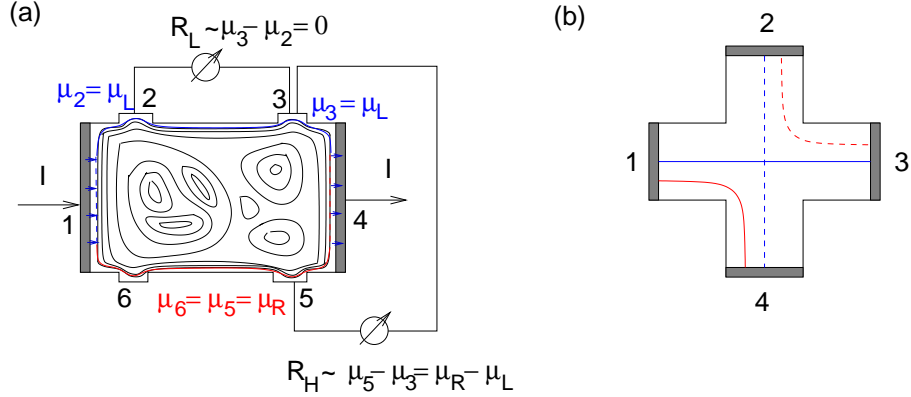,width=12cm,clip}
\end{center}
\caption{ {\sl (a)} Six-terminal measurement. The current $I$ is driven through the sample via the contacts 1 and 4.
Between these two contacts the chemical potential on the upper edge $\mu_L$ (blue) does not vary because the electrons
do not leak out or in at the contacts 2 and 3, where one measures the longitudinal resistance. In the same manner,
the chemical potential $\mu_R$ (red) remains constant between the contacts 5 and 6 on the lower edge. The longitudinal
resistance measured between 2 and 3 as well as between 5 and 6
is therefore $R_L=(\mu_2 - \mu_3)/eI=(\mu_5 - \mu_6)/eI = 0$. The Hall resistance is determined by the 
potential difference between the two edges and thus measured, e.g. between the contacts 5 and 3, where $\mu_5-\mu_3=\mu_R - \mu_L$,
and thus $R_H=(\mu_3-\mu_5)/eI$.
{\sl (b)} Four-terminal measurement in the van-der-Pauw geometry. In a Hall-resistance measurement, one drives a 
current through the sample via the contacts $1$ and $3$ (connected by the continuous blue line) and measures the Hall resistance
via the contacts $2$ and $4$ (dashed blue line). In a measurement of the longitudinal resistance, the current is driven
through the sample via the contacts 1 and 4 (continuous red line) and one measures a resistance between the contacts 2 and 3
(connected by the dashed red line).}
\label{fig14}
\end{figure}

A more sophisticated geometry that allows for the simulaneous measurement of a well-defined longitudinal and Hall resistance
is the six-terminal geometry, with two additional contacts at the upper and two at the lower edge [see Fig. \ref{fig14}(a)].
These additional contacts (2 and 3 at the upper and 5 and 6 at the lower edge, the left and the right contacts being labelled by 1 and
4, respectively) are used to measure a voltage, i.e. they have ideally an infinitely high internal resistance to prevent electrons 
to leak out of or into the sample. The chemical potential therefore remains constant at the upper edge $\mu_L=\mu_2=\mu_3$,
as well as that at the lower edge $\mu_R=\mu_5=\mu_6$, and one measures a zero-resistance, $R_L = (\mu_2 - \mu_3)/eI = 
(\mu_5 -\mu_6)/eI = 0$, as one expects from the calculation of the conductance through $n$ LLs (see Sec. \ref{LLcond}), which
is entirely transverse. The conductance matrix is thus off-diagonal, as well as the resistance matrix,
\beq\label{ResMatrix}
G = \left(\begin{array}{cc} 0 & n\frac{e^2}{h} \\ -n\frac{e^2}{h} & 0 \end{array}\right) \qquad {\rm and} \qquad
R = \left(\begin{array}{cc} 0 & -\frac{h}{e^2}\frac{1}{n} \\ \frac{h}{e^2}\frac{1}{n} & 0 \end{array}\right),
\eeq
and one precisely measures the diagonal elements of the resistance matrix, the longitudinal resistance, 
between the contacs 3 and 2 (or 6 and 5). The off-diagonal
elements, i.e. the Hall resistance, may e.g. be measured between the contacts 5 and 3 [as shown in Fig. \ref{fig14}(a)],
and one measures then the result $R_H=G_n^{-1}=h/e^2 n$ obtained from the calculation presented in Sec. \ref{LLcond} because of the 
voltage drop $V=(\mu_L - \mu_R)/e = (\mu_3 - \mu_3)/e$ between the upper and the lower edge. 

Finally we mention that
there exists an intermediate geometry that consists of four terminals (van-der-Pauw geometry), where the resistance measurements are
equally well defined [Fig. \ref{fig14}(b)]. 
If one labels the contacts from 1 to 4 in a clockwise manner, one may measure a Hall resistance between the 
contacts 2 and 3 while driving a current through the sample by the contacts 1 and 3 [blue lines in Fig. \ref{fig14}(b)]. 
In this case, one may use the clear topological definition mentioned at the beginning of the introduction. If one connects 
the contacts 2 and 3 by an imaginary line through the sample (dashed blue line) it necessarily crosses the imaginary line
(continuous blue) which connects the current contacts 1 and 3 through the sample. This is precisely the topological definition
of a Hall-resistance measurement. 

Similarly, 
one may measure the longitudinal resistance between the contacts 2 and 3 if one drives a current through the sample via the
contacts 1 and 4. In this case, the imaginary line (dashed red) which connects the contacts 2 and 3 where one measures a 
resistance does not need to cross the line (continuous red) between the contacts 1 and 4 at which one injects and collects
the current, respectively. As we have already mentioned at the beginning of the introduction, this defines 
topologically a measurement of the longitudinal resistance.

These considerations show that a resistance measurement, although it does not depend on the microscopic details of the sample,
depends nevertheless on the geometry in which the contacts are placed at the sample \cite{buett,butt}. This aspect is often 
not sufficiently appreciated in the literature, namely the fact that one measures, in a two-terminal geometry, a Hall resistance
between the contacts that are used to inject and collect the current and not a longitudinal resistance, as one may have expected
naively, when the system is in the IQHE condition.

\section{The Integer Quantum Hall Effect and Percolation}
\label{PercIQHE}

\markboth{Integer Quantum Hall Effect}{The Integer Quantum Hall Effect and Percolation}

\begin{figure}
\begin{center}
\epsfig{figure=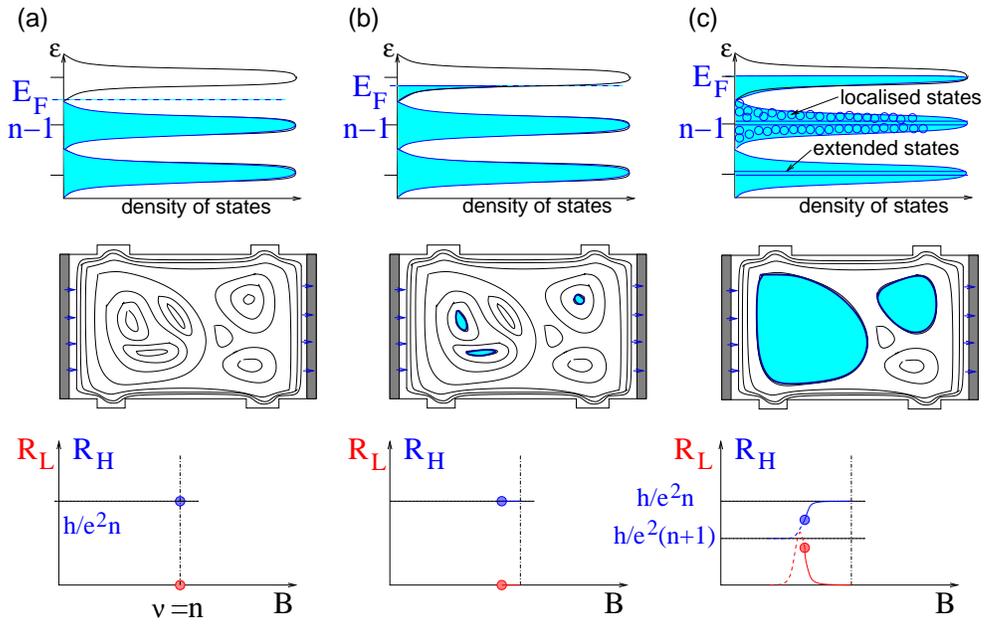,width=13cm,clip}
\end{center}
\caption{ Quantum Hall effect. The (impurity-broadened) density of states is shown in the first line for increasing fillings 
(a) - (c) described by the Fermi energy $E_F$. The second line represents the impurity-potential landscape the valleys of which
become successively filled with electrons when increasing the filling factor, i.e. when lowering the magnetic field at fixed 
particle number. The third line shows the corresponding
Hall (blue) and the longitudinal (red) resistance measured in a six-terminal geometry, as a function of the magnetic field.
The first figure in column (c) indicates that the bulk extended states are in the centre of the DOS peaks, whereas the
localised states are in the tails.}
\label{fig15}
\end{figure}

Until now we have shown that the Hall resistance is quantised [Eq. (\ref{Hall2Term})] when $n$ LLs are completely filled, i.e. 
when the filling factor is exactly $\nu=n$. However, we have not yet explained the occurence of plateaus in the Hall 
resistance, i.e. a Hall resistance that remains constant even if one varies the filling factor, 
e.g. by sweeping the magnetic field, around $\nu=n$.\footnote{Strictly speaking,
we have not gained anything because the quantum treatment allows us only to determine the Hall resistance at certain
points of the Hall curve, those at the magnetic fields corresponding to $\nu=hn_{el}/eB = n$. If we substitute the filling
factor in Eq. (\ref{Hall2Term}), we see immediately that $R_H=h/e^2\nu= B/en_{el}$, i.e. one retrieves the {\sl classical} 
result for the Hall resistance.}
In order to explain the constance of the Hall resistance over a rather large interval of magnetic field around $\nu=n$,
we need to take into account the semi-classical localisation of additional electrons (or holes) described in Sec. \ref{ExtPot}.
This is shown in Fig. \ref{fig15}, where we represent the filling of the LLs
(first line), the potential landscape of the last partially-filled
level (second line) and the resistances as a function of the magnetic field, measured in a six-terminal geometry (third line).
We start with the situation of $n$ completely filled LLs [column (a) of Fig. \ref{fig15}], 
which we have extensively discussed above: the LL $n$ (and its potential landscape) is unoccupied.\footnote{Remember
that due to the label $0$ for the lowest LL, all LLs with $n'=0, ..., n-1$ are the completely filled and the LL $n$ is then
the lowest {\sl unoccupied} level.}
In a six-terminal measurement, one therfore measures the Hall
resistance $R_H=h/e^2 n$ and a zero longitudinal resistance, as we have seen in Eq. (\ref{ResMatrix}). 

In column (b) of
Fig. \ref{fig15}, we represent the situation where the LL $n$ gets moderately filled by electrons when 
the magnetic field $B$ is decreased.
These electrons in $n$ populate preferentially the valleys of the potential landscape, or more precisely the closed equipotential
lines that enclose these valleys. The electrons in the LL $n$ are thus (classically) localised somewhere in the 
bulk and do not affect the global transport characteristics, measured by the resistances, 
because they are not probed by the sample contacts. Therefore, the Hall resistance remains unaltered and the longitudinal resistance
remains zero despite the change of the magnetic field. This is the origin of the plateau in the Hall resistance. 

If one continues to lower the magnetic field, the regions of the potential landscape in the LL $n$
occupied by electrons become larger,
and they are eventually enclosed by equipotential lines that pass through the bulk and that
connect the opposite edges. In this case, an electron injected at 
the left contact and travelling a certain distance at the upper edge 
may jump into the state associated with this equipotential line and thus reach the lower edge. Due to its chirality, the electron
is then backscattered to the left contact, which causes an increase in the longitudinal resistance. Indeed, if one measures the
resistance between the two contacts at the lower edge, a potential drop is caused by the electron that leaks in from this 
equipotential connecting the upper and the lower edge. It is this potential drop that causes a non-zero longitudinal resistance.
At the same moment the Hall resistance is no longer quantised and jumps to the next (lower) plateau, a situation that is
called {\sl plateau transition}. This situation of electron-filled
equipotential lines connecting opposite edges, which are thus {\sl extended states} [see first line of Fig. \ref{fig15}(c)]
as opposed to the bulk {\sl localised states}, arises when the LL $n$ is approximately half-filled. Notice that these extended
states, which are found in the centre of the DOS peaks [see upper part of Fig. \ref{fig15}(c)],
are  bulk states in contrast to the above-mentioned  edge states, which are naturally also extended

The clean jump in the Hall resistance at the plateau transition accompanied by
a peak in the longitudinal one is only visible in the six- (or four-)terminal measurement. As we have argued in Sec. \ref{6term},
there is no clear cut between the longitudinal and the Hall resistivity in the two-terminal configuration, where the resistance
measured between the current contacts is indeed quantised in the IQHE. At the plateau transition, however, the chemical potential
at the edges is no longer constant because of backscattered electrons and the resistance is no longer quantised. One observes indeed
the resistance peak associated with the {\sl longitudinal} resistance in the six- or four-terminal configuration. As a consequence,
one measures, at the plateau transition, the superposition of the Hall and the longitudinal resistances.

If one increases even more
the filling of the LL $n$, the same arguments apply but now in terms of {\sl hole states}. The Hall resistance is quantised as
$R_H=h/e^2(n+1)$, and the holes (i.e. the lacking electrons with respect to $n+1$ completely filled LLs) get localised in states at
closed equipotential lines around the potential summits. As a consequence, the longitudinal resistance drops to zero again. 

\subsection{Extended and localised bulk states in an optical measurement}

\begin{figure}
\begin{center}
\epsfig{figure=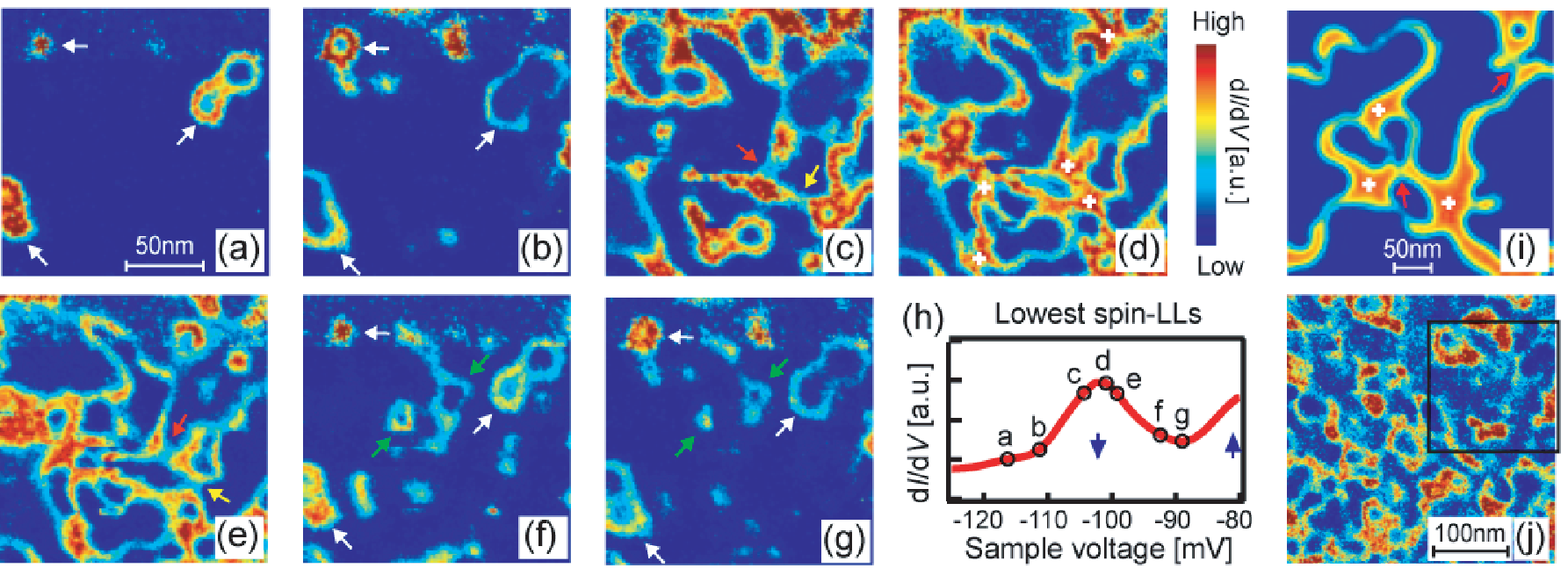,width=13cm,clip}
\end{center}
\caption{ STS measurements by Hashimoto {\sl et al.}, 2008, on a 2D electron system on a $n$-InSb surface. The figures (a) - (g)
show the local DOS at various sample voltages, around the peak obtained from a $dI/dV$ measurement (h). Figure (i) shows a calculated
characteristic LDOS, and figure (j) an STS result on a larger scale. }
\label{fig15b}
\end{figure}

The physical picture presented above, in terms of localised and extended bulk states, has recently been confirmed in scanning-tunneling 
spectroscopy (STS) of a 2D electron system that was prepared on an $n$-InSb surface
instead of the more common GaAs/AlGaAs heterostructure \cite{hashimoto}. 
Its advantage consists of its accessibility by an ``optical'' (surface) measurement that cannot be performed if
the 2D electron gas is buried deep in a semiconductor heterostructure. In an STS measurement one scans the sample and thus
measures the {\sl local} density of states at a certain energy that can be tuned via the voltage between the tip of the electron 
microscope and the sample. When measuring the differential conductance $dI/dV$, which is proportional to the DOS,
one observes a peak that corresponds to the centre of a LL [Fig. \ref{fig15b}(h)]
where the extended states are capable of transporting a current between the different electric contacts, as mentioned above.
Whereas the quantum states at energies corresponding to closed equipotential lines of the impurity landscape are clearly visible 
as closed orbits in Fig. \ref{fig15b}(a),(b) and (f),(g), the states in the vicinity of the peak are more and more extended, 
as shown by the spaghetti-like lines in Figs. \ref{fig15b}(c),(d) and (e), as one would expect from the arguments presented above.

\subsection{Plateau transitions and scaling laws}

The physical picture presented above suggests that the plateau transition in the Hall resistance is related to a {\sl percolation}
transition, where initially separated electron-filled valleys start to percolate between the opposite sample edges beyond 
a certain threshold of the filling. Because of the second-order character of a percolation transition, this scenario
suggests that the plateau transition is a {\sl second-order quantum phase transition} described by universal scaling laws,
where the control parameter is just the magnetic field $B$ \cite{sondhiRev,sachdev}. We finish this chapter on the IQHE with 
a brief overview over these scaling laws, and refer the interested reader to the literature \cite{sondhiRev,sachdev} and the 
class given by G. Batrouni at the same Singapore session of Les Houches Summer School 2009.\footnote{The  
lecture notes for this class are availabel on the School's program webpage: 

\noindent
http://www.ntu.edu.sg/ias/upcomingevents/LHSOPS09/Pages/programme.aspx
}

The phase transition occurs at the critical magnetic field $B_c$ and is characterised by an algebraically diverging correlation
length 
\beq\label{corrL}
\xi \sim \left|B - B_c\right|^{-\nu},
\eeq
where $\nu$ is called the {\sl critical exponent}.\footnote{Although we use the same Greek letter $\nu$ for the critical exponent,
it must not be confunded with the filling factor, which plays no role in this subsection.}
In the same manner, the temporal fluctuations are described by a correlation ``length'' $\xi_{\tau}$ that is related to the 
spatial correlation length $\xi$,
\beq\label{corrLtau}
\xi_{\tau} \sim \xi^z \sim \left|B - B_c\right|^{-z\nu},
\eeq
where $z$ is called {\sl dynamical critical exponent}. It is roughly a measure of the anisotropy between the spatial and
temporal fluctuations, which is often encountered in non-relativistic condensed-matter systems.\footnote{Notice that in
relativity, time is considered as the ``fourth'' dimension, and Lorentz invariance would require that spatial and temporal
fluctuations be equivalent, i.e. $z=1$.}

\begin{figure}
\begin{center}
\epsfig{figure=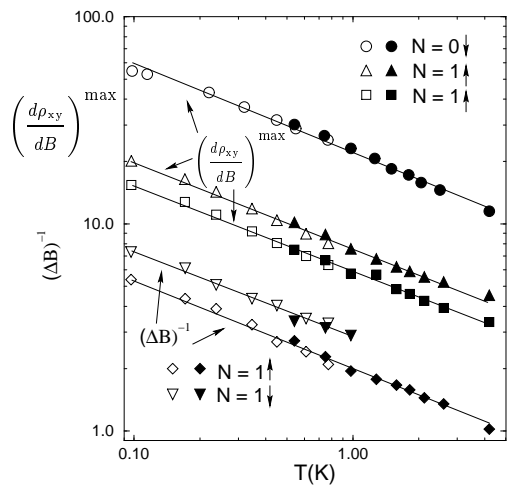,width=8cm,clip}
\end{center}
\caption{ Experiment by Wei {\sl et al.}, 1988. The width of the transition $\Delta B$ and of the derivative of the 
Hall resisitivity $\partial \rho_{xy}/\partial B$, measured as a function of temperature, reveals a scaling law 
with an exponent $1/z\nu=0.42\pm 0.04$, for the transition between the filling factors $1\rightarrow 2$ ($N=0\downarrow$),
 $2\rightarrow 3$ ($N=1\uparrow$) and $3\rightarrow 4$ ($N=1\downarrow$). }
\label{fig16}
\end{figure}

At the phase transition $B_c$, the longitudinal and transverse resistivities $\rho_{L/H}$ are described in terms of
{\sl universal} functions that are functions of the ratio $\tau/\xi_{\tau}$ between the (imaginary) time $\tau$,
which is proportional to the inverse temperature, $\hbar/\tau = k_B T$ \cite{sondhiRev,sachdev} and the
temporal correlation length $\xi_{\tau}$,
\beq\label{univFunc}
\rho_{L/H} = f_{L/H} \left(\frac{\tau}{\xi_{\tau}}\right) = f_{L/H}\left(\frac{\Delta B^{z\nu}}{T}\right), 
\eeq
where we have defined $\Delta B \equiv |B-B_c|$. In the case of an AC (alternating current) measurement
at frequency $\omega$, another 
dimensionless quantity, namely the ratio between the frequency and the temperature,
$\hbar\omega/k_BT$, needs to be taken into account such that the universal function reads 
$$
\rho_{L/H}^{AC} = f_{L/H} \left(\frac{\tau}{\xi_{\tau}},\frac{\hbar \omega}{k_BT}\right).
$$
However, we do not consider an alternating current here. Equation (\ref{univFunc}) then yields the scaling of the width of the 
peak in the longitudinal resistance (or else the plateau transition)
\beq\label{scalePW}
\Delta B \sim T^{1/z\nu}.
\eeq
A measurement of this width by Wei {\sl et al.} \cite{wei} has confirmed such critical behaviour with an exponent
$1/z\nu = 0.42\pm 0.04$ (see Fig. \ref{fig16}). 

Furthermore, one may distinguish between the two exponents $\nu$ and $z$ within a measurement of the plateau-transition 
width as a function of the electric field $E$ via current fluctuations. One may identify the energy fluctuation 
$eE\xi$ at the correlation length $\xi$ with the energy scale $\hbar/\xi_{\tau}\sim \hbar/\xi^z$
set by the temporal fluctuation $\xi_{\tau}$, which yields $E\sim \xi^{-(1+z)}\sim \Delta B^{\nu(1+z)}$, and thus
\beq\label{scalePWel}
\Delta B \sim E^{1/\nu(1+z)}.
\eeq
Other measurements by Wei {\sl et al.} \cite{wei2} have shown that these types of fluctuations yield $z\simeq 1$, i.e. 
$\nu\simeq 2.3$. The precision of the measured critical exponent has since been improved -- more recent experiments 
\cite {li1,li2} have revealed $\nu=2.38\pm 0.06$.

Theoretically one knows that 
the critical exponent for classical 2D percolation is $\nu_{\rm class}=4/3$ and thus much smaller than the measured
one. This discrepancy is due to the quantum nature of the percolation in quantum Hall systems. Indeed,
quantum-mechanical tunneling and the typical extension $\sim l_B$ of the wave functions associated with the equipotential
lines enhance percolation, i.e. the electron puddles in the potential valleys may percolate {\sl before} they touch each
other in the classical sense. A model that takes into account this effect has been proposed by Chalker and Coddington \cite{CC},
though with simplifying assumptions for the puddle geometry,\footnote{Notice, however, that due to the universality of the 
scaling laws and the fluctuations at all length scales, the results are expected to be independent on these {\sl microscopic}
assumptions.} 
and one obtains a critical exponent $\nu=2.5 \pm 0.5$ from numerical studies of this model \cite{CC,huckestein}, in 
quite a good agreement with the experimental data \cite{wei,wei2}. 

In spite of the good agreement with experimental findings, these theoretical
results need to be handled with care -- indeed, analytical calculations have shown that the dynamical exponent
should be exactly $z=2$ for non-interacting electrons, whereas the measured value $z\simeq 1$ is obtained when interactions are 
taken into account on the level of the Hartree-Fock approximation \cite{Huck}. Furtherrmore, very recent numerical calculations
within the Chalker-Coddington model have shown that the accurate value of the critical exponent is slightly larger 
($\nu\simeq 2.59$) than the measured one when interactions are not taken into account \cite{slevin}. 

\section{Relativistic Quantum Hall Effect in Graphene}
\label{relQHE}

\markboth{Integer Quantum Hall Effect}{Relativistic Quantum Hall Effect in Graphene}

\begin{figure}
\begin{center}
\epsfig{figure=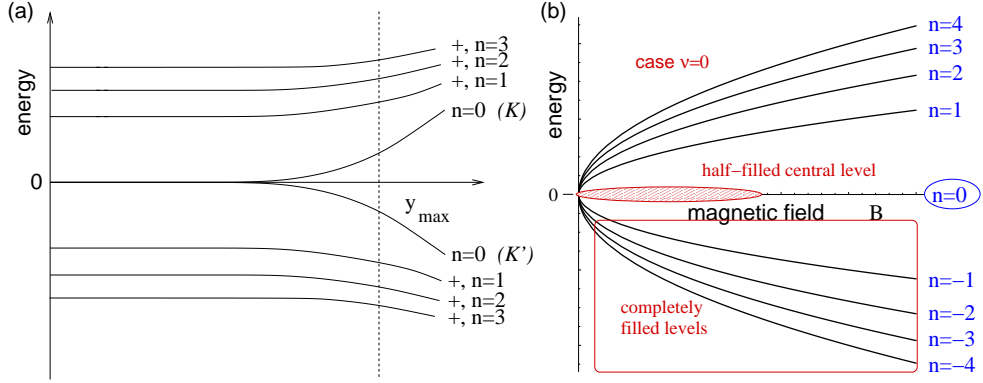,width=13cm,clip}
\end{center}
\caption{ {\sl (a)} Mass confinement for relativistic Landau levels. Whereas the electron-like LLs ($\lambda=+$) are bent upwards
when approaching the sample edge ($y_{max}$), the hole-like LLs ($\lambda=-$) are bent downwards. The fate of the $n=0$ LL
depends on the valley (parity anomaly) -- 
in one valley ($K$), the level energy decreases, whereas it increases in the other valley ($K'$). 
{\sl (b)} Filling of the bulk Landau levels at $\nu=0$. All electron-like LLs ($\lambda=+$) are unoccupied whereas
all hole-like LLs ($\lambda=-$) are completely filled. The $n=0$ LL is altogether half-filled.}
\label{fig17}
\end{figure}

We finish this chapter on the IQHE with a short presentation of the relativistic quantum Hall effect (RQHE) in graphene,
which is understandable in the same framework of LL quantisation and (semi-classical) one-particle localisation as the 
IQHE in a non-relativistic 2D electron system. Indeed, the above arguments also apply to relativistic electrons in graphene,
but we need to take into account the two different carrier types, electrons and holes, which carry a different charge.
This is not so much a problem in the case of the impurity potential with its valleys and summits: in a particle-hole
transformation, a valley becomes a summit and {\sl vice versa}.\footnote{The particle-hole transformed landscape corresponds
to an impurity distribution in which one interchanges negatively and positively charged impurities. }
Furthermore, the direction of the Hall drift changes in
this transformation. 
Because of the universality of the quantum Hall effect, both types of impurity distributions related by particle-hole
symmetry yield the {\sl same} quantisation of the Hall resistance. 
The picture of semi-classical localisation therefore applies also in the case of relativistic 
electrons in graphene.

The situation is different for the confinement potential. An ansatz of the form $V(y)\bone$ -- remember that the
Hamiltonian of electrons in relativistic graphene is a $2\times 2$ matrix that reflects the two different sublattices A and B --
has the problem that an increase $V(y-y_{max/min})\rightarrow \infty$ at the sample edge confines electrons but not the holes
of the valence band for which we would need $V(y-y_{max/min})\rightarrow -\infty$ for an efficient confinement. 
A possible confinement potential may be formed with the Pauli matrix $\sigma^z$,
\beq\label{MassTerm}
V_{\rm conf}(y) = V(y)\, \sigma^z=
\left(\begin{array}{cc} V(y) & 0 \\ 0 & - V(y) \end{array}\right),
\eeq
which, together with the Hamiltonian (\ref{BHamD}), yields the Hamiltonian which corresponds to the non-relativistic model
(\ref{HamLGconf}). For a constant term $M=V(y)$ the contribution (\ref{MassTerm}) plays the role 
of a mass of a relativistic particle (see also Appendix \ref{MassLL}). 
Therefore, the confinement (\ref{MassTerm}) is sometimes also called {\sl mass
confinement}. The corresponding energy spectrum, which one obtains within the same approximation as in Sec. \ref{ExtPot}
via the replacement $y\rightarrow y_0=kl_B^2$ in the Landau gauge, reads [c.f. Eq. (\ref{MassSpecA}) in Appendix \ref{MassLL}]
\beq\label{MassSpec}
\epsilon_{\lambda n,y_0} = \lambda \sqrt{M^2(y_0) + 2\frac{\hbar^2 v^2}{l_B^2} n}, 
\eeq
and is schematically represented in Fig. \ref{fig17}(a). Notice that Eq. (\ref{MassSpec}) is  only valid for
$n\neq0$ -- indeed, the $n=0$ acquires a non-zero energy $M(y_0)$, which is negative for our particular choice
(see Appendix \ref{MassLL}). This 
feature is sometimes called {\sl parity anomaly} in high-energy physics. Remember that in the case of graphene, one 
has two inequivalent low-energy points in the first BZ which give rise to a relativistic energy spectrum.
The Dirac Hamiltonians 
(\ref{0BHams}) and (\ref{BHamD}) for the zero-$B$ and magnetic-field case, respectively, applies principally only to
one of the two valleys (say $K$), whereas that for the other valley is given by $-H_D$ (or $-H_D^B$) if one interchanges
the $A$ and $B$ components [c.f. Eq. (\ref{DirHamBis}) in Appendix 
\ref{TBgraphene}]. The confinement term (\ref{MassTerm}) therefore reads $-V_{\rm conf}(y)$ in the other valley,
i.e. with a negative mass. The $n=0$ LL thus shifts to positive energies in the second valley, and the two-fold valley
degeneracy is lifted in this level. A more detailed discussion of the mass confinement (\ref{MassTerm}) in graphene may be found 
in the Appendix \ref{MassLL}.

This type of confinement may seem to be somewhat artificial, whereas the confinement in the non-relativistic 
case is easier to accept. Notice, however, that the whole model of massless Dirac fermions [second Hamiltonian in Eq. (\ref{0BHams})]
only describes the physical properties at length scales that are large compared to the lattice spacing (in graphene). In
the true lattice model, the electrons are naturally confined because one does not allow for hopping from a lattice site at
the edge into free space. The expression (\ref{MassTerm}) is therefore only an {\sl effective} model to describe confinement.
We notice that, although the effective model yields a qualitatively correct picture, 
the fine structure of the dispersion at the edge depends on the edge geometry \cite{BreyFertig}.
For further reading, we refer the interested reader to the literature \cite{antonioRev}.

\begin{figure}
\begin{center}
\epsfig{figure=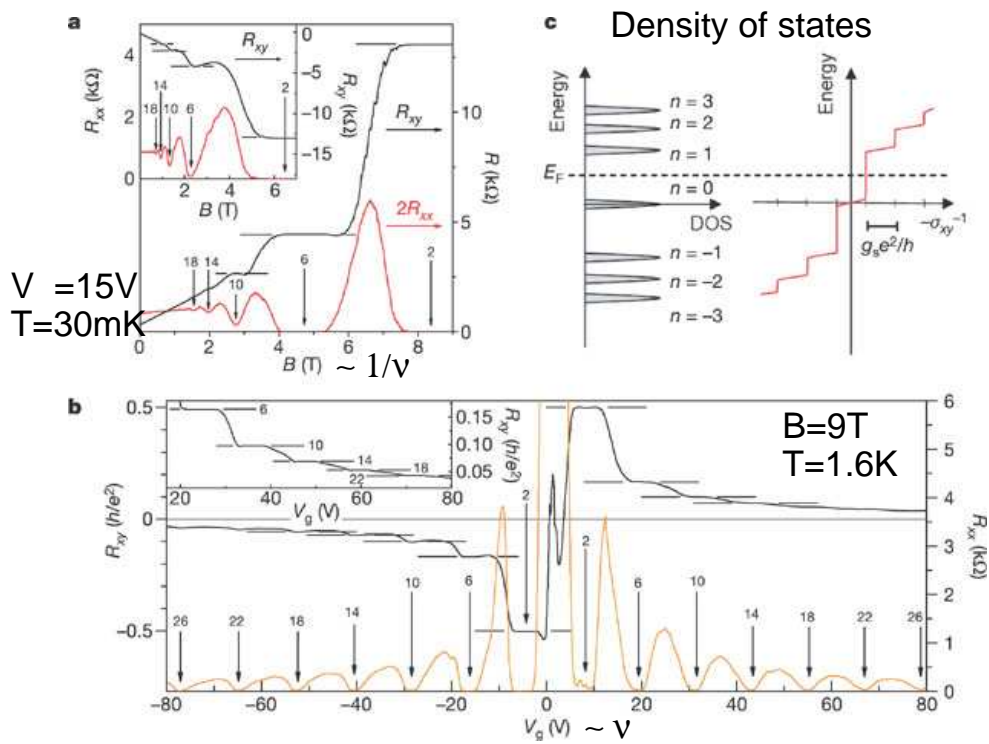,width=13cm,clip}
\end{center}
\caption{ Measurement of the relativistic quantum Hall effect (Zhang {\sl et al.}, 2005). {\sl (a)} RQHE at fixed carrier
density ($V_G=15$ V) at $T=30$ mK. The filling factor is varied by sweeping the magnetic field. 
{\sl (b)} Sketch of the DOS with the Fermi energy between the LLs $n=0$ and $+,n=1$.
{\sl (c)} RQHE at fixed magnetic field ($B=9$ T) at higher temperatures, $T=1.6$ K. The filling factor is now varied by 
changing the gate voltage.}
\label{fig18}
\end{figure}

With the help of these preliminary considerations, we are now prepared to understand the RQHE in graphene -- the semi-classical
localisation is the same as in the non-relativistic case, and the confinement, which needed to be adopted to account for the
simultaneous presence of electron- and hole-like LLs, yields the edge states which are responsible for the quantum transport
and, thus, the resistance quantisation. The RQHE was indeed discovered in 2005 by two different groups \cite{graph1,graph2},
and the results are shown in Fig. \ref{fig18} \cite{graph2}. The phenomenology of the RQHE is the same as that of the IQHE in
non-relativistic LLs: one observes plateaus in the Hall resistance while the longitudinal resistance vanishes. Notice that
one may vary the filling factor either by changing the $B$-field at fixed carrier density [Fig. \ref{fig18}(a)] or
one keeps the $B$-field fixed while changing the carrier density with the help of a gate voltage [Fig. \ref{fig18}(c)]. The latter
measurement is much easier to perform in graphene than in non-relativistic 2D electron gases in semiconductor heterostructures.

In spite of the similarity with the non-relativistic IQHE, one notices, in Fig. \ref{fig18}, an essential difference: the quantum
Hall effect is observed at the filling factors 
\beq\label{RelLLseries}
\nu=\pm 2(2n+1),
\eeq
in terms of the LL quantum number $n$, whereas the IQHE
is observed at $\nu=n$ (or $\nu=2n$ if the LLs are spin-degenerate). The step in units of 4 is easy to understand: each
relativistic LL in graphene is four-fold degenerate (in addition to the guiding-centre degeneracy), due to the two-fold spin and
the additional two-fold valley degeneracy. However, there is an ``offset'' of 2. This is due to the fact that the filling 
factor $\nu=0$ corresponds to no carriers in the system, i.e. to a situation where the Fermi energy is exactly at the Dirac
point (undoped graphene). In this case, one has a perfect electron-hole symmetry, and the $n=0$ LL must therefore be {\sl half-filled}
[see Fig. \ref{fig17}(b)], or else: there are as many electrons as holes in $n=0$. According to the considerations presented
in Sec. \ref{PercIQHE}, this does not correspond to a situation where one observes a quantum Hall effect due to percolating extended
states. Indeed, the system turns out to be {\sl metallic} at $\nu=0$ with a finite non-zero longitudinal resistance 
\cite{graph1,graph2}. 
A situation, where one would expect a quantum Hall effect, arises when the central LL $n=0$ is completely filled (or completely empty).
As a consequence of the four-fold level degeneracy, one obtains the quantum Hall effect at $\nu=2$ (or $\nu=-2$) observed in
the experiments (see Fig. \ref{fig18}). This is the origin of the
particular filling-factor sequence (\ref{RelLLseries}) of the RQHE in graphene.

\chapter{Strong Correlations and the Fractional Quantum Hall Effect}
\label{FQHE}

\markboth{Strong Correlations and the Fractional Quantum Hall Effect}{Strong Correlations and the Fractional Quantum Hall Effect}

In the preceding chapter, we have seen that one may understand the essential featues of 
the IQHE within a one-particle picture, i.e. in terms
of Landau quantisation; at integer filling factors $\nu=n$, which correspond to $n$ completely filled LLs,\footnote{As before,
we neglect the electron spin to render the discussion as simple as possible. The role of spin will be discussed briefly in
the last chapter on multi-component systems.} an additional electron is forced, as a result of the Pauli principle, to 
populate the next higher (unoccupied) LL [see Fig. \ref{fig19}(a)]. It therefore, needs to ``pay'' a finite amount of energy 
$\hbar\omega_C$ [or $\sqrt{2}(\hbar v/l_B)(\sqrt{n} - \sqrt{n-1})$ in the case of the RQHE in graphene] and is localised 
by the impurities in the sample, due to the classical Hall drift which forces the electron to move on closed equipotential lines. 
The system is said to be {\sl incompressible} because one may not vary the filling factor and pay only an infinitesimal amount
of energy -- indeed in the case of a fixed particle number, consider an infinitesimal decrease of the magnetic field 
which amounts to an infinitesimal change of the surface $2\pi l_B^2$ occupied by each quantum state. 
Since the total surface of the system remains constant, the infinitesimal increase of $2\pi l_B^2$ may not be accomodated by
an infinitesimal change in energy, due to the gap between the LL $n-1$ and $n$ where at least one electron must be promoted
to. This gives rise to a zero compressibility.

In view of this picture of the quantum Hall effect, it was therefore a big surprise to observe a FQHE at a filling factor
$\nu=1/3$, with the corresponding Hall quantisation $R_H = h/e^2\nu= 3 h/e^2$ \cite{TSG}, and, later, at a large set of other 
fractional filling factors. Indeed, if only the kinetic energy is taken into account, the ground state at $\nu=1/3$ is 
highly degenerate and there is no evident gap present in the system: the Pauli principle no longer prevents an additional 
electron to populate the next higher LL, but it finds enough place in the lowest LL which is only one-third filled. 

Notice that we have neglected so far the mutual Coulomb repulsion between the electrons, which happens to be responsible for the
occurence of the FHQE. The relevance of electronic interactions is discussed in the next section (Sec. \ref{Coul}). In Sec. \ref{FQHE1},
we present the basic results of Laughlin's theory of the FQHE, such as the ground-state wave functions, fractionally
charged quasi-particles and the interpretation of Laughlin's wave function in terms of a 2D one-component plasma.
The related issue of fractional statistics is introduced in a section apart (Sec. \ref{FQHE2}), and
we close this chapter with a short discussion of different generalisations of Laughlin's wave function, such as CF
theory or the Moore-Read wave function in half-filled LLs.

\section{The Role of Coulomb Interactions}
\label{Coul}

\markboth{Strong Correlations and the Fractional Quantum Hall Effect}{The Role of Coulomb Interactions}

\begin{figure}
\begin{center}
\epsfig{figure=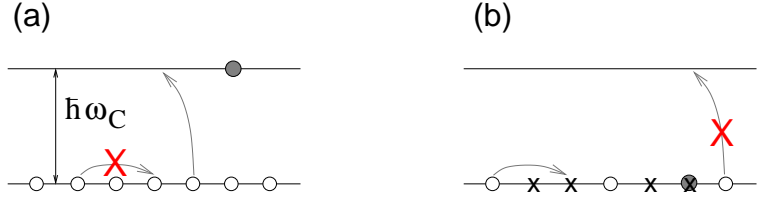,width=10cm,clip}
\end{center}
\caption{ {\sl (a)} Sketch of a completely occupied LL. An additional electron (grey circle) is forced to populate the next higher
LL because of the Pauli principle. 
{\sl (b)} Sketch of a partially filled LL. Because of the presence of unoccupied states in the LL (crosses), the Pauli principle 
does not prevent an additional electron (grey circle) to populate the next higher LL. The low-energy dynamical properties of
the electrons are described by excitations within the same LL (no cost in kinetic energy), and inter-LL excitations are now
part of the high-energy degrees of freedom.}
\label{fig19}
\end{figure}

As already mentioned above, the situation of a partially filled LL is somewhat opposite to that of $n$ completely occupied
levels, where one observes the IQHE. This difference is summarised in Fig. \ref{fig19} and it is also the origin of the 
different role played by the Coulomb repulsion between the electrons. In the case of $n$ completely filled
LLs, one has a non-degenerate (Fermi-liquid-like) ground state, where the interactions may be treated within a perturbative
approach. Indeed, any type of excitation involves a transition between two adjacent LLs that are separated by an energy
gap of $\hbar\omega_C$ [see Fig. \ref{fig19}(a)],\footnote{In order to simplify the discussion, 
we consider only the IQHE in non-relativistic quantum
Hall systems, but the arguments apply also to the RQHE in graphene.}
and we need to compare the Coulomb energy at the characteristic length scale $R_C=l_B\sqrt{2n+1}$ 
to this gap,
$$
\frac{V_C}{\hbar \omega_C}\sim \frac{m e^{3/2}}{\epsilon \hbar^{3/2}}(Bn)^{-1/2},
$$
which turns out to be nothing other than the usual dimensionless coupling constant
$$r_s=\frac{me^2}{\epsilon \hbar^2}n_{el}^{-1}
$$
for the 2D Coulomb gas in Fermi liquid theory \cite{mahan,GV}. 
The last expression is obtained by identifying the Fermi energy $E_F=\hbar^2k_F^2/2m$, in terms
of the Fermi wave vector $k_F$, with the energy of the last occupied LL $\hbar\omega_C n$.
The perturbative approach allows one, e.g., to describe collective electronic excitations in the IQHE, such as magneto-plasmon
modes (the 2D plasmon in the presence of a magnetic field) or magneto-excitons (inter-LL excitations that acquire a dispersion
due to the Coulomb interaction) \cite{KH}, or else the corresponding modes of the RQHE in graphene \cite{iyengar,RFG}.

In the case of a partially filled LL $n$ the situation is inverted: for an electronic excitation,
there are enough unoccupied states in the LL $n$ for an electron of the same level to hop to. 
From the point of view of the kinetic energy, there is no energy cost associated with such an excitation
({\sl low-energy} degrees of freedom) whereas an excitation to
the next higher (unoccupied) LL costs an energy $\hbar\omega_C$. Inter-LL excitations may then be neglected as belonging 
to {\sl high-energy} degrees of freedom [Fig. \ref{fig19}(b)]. Notice that all possible
distributions of $N$ electrons within the same partially filled LL $n$ therefore have the same kinetic energy, which
effectively drops out of the problem. The macroscopic degeneracy may be lifted by phenomena due to other energy scales, such
as those associated with the impurities in the sample or else the electron-electron interactions. The first hypothesis (impurities)
may be immediately discarded as the driving mechanism of the FQHE because, in contrast to the IQHE, the FQHE only occurs in
high-quality samples with low impurity concentrations. 
Indeed, the hierarchy of energy scales in the FQHE may be characterised by the succession
\beq\label{EnScales}
\hbar\omega_C\gtrsim V_C \gg V_{imp},
\eeq
and we therefore need to consider seriously the Coulomb repulsion, which govern the low-energy electronic properties in a
partially filled LL.\footnote{As for the IQHE, impurities play nevertheless 
an important role in the localisation of quasi-particles, which we need to invoke later in this chapter in order to explain the 
transport properties of the FQHE.}
Notice that we thus obtain a system of {\sl strongly-correlated} electrons for the description of which all perturbative approaches 
starting from the Fermi liquid are doomed to fail. The only hope one may have to describe the FQHE is then a well-educated guess
of the ground state. 

The most natural guess would be that the electrons in a partially filled LL behave as classical charged particles that form
a crystalline state in order to minimise their mutual Coulomb repulsion. Such a state  is also called Wigner crystal (WC) because it
was first proposed by Wigner in 1934 \cite{wigner}. A WC has indeed been thought -- before the discovery of the FQHE -- to be
the ground state of electrons in a partially filled LL \cite{FPA}. Even if the WC is the ground state at very low
filling factors, as it has been shown experimentally \cite{glatt}, this state may not allow for an explanation of the FQHE. Indeed,
the WC is a state that breaks a continuous spatial symmetry (translation invariance) and any such state has gapless
long-wave-length excitations ({\sl Goldstone modes}). The Goldstone mode of the WC (as of any other crystal) is the acoustic 
phonon the energy of which tends to zero at zero wave vector. One may thus compress the WC by changing the occupied surface 
in an infinitesimal manner or else by adding an electron without changing the macroscopic surface and pay only an infinitesimal
amount of energy. The ground state is therefore compressible, i.e. it is
not separated by an energy gap from its single-particle excitations, a situation
that is at odds with the FQHE.

\section{Laughlin's Theory}
\label{FQHE1}

\markboth{Strong Correlations and the Fractional Quantum Hall Effect}{Laughlin's Theory}

As a consequence of the above-mentioned considerations on the WC, one thus needs to search for a candidate for the ground state
that does not break any continuous spatial symmetry and that has an energy gap. Such a state is the {\sl incompressible quantum 
liquid} which was proposed by Laughlin in 1983 \cite{laughlin} the basic features of which we present in the present section.
We consider, here, only the FQHE in the lowest LL (LLL), for simplicity. There are different prescriptions to generalise 
the associated wave functions to higher LLs, e.g. with the help of Eq. (\ref{constrN}) (see MacDonald, 1984).
Experimentally, several FQHE states have been observed in the next higher LL $n=1$ although the majority of FQHE states
is found in the LLL.\footnote{There is even some slight indication for a 1/5 FQHE state in the next excited LL $n=2$
\cite{gervais}.}

\subsection{Laughlin's guess from two-particle wave functions}

In order to illustrate -- one cannot speak of a derivation -- Laughlin's wave function, we first need to remember the 
one-particle wave function of the LLL and then consider the corresponding two-particle wave function. We have already seen
in Sec. \ref{WFsym} that a one-particle wave function in the LLL is described in terms of an analytic function times a
Gaussian,\footnote{We neglect the numerical prefactors here that account for the normalisation of the wave functions.}
$$
\psi \sim z^{m'} e^{-|z|^2/4},
$$
in terms of the integer $m'=0, ..., N_{B}-1$, 
where we have absorbed now (and in the remainder of these lecture notes) the magnetic length in the definition of the complex
position, $z=(x-iy)/l_B$. 

Consider, in a second step, an arbitrary two-particle wave function. 
This wave function must also be an analytic function of both postions $z_1$ and
$z_2$ of the first and second particle, respectively, and may be a superposition of polynomials, such as e.g. of the basis states
\beq\label{2partWF}
\psi^{(2)}(z,Z)\sim Z^M z^m e^{-(|z_1|^2+|z_2|^2)/4},
\eeq
where we have defined the centre of mass coordinate $Z=(z_1+z_2)/2$ and the relative coordinate $z=(z_1-z_2)$.
The quantum number $m$ plays the role of the {\sl relative} angular momentum between the two particles, and $M$ is
associated with the {\sl total} angular momentum of the pair. Because of the analyticity of the LLL wave functions, $m$ must be an 
integer, and the exchange of the positions $z_1$ and $z_2$ imposes on $m$ to be {\sl odd} because of the electrons' 
fermionic nature. 

Laughlin's wave function \cite{laughlin} is a straight-forward 
$N$-particle generalisation of the two-particle wave function (\ref{2partWF}),
\beq\label{LaughlinWF}
\psi_m^L\left(\left\{z_j,z_j^*\right\}\right) = \prod_{k<l}\left(z_k - z_l\right)^m e^{-\sum_j|z_j|^2/4},
\eeq
where we have omitted the normalisation constants in order to simplify the notation
and where all indices run from 1 to 
the total number of particles $N$. Notice that there is no 
dependence on the centre of mass, but only on the relative coordinates between the particle pairs. Had there been 
such a dependence, described by a non-zero value of the total angular momentum quantum number $M\neq 0$, one would have
broken a continuous spatial symmetry, in which case the state would  describe a compressible rather than an 
incompressible state required for the FQHE, as we have mentioned above. We emphasize once again that Laughlin's wave function
is not based on a mathematical derivation, although we will see below that there exist some mathematical models for which
it describes the {\sl exact ground state}, but it is more appropriately characterised as a {\sl variational} wave function.

\subsubsection{Variational parameter}

The variational parameter in Laughlin's wave function (\ref{LaughlinWF})
is nothing other than the exponent $m$, with respect to which we would, in principle, need to optimised 
the wave function in order to
approximate the true ground state of the system. Notice, however, that due to the LLL analyticity condition and fermionic 
statistics, the exponent is restricted to odd integers, $m=2s+1$, in terms of the integer $s$.
Furthermore, this variational parameter turns
out to be fully determined by the filling factor $\nu$, as we will show with the following argument.\footnote{We are therefore
confronted with the somewhat bizarre situation where we dispose of a variational wave function with no possible variation.}

Consider Laughlin's wave function as a function of the position $z_k$ of some arbitrary but fixed electron $k$. There are $N-1$ factors
of the type $(z_k - z_l)^m$, one for each of the remaining $N-1$ electrons, $l$, 
occuring in the ansatz (\ref{LaughlinWF}), such that the highest power of $z_k$ is $m(N-1)$,
$$
\prod_{k<l}\left(z_k - z_l\right)^m \sim z_k^{m(N-1)}.
$$
Now, remember from Sec. \ref{WFsym} [see Eq. (\ref{NBsym})] that the highest power of the complex particle position is fixed by 
the number of states $N_B$ in each LL. This yields the relation
\beq\label{N-M}
mN - \delta =N_B
\eeq
between the number of particles $N$ and the number of flux quanta $N_B$ threading the system. Here, $\delta$ is some {\sl shift}
that is on the order of unity and that plays no role in the thermodynamic limit $N,N_B\rightarrow \infty$.\footnote{Notice, however,
that this shift plays an important role in numerical calculations, such as exact diagonalisation, when performed on special
geometries, such as on a sphere.}
Because the ratio between the number of particles and that of flux quanta is nothing other than the LL filling factor (\ref{filling}),
$\nu=N/N_B$, one notices that, in the thermodynamic limit, the ``variational parameter'' is entirely fixed by the filling factor,
i.e. 
\beq\label{LaughlinFF}
m=2s+1=\frac{1}{\nu} \qquad \Leftrightarrow \qquad \nu = \frac{1}{m} = \frac{1}{2s+1},
\eeq
and Laughlin's wave function is therefore a candidate wave function for the ground state at the filling factors
$$\nu=1, 1/3,1/5, ...$$

Remember that the odd value $m=2s+1$ is required by the fermionic nature of the electrons. Formally, one may though lift
this restriction and generalise Laughlin's wave function to {\sl bosonic} particles by choosing an even exponent $2s$.
Such bosonic Laughlin wave functions have been studied theoretically in the context of rotating cold Bose gases in an
optical trap \cite{cooper}.

\subsubsection{Laughlin's wave function at $\nu=1$}
\label{FQHEnu1}

It may seem, at first sight, astonishing that also the case of a completely filled LL for $\nu=1$ is described in terms of a 
Laughlin wave function with $m=1$ (or $s=0$). Indeed, the state 
$$
\psi\left(\left\{z_j\right\}\right)=f_N\left(\left\{z_j\right\}\right)
e^{-\sum_j|z_j|^2/4}
$$
would be non-degenerate and could thus be described in terms of
a Slater determinant,
\beq\label{eqSlater}
f_N(\{z_j\})=\det\left(\begin{array}{cccc}
z_1^0 & z_1^1 & \hdots & z_1^{N-1}\\
z_2^0 & z_2^1 & \hdots & z_2^{N-1}\\
\vdots & \vdots & & \vdots\\
z_N^0 & z_N^1 & \hdots & z_N^{N-1}
\end{array}\right),
\eeq
where we have omitted the ubiquitous Gaussian factor $\exp(-\sum_j|z_j|^2/4)$. Notice that the $j$-th line
in this determinant corresponds to all LLL states of the $j$-th particle
described in terms of the polynomials $z_j^m$. The determinant takes into account all permutations
of the $N$ particles over the $N$ particle positions, $z_1, ..., z_N$, and may 
be rewritten in a compact manner with the help of the co-called Vandermonde determinant,
\beq\label{eqVandermonde}
f_N(\{z_j\})=\prod_{i<j}\left(z_i-z_j\right),
\eeq
which is indeed nothing other than the polynimial prefactor in Laughlin's wave function (\ref{LaughlinWF}) with $m=1$.

Until now we have obtained an $N$-particle wave function from some very general symmetry considerations (LLL analyticity condition,
fermionic statistics, no broken continuous spatial symmetries), but we have not at all shown that it describes indeed the 
ground state responsible of the FQHE. In the following parts, we will therefore discuss the basic physical properties 
of this, for the moment rather abstract, mathematical entity. In a first step, we will discuss some energy properties of the
ground state and show that Laughlin's wave function is the exact ground state of a certain class of models that are qualitatively
compared to the physical one (Coulomb interaction). We will then discuss the fractionally charged quasi-particle excitations 
of this wave function.

\subsection{Haldane's pseudopotentials}
\label{HaldanePPsec}

\begin{figure}
\begin{center}
\epsfig{figure=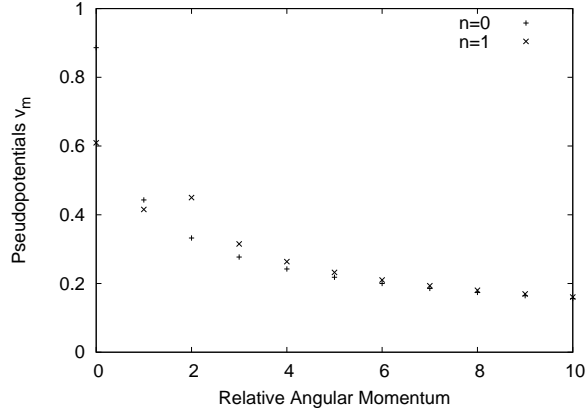,width=8cm,clip}
\end{center}
\caption{ Haldane's pseudopotentials for the Coulomb interaction in the LLs $n=0$ and $n=1$. Notice that we have plotted 
the pseudopotentials for both odd and even values of the relative angular momentum $m$ even though only odd values
matter in the case of fermions.}
\label{fig19bis}
\end{figure}

In order to describe the energetic properties of Laughlin's wave function (\ref{LaughlinWF}), we consider again the
two-particle wave function (\ref{2partWF}). Notice that this wave function is an {\sl exact} eigenstate for any central
interaction potential that depends only on the relative coordinate $z$ between particle pairs, such as it is the case for the 
Coulomb interaction, $V=V(|z|)$. One may therefore decompose the interaction potential in the relative angular momentum quantum
numbers $m$,
\beq\label{HaldanePP}
v_m\equiv \frac{\langle m,M|V|m,M\rangle}{\langle m,M|m,M\rangle},
\eeq
where the denominator takes into account the fact that we have not properly normalised the two-particle wave functions 
(\ref{2partWF}), $\psi^{(2)}(z,Z)=\langle z,Z|m,M\rangle$.\footnote{In order to simplify the notations,
we have omitted the LL quantum number $n=0$, which is 
the same for both particles in this wave function.}
The fact that there is no dependence on $M$ is a direct consequence of the assumption that we deal with 
a central interaction potential, i.e. $\langle z,Z|V|z',Z'\rangle = V(|z|)\delta_{z,z'}\delta_{Z,Z'}$. Furthermore, there 
are no off-diagonal terms of the form $\langle m,M|V|m',M\rangle$, with $m'\neq m$, as one may show explicitly in 
the polar representation $z=\rho \exp(i\phi)$,
$$\langle m,M|V|m',M\rangle \propto \int_{0}^{2\pi}d\phi\,\int_0^{\infty}d\rho\, \rho^{m+m'+1} V(\rho)e^{-i(m-m')\phi}
\propto \delta_{m,m'},$$
due to the integration over the polar angle.
The potentials $v_m$ obtained from the decomposition into relative angular momentum states are also called {\sl Haldane's
pseudopotentials} \cite{haldane}. They fully characterise the two-particle energy spectrum because the
kinetic energy is the same for all two-particle states $|m,M\rangle$, as described above. Notice that this is a very
special case: normally any repulsive interaction potential yields unbound states with a continuous energy spectrum, such
as the plane-wave states in scattering theory. Here,
the energy spectrum is discrete even if the interaction is repulsive, due to the presence of a quantising magnetic field.
Notice further that Haldane's pseudopotentials are an image of the real-space form of the interaction potential. Indeed,
if a pair of electrons is in a quantum state with relative angular momentum $m$, the average distance between the
electrons is $|z| \sim l_B\sqrt{2m}$.\footnote{This is similar to the average value of the radius at which the electron's
guiding centre is placed in the symmetric gauge (see Sec. \ref{WFsym}). Remember (e.g. from classical mechanics)
that the decomposition of a two-particle wave function
in relative and centre-of-mass coordinates maps the two-body problem to an effective one-body problem.}
Haldane's pseudopotential $v_m$ is therefore roughly the value of the original interaction potential at the relative 
distance $l_B\sqrt{2m}$, 
\beq\label{PPreal}
v_m \simeq V\left(|z|=l_B\sqrt{2m}\right),
\eeq
and the small-$m$ components of Haldane's pseudopotentials correspond to the short-range components of the underlying
interaction potential. Figure \ref{fig19bis} shows the pseudopotential expansion for the Coulomb interaction in the lowest 
($n=0$) and the first excited ($n=1$) LL.

Haldane's pseudopotentials are extremely useful in the description of the $N$-particle state as well. Indeed, the $N$-particle
interaction Hamiltonian $V$ may be rewritten in terms of pseudopotentials as
\beq\label{IntHam}
V=\sum_{i<j} V(|z_i - z_j|) = \sum_{i<j}\sum_{m'=0}^{\infty}v_{m'} \Pmath_{m'}(ij),
\eeq
where the operator $\Pmath_{m'}(ij)$ projects the electron pair $ij$ onto the relative angular momentum state $m'$.
Notice that due to the factor $\prod_{k<l}(z_k - z_l)^m$ in Laughlin's wave function (\ref{LaughlinWF}), no particle pair
is in a relative angular momentum state $m'<m$. If one then chooses, though somewhat artificially, all pseudopotentials with
a $m<m'$ to be positive (say 1) and all others zero,
\beq\label{PPmodel}
v_m'=\left\{ 
\begin{array}{cc}
1 & {\rm for~} m'<m\\
0 & {\rm for~} m'\geq m
\end{array}\right.
\eeq
one obtains $V\psi_m^L = 0$, i.e. Laughlin's wave function is the zero-energy eigenstate of the model (\ref{PPmodel}). Since the 
model describes an entirely repulsive interaction, all possible states must have an energy $E\geq 0$. Therefore, Laughlin's
wave function is even the {\sl exact} ground state of the model (\ref{PPmodel}). Furthermore, it is the only zero-energy state
because if one keeps the total number of particles and flux fixed, any other state different from that described by Laughlin's 
wave function involves a particle pair in a state with an angular momentum quantum number different from $m$. If it is smaller than
$m$, this particle pair is affected by the associated non-zero pseudopotential $m'$ and thus costs an energy on the order of $v_{m'}>0$.
If the particle pair is in a momentum state with $m'>m$, there is at least another pair with $m''<m$ in order to keep the filling
factor fixed, and this pair raises the energy. These general arguments show that any excited state involves a finite (positive)
energy given by a pseudopotential $v_{m'}$, with $m'<m$, which plays the role of an {\sl energy gap}. In this sense, the 
liquid state described by Laughlin's wave function is indeed an {\sl incompressible} state that already hints at the possibility
of a quantum Hall effect if we can identify the correct quasi-particle of this $N$-particle state that becomes localised by the
sample impurities.

Notice that the above considerations are based on an extremely artificial model interaction (\ref{PPmodel}) that has, at
first sight, very little to do with the physical Coulomb repulsion. However, the model is often used to generate numerically
(in exact-diagonali-sation calculations) the Laughlin state, which may then be compared to the Coulomb potential decomposed in
Haldane's pseudopotentials. This procedure has shown that theL aughlin state generated in this manner has an overlap of more than
$99\%$ with the state obtained from the Coulomb potential \cite{HaldRez,FOC}, which is amazingly high for a wave function
obtained from a well-educated guess. This  high accuracy of Laughlin's wave function may be understood in the
following manner: when one decomposes the Coulomb interaction potential in Haldane's pseudopotentials, one obtains a
monotonically decreasing function when plotted as a function of $m$ (see Fig. \ref{fig19bis}).
Furthermore, the component $v_1$ is much larger than
$v_3$ and all other pseudopotentials $v_m$ with higher values of $m$.\footnote{One has $v_1/v_3\simeq 1.6$ in the LLL. Notice that
pseudopotentials with even angular momentum quantum number $m$ do not play any physical role because of the fermionic
nature of the electrons.} These higher terms may be treated in a perturbative manner and do not change the ground state which
is protected by the above-mentioned gap on the order of $v_1>v_m$, with $m>1$.

Furthermore, we mention that, apart from its successful verification by exact-diagonalisation calculations \cite{HaldRez,FOC},
Laughlin, in his original paper \cite{laughlin}, showed within a variational calculation that the quantum liquid described
by his wave function (\ref{LaughlinWF}) has indeed a lower energy than the previously proposed WC. Again the reason for this 
unexpected feature is the capacity of Laughlin's wave function, which varies as $r^{2m}$ when two particles $i$ and $j$ 
approach each other with $r=|z_i - z_j|$, to screen the short-range components of the interaction potential. Notice that for 
a WC of fermions, the corresponding $N$-particle wave function decreases as $r^2$, as dictated by the Pauli principle.

\subsection{Quasi-particles and quasi-holes with fractional charge}

Until now, we have discussed some ground-state properties of Laughlin's wave function. We have seen 
that the Laughlin state at $\nu=1/m$ is insensitive to the short-range components of the
interaction potential described by Haldane's pseudopotentials $v_{m'}$ with $m'<m$, whereas excited states
must be separated from the ground state by a gap characterised by these short-range pseudopotentials. However, we have not
characterised so far the nature of the excitations. 

\begin{figure}
\begin{center}
\epsfig{figure=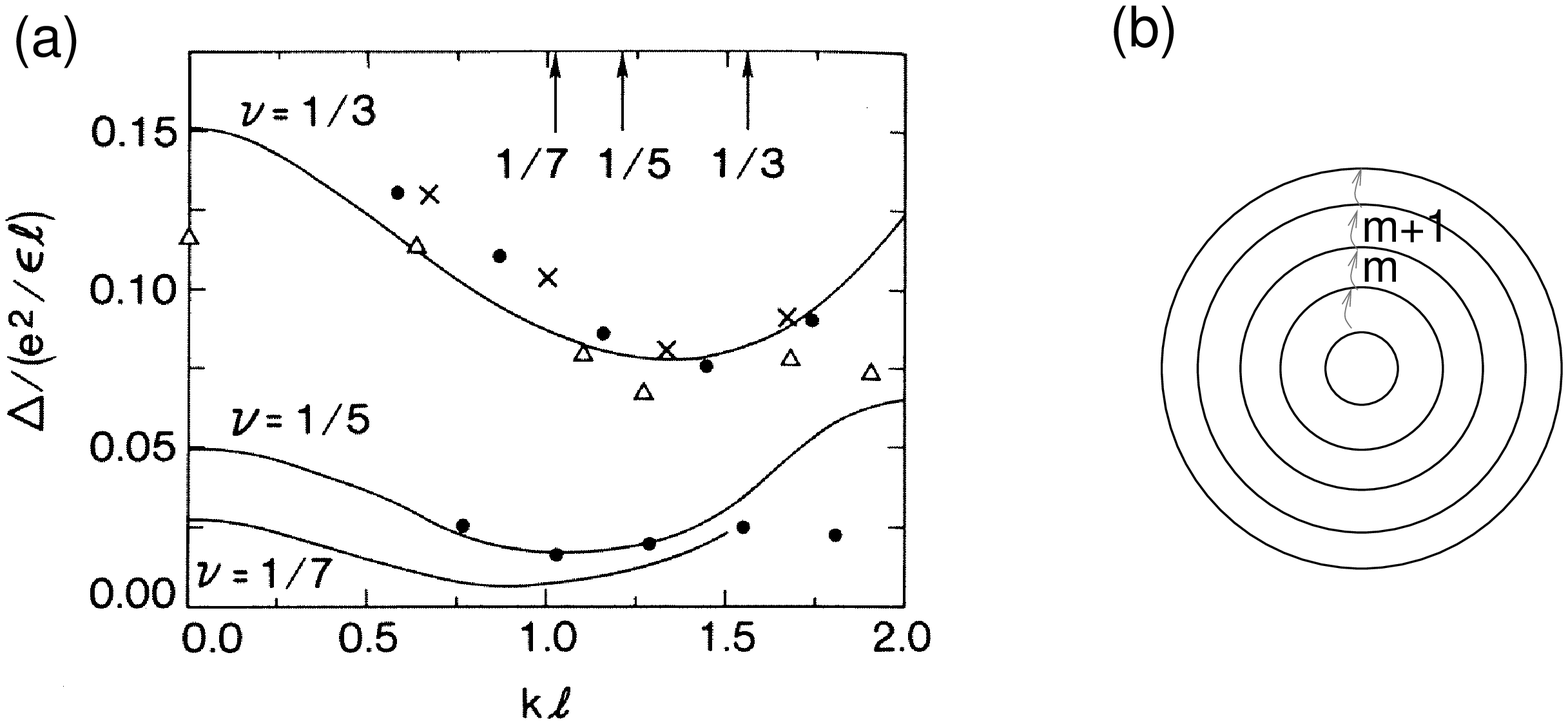,width=11cm,clip}
\end{center}
\caption{ {\sl (a)} Dispersion relation for collective charge-density-wave excitations (Girvin, MacDonald and Platzman, 1986; Girvin, 1999).
The continuous lines have been obtained in the so-called single-mode approximation (Girvin, MacDonald and Platzman, 1986) for
the Laughlin states at $\nu=1/3$, 1/5 and 1/7, whereas the points are exact-diagonalisation results (Haldane and Rezayi, 1985;
Fano, Ortolani and Colombo, 1986). The arrows indicated
the characteristic wave vector of the WC state at the corresponding densities.
{\sl (b)} Quasi-hole excitation. Each electron jumps from the state $m$ to the next-higher angular momentum state $m+1$.}
\label{fig20}
\end{figure}

There are two different sorts of excitations: (i) elementary excitations (quasi-particles or quasi-holes) that one obtains by adding
or removing charge from the system, and (ii) collective excitations at fixed charge. The latter are simply a charge-density-wave
excitation which consist of a superposition of particle-hole excitations at a fixed wave vector $\bq$ (the momentum of the pair) and
which may be shown to be gapped at all values of $\bq$. Its dispersion reveals a minimum
(called {\sl magneto-roton minimum})
at a non-zero value of the wave vector that indicates a certain tendency to form a ground state with modulated charge density,
such as a WC. The characteristic dispersion relation of these collective excitations is shown in Fig. \ref{fig20}(a).
However, we do not discuss collective excitations
here and refer the interested reader to the literature for a more detailed discussion \cite{GMP,PG,GirvinLH}, and concentrate here on
a presentation of the elementary excitations.

\subsubsection{Quasi-holes}

Elementary excitations are obtained when sweeping the filling factor 
slightly away from $\nu=1/m$. Remember that there are two possibilities
for varying the filling factor: adding charge to the system by changing the electronic density or adding (or removing) flux by
varying the magnetic field. Remember further [see Eq. (\ref{N-M})] that the number of flux is intimitely related to the number
of zeros in Laughlin's wave function. We therefore consider the ansatz
\beq\label{QH}
\psi_{qh}\left(z_0,\left\{z_j,z_j^*\right\}\right)=\prod_{j=1}^N (z_j - z_0)\,\psi_m^L\left(\left\{z_j,z_j^*\right\}\right)
\eeq
for an excited state.
Each electron at the positions $z_j$ thus ``sees'' an additional zero at $z_0$. In order to verify that 
this wave function adds indeed another flux quantum to the system, we may expand Laughlin's wave function (\ref{LaughlinWF}) 
formally in a polynomial,
$$\psi_m^L(\{z_j,z_j^*\})=\sum_{\{m_i\}} \alpha_{m_1, ...,m_N}z_1^{m_1}...\, z_N^{m_N}
e^{-\sum_j|z_j|^2/4},$$
where the $\alpha_{m_1, ...,m_N}$ describe the expansion coefficients. We now choose the position $z_0$ at the centre of
the disc, in which case the wave function of the excited state (\ref{QH}) simply reads
$$\psi_{qh}(\{z_j,z_j^*\})=\sum_{\{m_i\}} \alpha_{m_1, ...,m_N}z_1^{m_1+1}...\, z_N^{m_N+1}
e^{-\sum_j|z_j|^2/4},$$
i.e. each exponent is increased by one, $m_i\rightarrow m_i+1$. This may be illustrated in the following manner: each electron
jumps from the
angular momentum state $m$ to a state in which the angular momentum is increased by one (see Fig. \ref{fig20}), leaving behind
an empty state at $m=0$. The excitation is therefore called a {\sl quasi-hole} as we have already suggested by the subscript in
Eq. (\ref{QH}). This also affects the quantum state with highest angular momentum $M$, i.e. we have increased the sample
size by the surface occupied by one flux quantum, while keeping the number of electrons fixed.\footnote{Naturally, the total
surface of the quantum Hall system remains constant, but physically we have slightly {\sl increased} the $B$-field. Each
quantum state occupies then an infinitesimally smaller surface $2\pi l_B^2$, such that the system may accomodate for one more quantum state, $M=N_B\rightarrow N_B+1$.} 
Furthermore, this quasi-hole
is associated with a {\sl vorticity} if one considers the phase of the additional factor in Eq. (\ref{QH}),
$$\psi_{qh}(z_0 = 0,\{z_j,z_j^*\})\propto \prod_{j}^N e^{-i\theta_j}\times \psi_m^L\left(\left\{z_j,z_j^*\right\}\right),
$$
i.e. each particle that circles around the origin $z_0=0$ experiences an additional phase shift of $2\pi$ as compared
to the original situation described in terms of Laughlin's wave function (\ref{LaughlinWF}). This is 
reminiscent of the vortex excitation in a type-II superconductor \cite{tinkham}.

We have seen above that one can create a quasi-hole excitation at the postion $z_0$ by introducing one additional flux quantum,
$N_B\rightarrow N_B+1$,
which lowers the filling factor by a tiny amount. However, we have not yet determined the charge associated with this
elementary excitation. This charge may be calculated by considering the filling factor fixed, i.e. we need to add some (negative)
charge to compensate the extra flux quantum in the system. From Eq. (\ref{N-M}), we notice that the relation between
the extra flux $\Delta N_B$ and the compensating extra charge $\Delta N$ is simply given by 
\beq\label{fracQb}
m\Delta N = \Delta N_B \qquad \Leftrightarrow \qquad \Delta N = \frac{\Delta N_B}{m}\, .
\eeq
This very important result is somewhat unexpected: in order to compensate one additional flux quantum ($\Delta N_B=1$),
one would need to add the {\sl $m$-th fraction of an electron}. The charge deficit caused by the quasi-hole excitation is 
therefore
\beq\label{fracQ}
e^* = \frac{e}{m}\, ,
\eeq
i.e. the quasi-hole carries {\sl fractional charge}. 

\subsubsection{Quasi-particles}

In the preceding 
paragraph, we have  considered a quasi-hole excitation that is obtained by introducing an additional flux quantum in the system
[or, mathematically, an additional zero in the Laughlin wave function, see Eq. (\ref{QH})]. Naturally, one may also {\sl lower} the
number of flux quanta by one in which case one obtains a {\sl quasi-particle} excitation with opposite vorticity as compared
to that of the quasi-hole excitation. This opposite vorticity suggests that we use a prefactor $\prod_{j=1}^N(z_j^* - z_0^*)$, 
instead of $\prod_{j=1}^N(z_j - z_0)$ as in the expression (\ref{QH}), in
order to create a quasi-particle excitation at the position $z_0$. Remember, however, that the resulting wave function would
have unwanted components in higher LLs because the analyticity condition of the LLL is no longer satisfied. In order to heal
the quasi-particle expression, one formally projects it into the LLL,
\beq\label{QP}
\psi_{qp}\left(z_0,\left\{z_j,z_j^*\right\}\right)=\Pmath_{LLL}\prod_{j=1}^N (z_j^* - z_0^*)\,
\psi_m^L\left(\left\{z_j,z_j^*\right\}\right).
\eeq
There are several manners of taking into account this projection $\Pmath_{LLL}$. A common one consists of replacing each
occurence of the non-analytic variables $z_j^*$ (and powers of them) in the polynomial part of the wave function by 
a derivative with respect to $z_j$ in the same polynomial \cite{jach}. By partial integration, this amount to deriving
the Gaussian factor by $(\partial_{z_j})^m$ which, up to a numerical prefactor, yields exactly the non-analytic polynomial
factor ${z_j^*}^m$. We will encounter this projection scheme again in the discussion of the CF generalisation of
Laughlin's wave function (Sec. \ref{CF}).

\subsection{Experimental observation of fractionally charged quasi-particles}

\begin{figure}
\begin{center}
\epsfig{figure=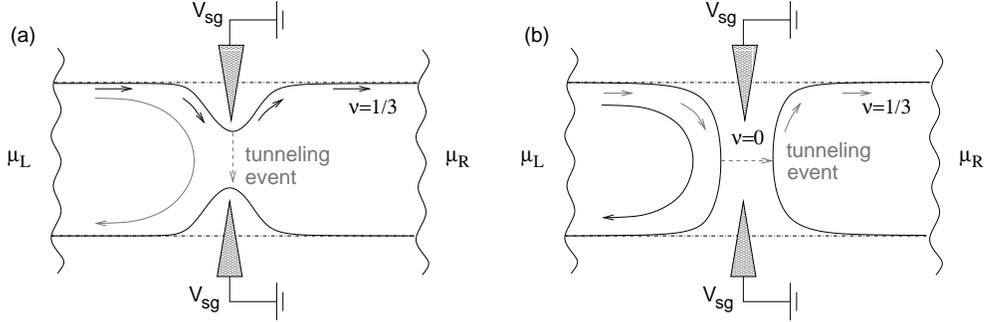,width=13cm,clip}
\end{center}
\caption{ Experimental setup for the observation of fractionally charged quasi-particles. In addition to the usual geometry,
one adds, at the upper and the lower edges, side gates that are used to deplete the region around the gates
by the application of a voltage $V_{sg}$. The filling
factor is chosen to be $\nu=1/3$. As a result,
the edge states at the opposite edges are brought into close vicinity. {\sl (a)} Weak-backscattering limit.
The incompressible liquid has a {\sl bottleneck}
at the side gates, i.e. the edges are so close to each other that a tunneling event between them has a finite probability. A particle
injected at the left contact may thus be backscattered (grey arrow) in a region filled by the incompressible Laughlin liquid, although
the majority of the particles reaches the right contact (black arrows).
{\sl (b)} Strong-backscattering limit.
If one increases the side-gate voltage $V_{sg}$, the 
incompressible $\nu=1/3$ liquid is eventually cut into two parts separated by a fully depleted region ($\nu=0$). In this case,
backscattering is the majority process (black arrow), 
and a tunneling may occur over the depleted region such that a particle injected at the 
left contact may still reach the right one (grey arrows).}
\label{fig21}
\end{figure}

That the fractional charge of Laughlin quasi-particles\footnote{From now on, we use the term ``(Laughlin) quasi-particles'' generically 
in order to denote quasi-particles {\sl and} quasi-holes.}
is not only a mathematical concept but a physical reality has been proven in a spectacular manner in so-called {\sl shot-noise}
experiments on the $\nu=1/3$ FQHE state \cite{SN1,SN2}.\footnote{Later this kind of experiment has been repeated for other FQHE states.}
In these experiments, one constrains the quantum Hall system with the help of side gates (see Fig. \ref{fig21}) that are used to deplete
the region in their vicinity via the application of a gate voltage $V_{sg}$. As a consequence of this depletion the quantum Hall 
system has a bottleneck where the corresponding edge states are brought into spatial vicinity [Fig. \ref{fig21}(a)]
or where the incompressible quantum liquid may even be cut into two parts separated by a completely depleted barrier 
[Fig. \ref{fig21}(b)]. In the first case, an injected charge may be backscattered in a tunneling event at the bottleneck over a 
region filled by the $\nu=1/3$ liquid (weak-backscattering limit).  If one increases the side-gate voltage, the incompressible
liquid becomes eventually cut into two parts separated by a completely depleted barrier, and one obtains the strong-backscattering
limit.

In a shot-noise measurement, one does not only measure the average current $\bar{I}$ (over a certain time interval) but simultaneously
the (square of the) current fluctuation $\bar{(\Delta I)^2}$ which is proportional to the carrier charge.
If the elementary charged excitations are $e^*=e/3$ quasi-particles and not electrons, one
may expect to measure this particular charge. The experiments \cite{SN1,SN2} have indeed shown
that the charge measured in the shot noise is $e^*=e/3$ if the tunneling process takes place at a bottleneck filled with 
the incompressible quantum liquid [Fig. \ref{fig21}(a)], whereas it is the usual elementary charge $e$ in the case of a
tunneling process over a depleted region [Fig. \ref{fig21}(b)].

\subsection{Laughlin's plasma analogy}
\label{PlasmaLaugh}

A compelling physical picture of Laughlin's wave function (\ref{LaughlinWF}) and the properties of its elementary
excitations (\ref{QH}) and (\ref{QP}) with fractional charge has been provided by Laughlin himself \cite{laughlin},
in terms of an analogy with a {\sl classical 2D one-component plasma}. In the present subsection, we present the basic
ideas and results of this plasma analogy, for completeness and pedagogical reasons. However, no new results will come 
out of this analogy here, as compared to those derived above.

Remember from basic {\sl quantum mechanics} that the modulus square of a quantum-mechanical wave function may be interpreted as a 
statistical probability distribution. For Laughlin's wave function (\ref{LaughlinWF}), one obtains the
probability distribution
$$
\left|\psi_{m}^L\left(\left\{z_j\right\}\right)\right|^2=
\prod_{i<j}\left|z_i-z_j\right|^{2m} e^{-\sum_j|z_j|^2/2}.
$$
Now, remember from {\sl classical statistical mechanics} that a probability distribution in the canonical ensemble
is the Boltzmann weight, $\exp(-\beta\Hmath)$, of some Hamiltonian $\Hmath$ and that the classical partition function,
which encodes all relevant statistical information, is obtained from a sum over the Boltzmann weights of all possible
configurations $\Cmath$, $\Zmath=\sum_{\Cmath}\exp[-\beta\Hmath(\Cmath)]$. Laughlin's plasma analogy consists precisely
of the identification of the modulus square of his wave function with the Boltzmann weight of some {\sl mock} Hamiltonian
$U_{cl}$.\footnote{mock: {\sl singlish} for fake; mainly used the description of Singaporean catering food.}
The mock Hamiltonian may be obtained exactly from this identification,
\beq\label{PlasmaAnal}
-\beta U_{cl}=\ln\left|\psi_{m}^L\left(\left\{z_j\right\}\right)\right|^2,
\eeq
and one obtains, by choosing somewhat artificially $\beta=2/q$,\footnote{Notice that Laughlin's wave function describes 
a system at $T=0$, such that temperature does not intervene in the expressions. The choice is purely formal.}
\beq\label{eqPlasma}
U_{cl}=-q^2\sum_{i<j}\ln\left|z_i-z_j\right|+q\sum_j \frac{|z_j|^2}{4}.
\eeq
This is nothing other than the classical Hamiltonian of a 2D one-component plasma, in terms of the charge
\beq\label{plasmaQ}
q=m=2s+1
\eeq
of the plasma particles. The first term of Eq. (\ref{eqPlasma}) reflects the interactions between the charged plasma particles,
whereas the second term describes their interaction with a neutralising background of positive charge, as in the case of the
jellium model of the Coulomb gas \cite{mahan,GV}. This may be seen best with the help of Poisson's equation, $-\Delta \phi=
2\pi q n_q(\br)$, for an electrostatic 2D potential due to the charge density $qn_{q}$. The first term describes then indeed
particles with charge $q$ interacting via the 2D Coulomb interaction potential $\phi(\br)= - \ln(|\br|/l_B)$, and the second
term is the interactions with the neutralising background because $\Delta |\br|^2/4l_B^2 = 1/l_B^2=2\pi n_{B}$, where the 
flux density $n_B$ may thus be viewed as the charge density of the positively charged background. 

In order to minimise the energy of the mock Hamiltonian $U_{cl}$, which corresponds to a distribution of highest weight, 
the 2D plasma thus needs to be charge-neutral, i.e. the charge density of the plasma particles $q n_{el}$ must be compensated
by that of the background $n_B$,
\beq\label{eqNeutrCharge}
n_B-qn_{el}=0,
\eeq
which, together with Eq. (\ref{plasmaQ}), yields nothing other than the relation between the filling factor $\nu$ and
the exponent in Laughlin's wave function (\ref{LaughlinFF}), $\nu=n_{el}/n_B=1/m$.

The plasma analogy does not only apply to the ground-state wave function (\ref{LaughlinWF}) but also to the quasi-hole excitation
(\ref{QH}). The additional factor $\prod_{j=1}^N(z_j - z_0)$ in the quasi-hole wave function (\ref{QH}) yields, within
the plasma analogy (\ref{PlasmaAnal}), an additional term 
\beq\label{PlasmaImp}
V=-q\sum_{j=1}^N \ln \left|z_j-z_0\right|
\eeq
to the mock Hamiltonian (\ref{eqPlasma}), $U_{cl}\rightarrow U_{cl}+V$. This additional term may be interpreted as the interaction
of the plasma particles with an ``impurity'' of unit charge at the position $z_0$. In order to maintain  charge 
neutrality, the impurity needs to be screened by the plasma particles. Since the charge of each plasma particle is 
$q=m=2s+1$ and thus greater than unity, one needs $1/q$ plasma particles to screen the impurity of charge one. 
Remember that each plasma particle represents one electron of unit charge in the original Laughlin liquid. One therefore obtains
the same charge fractionalisation of the Laughlin quasi-particle (\ref{fracQ}), $e^*=e/m$, as in the original quantum model.

\section{Fractional Statistics}
\label{FQHE2}

\markboth{Strong Correlations and the Fractional Quantum Hall Effect}{Fractional Statistics}

\subsection{Bosons, fermions and anyons -- an introduction}

\begin{figure}
\begin{center}
\epsfig{figure=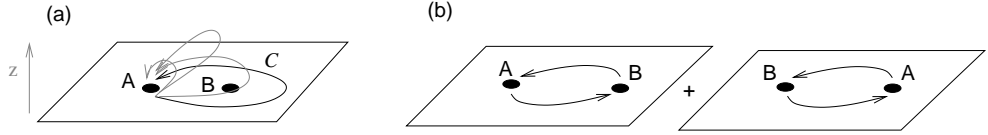,width=13cm,clip}
\end{center}
\caption{ {\sl (a)} Process in which a particle A moves on a path $\Cmath$ around another particle B. In three space dimensions, one
may profit from the third direction ($z$-direction) to lift the path over particle B and thus to shrink the path into a single point. 
{\sl (b)} Process equivalent to moving A on a closed path around B which consists, apart 
from a topologically irrelevant translation, of two successive exchanges of A and B.}
\label{fig22}
\end{figure}

One of the most exotic consequences of charge fractionalisation in 2D quantum mechanics, exemplified by Laughlin quasi-particles, 
is {\sl fractional statistics}. Remember that, in three space dimensions, the quantum-mechanical treatment of two and more particles yields
a {\sl superselection rule} according to which quantum particles are, from a statistical point of view, 
either {\sl bosons} or {\sl fermions}. 
This superselection rule is no longer valid in 2D (two space dimensions), and one may find intermediate
statistics between bosons and fermions. The corresponding particles are called {\sl anyons}, because the statistics may be {\sl any}.
The present section is meant to illustrate these amazing aspects of 2D quantum mechanics, and we try to avoid a too formal or
mathematical treatment. We refer, again, the interested reader to the more detailed literature \cite{nayak}.

In order to illustrate the different statistical (i.e. exchange) properties of two quantum particles in three and two space dimensions,
consider a particle A that moves adiabatically 
on a closed path $\Cmath$ in the $xy$-plane around another one B of the same species (see Fig. \ref{fig22}).
We choose the path to be sufficiently far away from particle B and the 
two particles to be sufficiently localised such that we can neglect corrections due to the overlap
between the two corresponding wave functions. Notice first that such a process $\Tmath$ is equivalent, apart from a topologically
unimportant translation, to two successive exchange processes $\Emath$, in which one exchanges the positions of A and B.
Algebraically, this may be expressed in terms of the corresponding operators as
\beq\label{StatExch}
\Emath^2 = \Tmath \qquad {\rm or} \qquad \Emath = \pm \sqrt{\Tmath},
\eeq
modulo a translation. 

Let us discuss first the three-dimensional case. Because of the presence 
of the third direction ($z$-direction), one may elevate the closed
path in this direction while keeping the position of particle A fixed in the $xy$ plane. We call the elevated path $\Cmath'$.
Furthermore, one may now shrink the closed loop $\Cmath'$ into a single point at the position A without passing by the position
of particle B which remains in the $xy$-plane. This final (point-like) path is called $\Cmath''$.
Although this procedure may seem somewhat formal, a quantum-mechanical exchange process does principally not specify 
the exchange path in order to define whether a particle is a boson or a fermion, but only its {\sl topological} properties.
From a topological point of view, all paths that may be continuously deformed into each other define a {\sl homotopy class}
\cite{mermin}. 
Equation (\ref{StatExch}) must therefore be viewed as an equation for homotopy classes in which a simple translation 
and an allowed deformation are irrelavant.
As a consequence of these considerations, the simple point-like path $\Cmath''$ 
at the position of particle A, which may be formally described by 
$\Cmath''=1$, is in the same homotopy class as the original path $\Cmath$. Therefore, the associated processes
are the same, and one has 
\beq\label{BosFermStat}
\Tmath=\Tmath(\Cmath)=\Tmath(\Cmath'')=\bone \qquad {\rm and~thus}\qquad \Emath=\sqrt{\bone},
\eeq
where the last equation is symbolic in terms of the one operator. It indicates that the quantum-mechanical operator
$\Emath$, corresponding to particle exchange,
has two eigenvalues that are the two square roots of unity, $e_B=\exp(2i\pi)=1$ and $e_F=\exp(i\pi)=-1$. This
is precisely the above-mentioned superselection rule, according to which all quantum particles in three space dimensions
are either {\sl bosons} ($e_B=1$) or {\sl fermions} ($e_F=-1$). 

In two space dimensions, this topological argument yields a completely different result. It is not possible to shrink a path $\Cmath$ 
enclosing the second particle B into a single point at the position of A, without passing by B itself. This means that the position of
B must be an element of the path at a certain moment of the shrinking process, which cannot profit from a third dimension
in order to elevate the loop on which it moves 
above the $xy$-plane. The single point still represents a homotopy class of paths, but these paths
do not enclose another particle, and $\Cmath$ is therefore an element of another homotopy class, i.e. the one of all
paths starting from A and enclosing only the particle B. If there are more than two particles present, the homotopy classes are described
by the integer number of particles enclosed by the paths in this class. From an algebraic point of view, the exchange processes
are no longer described by the two roots of unity, $1$ and $-1$, but by the so-called {\sl braiding group}, and the classification
in bosons and fermions is no longer valid. In the simplest case of Abelian statistics,\footnote{There are more complicated
cases of non-Abelian statistics, in which the exchange processes of more than two different particles no longer commute, but we do not
discuss this case here and refer the reader to the review by Nayak {\sl et al.} \cite{nayak}.}
one needs to generalise the commutation relation
\beq\label{PauliStat}
\psi(\br_1)\psi(\br_2)=\pm \psi(\br_2)\psi(\br_1),
\eeq
for bosons and fermions, respectively, to 
\beq\label{AnyStat}
\psi(\br_1)\psi(\br_2)=e^{i\alpha\pi} \psi(\br_2)\psi(\br_1),
\eeq
where $\alpha$ is also called the {\sl statistical angle}. One has $\alpha=0$ for bosons and $\alpha=1$ for fermions,
and all other values of $\alpha$ in the interval between 0 and 2 for {\sl anyons}. Sometimes anyonic statistics is also called
{\sl fractional statistics} -- indeed all physical quasi-particles, such as those relevant for 
the FQHE, have an angle that is a fractional 
(or rational) number, but there is no fundamental objection that irrational values of the statistical angle should be excluded.

Before discussing the anyonic nature of Laughlin quasi-particles, we need to mention an important issue in these statistical
considerations. We know that fermions are forced to satisfy Pauli's principle which excludes double occupancy of a single
quantum state, whereas the number of bosons per quantum state is unrestricted. What about anyons then? In the context of
quantum fields the Pauli principle yields, via Eq. (\ref{PauliStat}) for $\br=\br_1=\br_2$,
$$
\psi(\br)\psi(\br)=0.
$$
For an arbitrary statistical angle, one obtains in the same manner, from Eq. (\ref{AnyStat}),
\beq\label{PauliGen}
\left(1-e^{i\alpha\pi}\right)\psi(\br)\psi(\br)=0,
\eeq
which may be viewed as a {\sl generalised Pauli principle for 2D anyons} \cite{haldane2}.
Only if $\alpha=0$ modulo 2, we may have $\psi(\br)\psi(\br)\neq 0$ in order to satisfy Eq. (\ref{PauliGen}).
Otherwise, when $\alpha\neq 0$ modulo 2, we necessarily have $\psi(\br)\psi(\br)=0$. Anyons are, thus, from
an exclusion-principle point of view more similar to fermions than to bosons.

\subsection{Statistical properties of Laughlin quasi-particles}

We may now apply the above general statistical considerations to the case of Laughlin quasi-particles. The basic
idea is to describe the statistical angle as an Aharonov-Bohm phase due to some gauge field that is generated by the
flux bound to the charges included in a closed loop $\partial \Sigma$. This closed loop, around which a quasi-particle
moves adiabatically, encloses a surface $\Sigma$. 
The gauge field is not to be confunded with the one which generates the true magnetic field $B$ -- it 
is rather a {\sl mock} (or fake) field $\bA_M$ (with $\bB_M=\nabla\times \bA_M$)
that generates the flux bound, e.g., by the electrons in the Laughlin liquid via the relation (\ref{N-M}).
We consider the case where the area $\Sigma$ is filled with $N_{el}(\Sigma)$ electrons condensed in 
an incompressible quantum liquid described by Laughlin's
wave function (\ref{LaughlinWF}) and $N_{qh}(\Sigma)$ quasi-hole excitations (\ref{QH}), 
such that there are two contributions to $B_M=|\bB_M|$,
\beq\label{Bmock}
B_M\Sigma = N_{\rm flux}\frac{h}{e} = \left[m N_{el}(\Sigma) + N_{qh}(\Sigma)\right]\frac{h}{e}\ .
\eeq

The corresponding Aharonov-Bohm phase, which the quasi-particle picks up when turning around the area $\Sigma$ on 
the boundary path $\partial\Sigma$, is given by
$$\Gamma_{A-B}=2\pi\frac{e^*}{h}\oint_{\partial\Sigma} d\br\cdot\bA_M(\br)
 = 2\pi\frac{e^*}{h}\int_{\Sigma} d^2r\, B_M(\br),$$
where $e^*=e/m$ is the charge of the quasi-particle and where we have used Stoke's theorem to convert the line integral
of $\bA_M$ on $\partial \Sigma$ into a surface integral of $B_M$ over the area $\Sigma$. The Aharonov-Bohm phase has therefore
two contributions, one $\Gamma_{el}$
that stems from the electrons condensed in the Laughlin liquid and the other one $\Gamma_{qh}$ that is due to the 
enclosed quasi-holes.
One obtains from Eq. (\ref{Bmock})
\beq\label{ABel}
\Gamma_{el}=2\pi\frac{e^*}{e} m N_{el}=2\pi N_{el},
\eeq
for the enclosed electrons, i.e. an integer times $2\pi$. Notice that this contribution to the Aharonov-Bohm phase may not
be interpreted in terms of a statistical angle because it
does not describe a true exchange process: the involved particles are not 
of the same type -- we have chosen a quasi-particle to move on a path enclosing condensed electrons. 
However, had we chosen an {\sl electron} rather than a {\sl quasi-hole} to move along the path $\partial \Sigma$,
the Aharonov-Bohm phase,
$$\Gamma_{el-el}=2\pi\frac{e}{e} mN_{el}(\Sigma),
$$
would give rise to a statistical angle $\alpha = mN_{el}(\Sigma)$.\footnote{Remember that the statistical angle 
is defined with respect to an exchange process $\Emath$ which
is the square root of the process $\Tmath$ considered here [Eq. (\ref{StatExch})]. 
The relation between the statistical angle and the Aharononv-Bohm
phase is therefore $\Gamma=2\pi \alpha$ and not $\pi\alpha$.} 
If we have only one electron enclosed by the path, $N_{el}(\Sigma)=1$, the statistical angle is simply the odd
integer $m$, which is equal to 1 (modulo 2), as it should be for fermions.

A more interesting situation arises when the path encloses Laughlin quasi-holes, in which case the Aharonov-Bohm phase reads
\beq\label{ABqh}
\Gamma_{el}=2\pi\frac{e^*}{e}  N_{qh}=2\pi \frac{N_{qh}}{m}.
\eeq
Consider a single quasi-hole in the area $\Sigma$, $N_{qh}=1$: one encounters the rather unusual situation in which the Aharonov-Bohm
phase is a {\sl fraction} of $2\pi$, and the associated statistical angle is $\alpha=1/m$. This illustrates that
Laughlin quasi-holes are indeed anyons with fractional statistics, as we have argued above.

\section{Generalisations of Laughlin's Wave Function}
\label{FQHE3}

\markboth{Strong Correlations and the Fractional Quantum Hall Effect}{Generalisations of Laughlin's Wave Function}

Although Laughlin's wave function (\ref{LaughlinWF}) has been extremely successful in the description of the FQHE
at $\nu=1/3$ and $1/5$, it is not capable of describing all observed FQHE states. Indeed, there are e.g. FQHE 
states at $\nu=2/5,\,3/7,\,4/9, ...$ corresponding to the series $p/(2p+1)$, or more generally to $p/(2sp+ 1)$, 
in terms of the integers $s$ and $p$,
which may be accounted for within composite-fermion (CF) theory, 
which we present below. Furthermore, even-denominator FQHE states have been 
observed at $\nu=5/2$ and $7/2$ \cite{willett}, 
in the first excited LL ($n=1$), and, in wide quantum wells or bilayer quantum Hall systems,
at $\nu=1/2$ and $\nu=1/4$ \cite{luhmann,shabani}. 
Whereas the latter may be understood within a multi-component picture, which we will briefly
introduce in Chap. \ref{MultiC}, the states at $\nu=5/2$ and $7/2$ may find their explanation in terms of a so-called
{\sl Pfaffian} wave function. Both the CF and the Pfaffian wave functions are sophisticated generalisations of Laughlin's
original idea.

\subsection{Composite Fermions}
\label{CF}

Soon after the discovery of the most prominent FQHE state at $\nu=1/3$, a lot of other states have been observed at the filling
factors $\nu=p/(2sp+1)$. 
In a first theoretical approach, these states were interpreted in the framework of a hierarchy scheme 
\cite{haldane,halperin} according to which the quasi-particles of the Laughlin (parent) state, such as $\nu=1/3$, condense
themselves into a Laughlin-type (daughter) state, due to their residual Coulomb repulsion -- remember that the Laughlin quasi-particles
are charged with charge $e^*=e/m$. In this picture, the $2/5$ state would be the daughter state formed of Laughlin
quasi-particle excitations of the $1/3$ state.

\begin{figure}
\begin{center}
\epsfig{figure=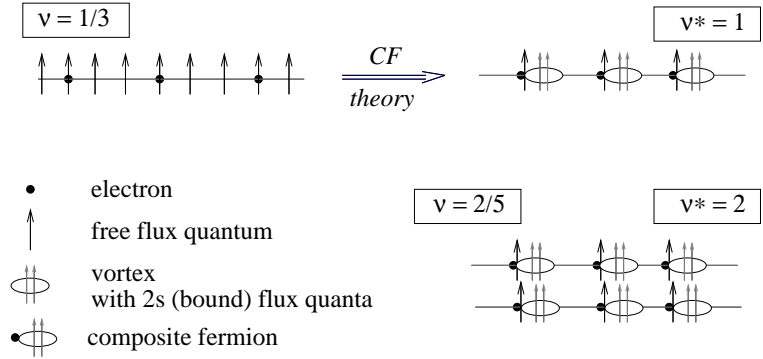,width=10cm,clip}
\end{center}
\caption{ Schematic view of composite fermions. The electronic state at $\nu=1/3$ may be interpreted as a CF state at an
integer CF filling factor $\nu^*=1$, where each vortex bound to an electron carries $2s$ (here $s=1$) flux quanta. In 
the same manner a CF filling factor $\nu^*=2$ gives rise to an (electronic) FQHE state at $\nu=2/5$.}
\label{fig23}
\end{figure}

An alternative picture, though related to the above-mentioned hierarchy scheme, was proposed by Jain in 1989 \cite{Jain1,Jain2}. 
The basic idea
consists of a reinterpretation of Laughlin's wave function (\ref{LaughlinWF}): consider only the polynomial part, the Gaussian
$\exp(-\sum_j^N|z_j|^2/4)$ being an ubiquitous factor which finally needs to be multiplied with the polynomial wave function,
\beq\label{LaughlinDec}
\psi_{m}^L\left(\left\{z_j\right\}\right) = \prod_{k<l}(z_k - z_l)^{2s+1} = \prod_{k<l}(z_k - z_l)^{2s}\prod_{k<l}(z_k - z_l).
\eeq
In the last step of this equation, we have split the product into two parts, one with the exponent $2s$, which
we call the {\sl vortex} part, and another one with the exponent $1$. 

Before introducing Jain's generalisation, let us interpret the above wave function in terms of the statistical properties
introduced in the last section. Quite generally, one may express any LLL $N$-particle wave function $\psi_{LLL}$
as a product of such
a vortex factor and another (residual) wave function $\psi_{res}$, 
$$
\psi_{LLL}\left(\left\{z_j\right\}\right) = \prod_{k<l}(z_k - z_l)^{m'}\psi_{res}\left(\left\{z_j\right\}\right).
$$
If the original wave function is fermionic, i.e. anti-symmetric with respect to an exchange process of an arbitrary particle
pair, the symmetry properties of $\psi_{res}$ depend on the parity of $m'$. If it is odd, $\psi_{res}$ must be a symmetric (bosonic)
wave function, and if it even, both the original and the residual wave functions are 
anti-symmetric (fermionic). In terms of the above-mentioned
gauge field $\bA_M(\br)$, the statistical angle associated with the vortex factor is just given by the parity of $m'$, which
may be viewed as the number of {\sl flux quanta attached to each particle} at the positions $z_j$. Flux attachment may thus
be used, in 2D quantum mechanics, to transform fermions into bosons and {\sl vice versa}. 

In the case of the above decomposition (\ref{LaughlinDec}) of Laughlin's wave function,
the vortex part attaches $s$ pairs of flux quanta
to each particle position and therefore does not affect the statistical properties of the wave function. 
The second factor 
$$\chi_{\nu^*=1}(\{z_j\})=\prod_{k<l}(z_k - z_l)$$ 
is indeed fermionic and
corresponds, as we have mentioned in Sec. \ref{FQHEnu1}, to a completely filled
LL at a virtual (CF) filling factor of $\nu^*=1$, the true filling factor being still $\nu=1/(2s+1)$. This is schematically
represented in Fig. \ref{fig23}.

Jain's generalisation consists of replacing the term $\prod_{k<l}(z_k - z_l)$ by any other Slater determinant 
$\chi_{\nu^*=p}(\{z_j, z_j^*\})$
of $p$ completely 
filled LLs, with a CF filling factor $\nu^*=p$,
\beq\label{JainWF}
\psi^J(\{z_j,z_j^*\})=\mathcal{P}_{LLL}\prod_{k<l}\left(z_k-z_l
\right)^{2s}\chi_{\nu^*=p}(\{z_j,z_j^*\}),
\eeq
where we need to take into account the same projection $\Pmath_{LLL}$ to the LLL as in the case of quasi-particle excitations
(\ref{QP}) because, contrary to the $\nu^*=1$ case, the wave function $\chi_{\nu^*=p}(\{z_j,z_j^*\})$ has 
by construction non-analytic components, i.e. components in higher (CF) LLs. 

Jain's wave function (\ref{JainWF}) may be illustrated in the following manner. Via the first factor $\prod_{k<l}(z_k - z_l)^{2s}$,
we have effectively bound $2s$ flux quanta to each of the electrons, as we 
have already mentioned above. This novel type of particle is what we call the {\sl composite
fermion} (CF).
The residual (free) flux quanta effectively determine
the effective number of states per (CF) LL,
$$N_B\rightarrow N_B^*= N_B - 2sN_{el},$$
which correspond to a renormalised magnetic field 
\beq\label{Bstar}
B\rightarrow B^*= B - 2s\left(\frac{h}{e}\right)n_{el}.\eeq
Similarly the CF filling factor is defined with respect to the renormalised number of flux quanta,
\beq\label{CFFF}
\nu^*=\frac{N_{el}}{N_B^*}\qquad \Rightarrow \qquad {\nu^*}^{-1}=\nu^{-1} - 2s,
\eeq
which leads to the relation
\beq\label{nu-nu}
\nu=\frac{\nu^*}{2s\nu^*+1}
\eeq
between the CF filling factor and the usual one $\nu$ [Eq. (\ref{filling})]. For completely filled LLs, $\nu^*=p$,
this yields the above-mentioned series 
\beq\label{JainSer}
\nu=\frac{p}{2sp+1}
\eeq
for the FQHE states which may thus be interpreted as IQHE states of CFs. 
To be explicit, the physical picture of CF theory is
the following: the ground state is described by the wave function (\ref{JainWF}), which describes an incompressible quantum 
liquid in the same manner as Laughlin's wave function does. The elementary excitation in the CF theory consists of a CF promoted
to the next higher CF LL, which is separated from the ground state by an energy gap, in analogy with the electron as compared to 
$n$ completely filled (electronic) LLs in the IQHE.\footnote{Remember, however, that the energy scale of this gap is not given
in terms of a kinetic energy $\hbar eB/m$, but in terms of the Coulomb interaction $e^2/\epsilon l_B$.} 
Again, these elementary CF excitations become localised by the sample 
impurities, and one therefore obtains a plateau in the Hall resistance which is thus quantised. 

Numerically, Jain's CF wave function (\ref{JainWF}) has been successful in the description of the series (\ref{JainSer}) of FQHE
states: even if the overlap with the exact ground states decreases when the quantum number $p$, which describes the number
of completely filled CF LLs, increases, the overlap is still reasonably high (above $95\%$) 
for the number of particles accessible in state-of-the-art exact diagonalisation calculations. Notice, however, that the
physical interpretation is more involved as compared to Laughlin's wave function, because of the LLL projection $\Pmath_{LLL}$,
which is rather complicated to implement in analytical as well as numerical calculations. For a further review of CF theory, 
we refer the interested reader to the literature. The above-mentioned wave-function approaches are thoroughly reviewed 
in Jain's recent book \cite{JainBook}. Furthermore, there have been field-theoretical approaches beyond the numerical wave-function
description presented above, such as in terms of Chern-Simons theories \cite{LF,HLR} or in terms of a Hamiltonian theory
\cite{MS}. For a review of these complementary theories we refer the reader to the book edited by Heinonen \cite{Heinonen} or
the excellent pedagogical review by Murthy and Shankar \cite{MS}.


\subsection{Half-filled LLs and Pfaffian states}

Within the CF picture, we have seen that the effective magnetic field becomes renormalised due to flux attachment [Eq. (\ref{Bstar})].
An interesting situation arises when the filling factor is $\nu=1/2$, which corresponds to the limit $p\rightarrow\infty$ in
Eq. (\ref{JainSer}). In this limit the effective magnetic field (\ref{Bstar}) vanishes, $B^*=0$, and one may then expect the corresponding
phase to be described in terms of a metallic state, such as a Fermi liquid that one would obtain for electrons when the magnetic
field vanishes. A natural ansatz for the $N$-particle wave function of such a
Fermi-liquid state is given by the Slater determinant
$$
\psi_{FL}=\det\left(e^{i\bk_i\cdot\br_j}\right),
$$
where the $N$ electrons occupy the states described by the wave vectors $\bk$, $i=1,...,N$, the modulus of which is delimited
by the Fermi wave vector $|\bk_i|\leq k_F$, and $\br_j$ is the position of the $j$-th particle. 
Notice that this state 
is nevertheless unappropriate in the description 
of a state in the LLL. Indeed, if the scalar product in the exponent is rewritten in terms of complex variables,
$\bk_i\cdot\br_j=(k_iz_j^*+k_i^*z_j)/2$, one realises that the Fermi-liquid state violates the LLL condition of
analyticity. Formally, one may again avoid this problem by projecting the Fermi-liquid state into the LLL, and one obtains
indeed a state,
\beq\label{FL_FC}
\psi_{FL}^{\nu=1/2}=\mathcal{P}_{LLL}\prod_{k<l}\left(z_k-z_l\right)^2\det\left(e^{i\bk_i\cdot\br_j}\right),
\eeq
that was proposed by Rezayi and Read for the description of a compressible metallic state at $\nu=1/2$ \cite{RR}.
The first term is the same factor as in CF theory, which attaches 2 flux quanta to each particle and which cancels thus
the external magnetic field, $B^*=B-2(h/e)n_{el}=0$.

Because the wave function (\ref{FL_FC}) describes a compressible state, one should not observe a quantised Hall resistance, in
agreement with most experimental data. A FQHE at $\nu=1/2$ (and 1/4) has only been observed in very wide 
quantum wells \cite{luhmann,shabani}, which are likely to be described by two-component wave functions \cite{papic} that
we will briefly introduce in Chap. \ref{MultiC}. 

In contrast to the LLL, the half-filled LL $n=1$ reveals, in both spin branches,
a FQHE (5/2 and 7/2 states). The difference between the half-filled LL $n=0$ and $n=1$ is due to a different {\sl effective}
interaction potential that takes into account the wave function overlap between two (interacting) particles, which we do not
discuss in detail here. Indeed, the Fermi-liquid-like state (\ref{FL_FC}) turns out to be quite unstable with respect to 
particle pairing. This is reminiscent of the BCS (Bardeen-Cooper-Schrieffer) instability of a conventional Fermi liquid that
gives rise to superconductivity \cite{mahan,tinkham}, although the glue between the particles is no longer a phonon-mediated
attractive interaction, but only the {\sl repulsive} Coulomb interaction in a strong magnetic field. As we have already
mentioned in Sec. \ref{HaldanePPsec}, such an interaction may yield a discrete two-particle spectrum, in contrast to a 
repulsive interaction in the absence of a magnetic field. As a consequence, pairing may occur at certain relative angular
momenta for particular pseudopotential sequences and for sufficiently high filling factors.\footnote{There have been attempts
in the literature to formalise this point \cite{HaldRez,Wojs}.} In the present case, one may exclude $s$-wave pairing,
i.e. in the relative angular momentum state with $m=0$ due to the Pauli principle, and the most natural candidate would
therefore be $p$-wave pairing in the relative angular momentum state $m=1$ \cite{GWW}. 

A wave function that accounts for $p$-wave pairing was proposed by Moore and Read in 1991 \cite{MR},
\beq\label{MRwave}
\psi_{MR}\left(\left\{z_j\right\}\right)={\rm Pf}\left(\frac{1}{z_i - z_j}\right)\prod_{k<l}(z_k - z_l)^2,
\eeq
where we have again omitted the ubiquitous Gaussian factor. As for the CF wave functions (\ref{JainWF}) and 
the Rezayi-Read wave function (\ref{FL_FC}), the factor $\prod_{k<l}(z_k - z_l)^2$ attaches two flux quanta 
to each electron and therefore does not change the statistical properties of the wave function. If the wave function
consisted only of this factor (times the Gaussian), one would have a {\sl bosonic} Laughlin wave function that
describes an incompressible quantum liquid at the desired filling factor $\nu=1/2$. However, it does not have the 
correct statistical properties. This problem is healed by the first factor ${\rm Pf}[1/(z_i - z_j)]$ which represents
the {\sl Pfaffian} of the $N\times N$ matrix $\Mmath_{ij}=1/(z_i-z_j)$. The Pfaffian may be viewed as the square root
of the more familiar determinant, ${\rm Pf}(\Mmath) = \sqrt{\det(\Mmath)}$, and has the same anti-symmetric properties
as the determinant in an exchange of two particles $i$ and $j$, such that it generates a fermionic wave function.
Notice, furthermore, that this Pfaffian seems, at first sight, to take away some of the zeros such that one could
expect the filling factor to increase. However, the function $\prod_{k<l}(z_k - z_l)^2$ is a product of
$N(N-1)\sim N^2$ terms, whereas the Pfaffian is a sum of products of $N/2\sim N$ terms. Therefore, the number of zeros,
and thus the filling factor, is unchanged in the thermodynamic limit, $N\rightarrow \infty$. 

A particularly interesting
feature of the Pfaffian state are the quasi-particle excitations of charge $e/4$ which satisfy non-Abelian anyonic
statistics \cite{MR}, in contrast to the corresponding excitations of Laughlin's (\ref{LaughlinWF}) or Jain's 
(\ref{JainWF}) wave functions. These non-Abelian quasi-particles are currently investigated in detail within the proposal
of topologically-protected quantum computation \cite{kitaev}. A more detailed discussion of this issue is beyond
the scope of these lecture notes, and we refer the reader to the review article by Nayak {\sl et al.} \cite{nayak}.



\chapter{Brief Overview of Multicomponent Quantum-Hall Systems}
\label{MultiC}

\section{The Different Multi-Component Systems}

\markboth{Brief Overview of Multicomponent Quantum-Hall Systems}{The Different Multi-Component Systems}

\subsection{The role of the electronic spin}

In the preceding chapters, we have completely neglected the physical consequences of possible internal degrees of
freedom, apart from an occasional degeneracy factor that has been smuggled in to account for experimental data.
This choice has been made simply for pedagogical reasons, but it is clear that one prominent internal degree of freedom --
the electronic spin -- may not be put under the carpet so easily. Naively, one may expect that each LL
is split into two distinct spin-branches separated by the energy gap $\Delta_Z$ due to the Zeeman effect. If this gap 
is large, one may use the same one-particle arguments as in the case of the IQHE, but now for each spin branch separately:
once the lowest spin branch of a paticular LL is completely filled, additional electrons must overcome an energy
gap that is no longer given by the LL separation but by $\Delta_Z$. This would indeed not change the presented 
explanation of the IQHE -- instead of a localised electron in the next higher LL, one simply needs to invoke localisation
in the upper spin branch. 

Also in the case of the FQHE, the explanation would need to be modified only in the fine structure
if the Zeeman gap is sufficiently large. If the electrons fill partially the lower spin branch of the lowest (or any) LL,
one may omit all transitions to the upper spin branch and argue that they constitute the high-energy degrees of freedom,
in the same manner as inter-LL excitations in the case of the ``spinless'' fermions which we have discussed in Sec.
\ref{Coul}.

However, the situation is not so easy as the above picture might suggest. Indeed, already in 1983 Halperin pointed out
\cite{halperin83}
that the Zeeman gap in GaAs, with a $g$-factor of $g=-0.4$, is $\Delta_Z =g\mu_B B=g(\hbar e/2m_0)B\simeq 0.33 B{\rm [T]}$ K 
and therefore much smaller
than both the LL separation $\hbar\omega_C=(\hbar e/m) B\simeq 24 B {\rm [T]}$ K, 
due to the rather small band mass ($m=0.068 m_0$, in terms
of the bare electron mass $m_0$, in GaAs), and the Coulomb energy scale $V_C
=e^2/\epsilon l_B\simeq 50 \sqrt{B {\rm [T]}}$ K with a dielectric
constant of $\epsilon\simeq 13$. For a 
characteristic field of 6 T, for which one typically reaches the LLL condition $\nu=1$, one therefore has the energy scales
\beq\label{EnScalesGaAs}
\Delta_Z\simeq 2\,{\rm K} ~~ \ll ~~ \frac{e^2}{\epsilon l_B}\simeq 120\,{\rm K} ~~ \lesssim ~~ \hbar\omega_C \simeq 140\, {\rm K},
\eeq
in GaAs. The situation is qualitatively the same in graphene, where one finds for a field\footnote{Remember that
this field is somewhat arbitrary because the situation $\nu=1$ may also be obtained easily for other fields by varying
the gate voltage $V_G$.} of 6 T
\beq\label{EnScalesGraph}
\Delta_Z\simeq 7\, {\rm K} ~~ \ll ~~ \frac{e^2}{\epsilon l_B}\simeq 620\, {\rm K} ~~ \lesssim ~~ \sqrt{2}\frac{\hbar v}{l_B} 
\simeq 1000\, {\rm K},
\eeq
for $g\simeq 2$ and $\epsilon \simeq 2.5$, which are the appropriate values for graphene on a SiO$_2$ 
substrate.\footnote{Naturally, the dielectric constant depends on the dielectric environment around the graphene sheet
and thus also on the substrate.}

The inevitable consequence of these considerations is that, even if one may neglect the kinetic energy scale in a 
low-energy description of a partially filled LL, one cannot do so with the Zeeman energy scale. One must therefore
take into account the electron spin within a two-component picture in which each quantum state $|n,m\rangle$ is doubled,
$|n,m;\sigma\rangle$ with $\sigma=\ua$ and $\da$.

\subsection{Graphene as a four-component quantum Hall system}

Another multi-component system that we have already discussed is precisely graphene, not only because of the tiny
Zeeman gap which requires to take into account the electronic spin, but also because of its double valley degeneracy
due to the two inequivalent Dirac points situated at the corners $K$ and $K'$ in the first BZ. Each quantum state
$|n,m\rangle$ therefore occurs in {\sl four} copies, $|n,m;\sigma\rangle$ with $\sigma=(K,\ua)$, $(K,\da)$, $(K',\ua)$
and $(K',\da)$. Formally this four-fold degeneracy may be described with the help of an SU(4) spin, whereas 
the two-fold spin degeneracy in GaAs, e.g., is represented by the usual SU(2) spin. Notice that it is very 
difficult in graphene to lift the valley degeneracy, and the associated energy scale is expected to be on the same order
of magnitude as the Zeeman gap, i.e. it is tiny with respect to the one set by the Coulomb interactions. 

\subsection{Bilayer quantum Hall systems}

\begin{figure}
\begin{center}
\epsfig{figure=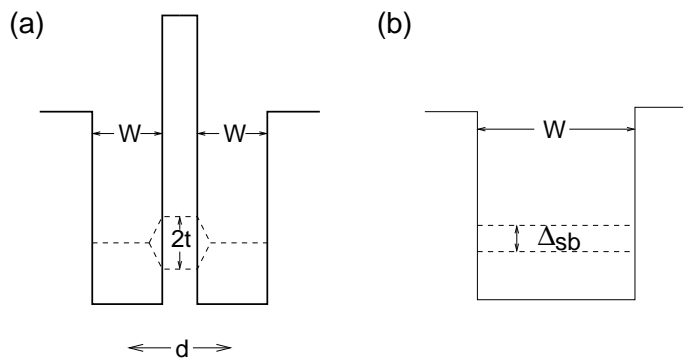,width=9cm,clip}
\end{center}
\caption{ {\sl (a)} Profile of a double quantum well. The two wells are separated by a distance $d$ that is typically on the same
order of magnitude as the well width $W$, $d\sim W\sim 10$ nm. In the presence of a tunneling term $t$ between the two
wells, the electronic subband is split into a symmetric and an anti-symmetric combination, separated by the 
energy scale $\Delta_{SAS}=2t$.
{\sl (b)} Wide quantum well. In a wide quantum well the energy gap between the occupied lowest electronic subband and the
unoccupied first excited subband, $\Delta_{sb}$, is decreaased as compared to a narrow quantum well.}
\label{fig24}
\end{figure}

A third multi-component system that we would like to mention consists of a double quantum well [see Fig. \ref{fig24}(a)]. 
These bilayer systems, which are fabricated by molecular-beam epitaxy, consist of two quantum wells spatially separated 
by an insulating barrier that is on the same order of magnitude as the width of each of the wells. Formally, each
of the wells (layers) may be described in terms of an SU(2) {\sl pseudo-spin}, $\sigma=\ua$ for an electron in the left well 
and $\sigma=\da$ for one in the right well. In contrast to the true electron spin, the Coulomb interaction does not respect
this SU(2) symmetry -- indeed, the repulsion is stronger between particles within the same layer (i.e. with the
same pseudo-spin orientation) than between particles in different layers (with opposite pseudo-spin orientation) because, in
the second case, electrons may not be brought together closer than the distance $d$ between the layers. In order to minimise
the interaction energy, it is therefore favourable to charge both layers equally. Alternatively,
this may be viewed as some capacitive energy, if one interprets the two-layer system in terms of a capacitor,
that favours an equal charge distribution between the two layers as compared to a charging of only one layer. Notice, furthermore,
that tunneling, with the tunneling energy $t$,
between the two quantum wells lifts the pseudo-spin degeneracy: whereas the symmetric superposition
$|+\rangle=(|\ua\rangle +|\da\rangle)/\sqrt{2}$ of the layer pseudo-spin lowers the energy, the anti-symmetric superposition
$|-\rangle=(|\ua\rangle -|\da\rangle)/\sqrt{2}$ describes anti-binding. The energy separation between the associated subbands
is given by $\Delta_{SAS}=2t$ [see Fig. \ref{fig24}(a)], but it may be strongly reduced experimentally with the help of 
a high potential barrier separating the two wells. The term $\Delta_{SAS}$, which plays the role of a Zeeman gap (though in the 
$x$-quantisation axis), may become the lowest energy scale in the system, such that the SU(2) pseudo-spin symmetry breaking
only stems from the difference in the Coulomb interaction between particles in the same and in different layers.

{\subsection{Wide quantum wells}}

Another quantum Hall system that may be characterised as a multi-component system is a wide quantum well [Fig. \ref{fig24}(b)]. 
Indeed, the 
samples which reveal the highest mobilities are those fabricated in wide quantum wells, where the well width $w$ is often
much larger than the magnetic length $l_B$. As compared to a narrow quantum well, the energy difference between the lowest 
and the first excited electronic subbands, which are the energy levels of the confinement potential in the $z$-direction, 
is strongly decreased. Although the Fermi level still resides in the lowest electronic subband (pseudo-spin $\sigma=\ua$), 
the energy gap to the next unoccupied one (pseudo-spin $\sigma=\da$) may then become smaller than the relevant Coulomb energy
scale. In the same manner as for the electronic spin, one must therefore no longer discard higher electronic subbands. In 
a first approximation one may restrict the calculations to these two lowest subbands \cite{albofath,papic}
although the next higher subbands also shift to lower energies and need eventually be taken into account. Similarly to 
the quantum-Hall bilayer, which is sometimes also used in the description of the large quantum well, the Coulomb interaction 
decomposed in these electronic subband states is not pseudo-spin SU(2)-symmetric.


\vspace*{0.5cm}

In the remainder of this chapter, we discuss some aspects of correlated states that one encounters in multi-component
quantum Hall systems in general, starting (Sec. \ref{Nu1}) 
with the completely spin-polarised state at $\nu=1$ (quantum Hall ferromagnet)
and its various manifestations in the different quantum Hall systems described above. We will not discuss, for 
reasons of space limitation, the amazing physical properties of the elementary excitations of the quantum Hall ferromagnet,
which is a topological spin-texture state {\sl (skyrmion)}, and refer the interested reader to the literature 
\cite{sondhi,moon,GirvinLH,ezawa}. In the line of the preceding chapter, we have chosen to discuss a generalisation of Laughlin's
wave function, which we owe to Halperin \cite{halperin83}, in order to account for the electronic spin (Sec. \ref{MCWF}).
These wave functions are further generalised to even more components than two, and we close this section with a discussion of their
possible use in the description of multi-component FQHE states.

\section{The State at $\nu=1$}
\label{Nu1}

\markboth{Brief Overview of Multicomponent Quantum-Hall Systems}{The State at $\nu=1$}

If one takes into account internal degrees of freedom, the state at $\nu=1$ is no longer simply a Slater determinant of all occupied
quantum states in the lowest LL, but one must take into account the macroscopic degeneracy due to the fact that each state $|n,m\rangle$
may now be occupied by 0, 1 or 2 particles. In this sense the situation at $\nu=1$ is much more similar to the FQHE in a partially
filled LL than to the IQHE which one obtains for completely filled LL \cite{sondhi}, and the macroscopic degeneracy is again
lifted by the mutual Coulomb interactions between the electrons.

\subsection{Quantum Hall ferromagnetism}
\label{SpinFM}

We first consider the generic case of electrons at $\nu=1$ in the conventional monolayer quantum Hall system while taking
into account their physical spin. In view of the above-mentioned energy arguments, we completely neglect the Zeeman effect, which
would otherwise trivially lift the macroscopic degeneracy at $\nu=1$ by polarising all electron spins. Because of the fact that
two electrons, with opposite spin, may now occupy the same quantum state $|n,m\rangle$, the electron pair may in principle be in
a relative angular momentum state with $m=0$ -- the Pauli principle, which only applies to fermions of the same species, does
no longer prevent this quantum number to be odd. Indeed, such an electron pair is described by a two-particle wave function with the
rather unspectacular polynomial factor $(z_{i,\ua}-z_{j,\da})^0=1$, where $z_{i,\ua}$ is the position of an arbitrarily chosen 
spin-$\ua$ electron and $z_{j,\da}$ that of a spin-$\da$ electron. Such an electron pair therefore interacts via the Haldane
pseudopotential $v_0$, which is the largest pseudopotential in the case of a repulsive Coulomb interaction because it
characterises the interaction at the shortest possible length scale
(see Fig. \ref{fig19bis}).\footnote{This pseudopotential, as well as any other
with an even value of $m$, does not play any physical role due to the Pauli principle 
if one considers only spinless electrons, as we have mentioned in Sec. \ref{HaldanePPsec}.}
Since $v_0\simeq 2v_1$, the system thus tends to avoid double occupancy, and the ground state is described by the fully anti-symmetric
(orbital) wave function (\ref{eqVandermonde}) regardless of whether the electron at the position $z_j$ is spin-$\ua$ or spin-$\da$.

Notice that, although both spinless and spin-1/2 electrons are described by the same wave function, the physical origin of
these ground states is different: in the case of spinless fermions, it is simply the non-degenerate wave function described
by a Slater determinant, whereas in the case of electrons with spin, the state is formed 
in order to minimise the mutual Coulomb repulsion.

Because the orbital wave function (\ref{eqVandermonde}) for electrons with spin at $\nu=1$ is fully anti-symmetric, the spin
wave function describing the internal degrees of freedom must be fully symmetric, e.g.
\beq\label{SpinWF}
\chi_{FM} = |\ua_1,\ua_2,\, ...,\ua_N\rangle, 
\eeq
in order to form an overall wave function that
is anti-symmetric. The subscript indicates the index of the particle that the spin is associated with. 
The global wave function, therefore, reads
\beq\label{ferro}
\psi_{\nu=1,FM} = \prod_{k<l}(z_k - z_l)\otimes |\ua_1,\ua_2,\, ...,\ua_N\rangle.
\eeq
This is nothing other than a (spin) wave function of a {\sl quantum ferromagnet}, similar to ferromagnetism
in a usual Fermi liquid. Indeed, the spontaneous spin polarisation in a Fermi liquid is also due to a minimisation of the Coulomb 
repulsion by the formation of an anti-symmetric orbital wave function. Notice, however, that the spin polarisation 
in a Fermi liquid comes along with an energy cost as a consequence 
of the mismatch between the Fermi energies of spin-$\ua$ and spin-$\da$ electrons. The competition
between the gain in interaction energy and the cost in kinetic energy determines the final polarisation of the system, which
is never perfect. In the case
of the quantum Hall ferromagnet, there is no cost in kinetic energy when the system is fully polarised because all quantum
states have the same kinetic energy, and the system is therefore {\sl fully} polarised. 

\subsubsection{Collective excitations} 

Because the spontaneous spin polarisation in the quantum Hall ferromagnet chooses, in the absence of a Zeeman effect, an arbitrary
direction in the three-dimensional spin space, one is confronted with a spontaneous SU(2) symmetry breaking. As a consequence
of this broken continuous symmetry, there exists a gapless collective excitation (Goldstone mode) the energy of which tends
to zero in the long wave-length limit. Indeed, even if we have chosen the ferromagnet in Eq. (\ref{SpinWF})
to be oriented in the $z$-direction, any other orientation, such as the one described by the wave function
$$|\da_1,\da_2,\, ...,\da_N\rangle \qquad {\rm or} \qquad \bigotimes_{j=1}^N|+_j\rangle=|+_1,+_2,\, ...,+_N\rangle,$$
where the $+_j$ sign indicates the symmetric superposition $|+_j\rangle = (|\ua_j\rangle + |\da_j\rangle)/\sqrt{2}$
of both spin orientations of the $j$-th electron, 
would also describe a ground state. The Goldstone mode in the large wave-length limit may then be viewed as a global 
rotation of all spins into another ground-state configuration, which naturally does not imply an energy cost.

In the case of a ferromagnet, the Goldstone mode is nothing other than the spin-density wave\footnote{Remember
that for a crystaline ground state (WC), the Goldstone mode is the acoustic phonon, as we have briefly discussed
in the previous chapter in Sec. \ref{Coul}.} that disperses as $\omega\propto q^2$
in the small wave vector limit, $ql_B\ll 1$. At first sight, this mode seems in contradiction 
with the observation of a quantum Hall effect
at $\nu=1$, even in the absence of a Zeeman effect, which requires a gap as we have seen above. Notice, however, that 
this gap needs to be a transport gap in which a quasi-particle moves independently from a quasi-hole in order to transport a current.
This is not the case in a spin wave with $ql_B\ll 1$, but one obtains freely moving quasi-particles and quasi-holes in the 
limit $q l_B\gg 1$. In this limit, the spin-wave dispersion tends to a finite value that is given by the exchange energy between particles
of different spin orientation and that is proportional to the interaction energy scale $e^2/\epsilon l_B$, as in the case
of the FQHE \cite{moon}.

There are more exotic spin-texture excitations (skyrmions), which are described by a topological quantum number associated with
the winding of the spin-texture. These are gapped excitation which carry an electric charge related to this topological quantum
number. As mentioned above, a detailed discussion of these amazing excitations is beyond the scope of the present lecture notes.

\subsection{Exciton condensate in bilayer systems}
\label{BilayerFM}

The $\nu=1$ in a bilayer system is remarkably different from the quantum Hall ferromagnet described in the preceding subsection.
Although the electronic interactions still favour a fully anti-symmetric orbital wave function (\ref{eqVandermonde})
and thus a symmetric, i.e. ferromagnetic, pseudo-spin wave function,
the interaction potential is no longer SU(2) symmetric in the pseudo-spin degree of freedom.\footnote{
Naturally, such an anti-symmetric orbital wave function is only physical if the layer separation $d$ is not too large
(as compared to the magnetic length) -- otherwise one would simply have completely decoupled layers.}
As we have already mentioned above,
a charge imbalance $Q$ between the two quantum wells (layers) is penalised by a charging energy, $E_C=Q^2/2C$, in terms of the 
capacitance $C=\epsilon \mathcal{A}/d$, where $\Amath$ is the area of the 2D system. Because $Q=-e\nu n_{el}\Amath=-e\nu n_B\Amath=
-e\nu \Amath/2\pi l_B^2$ when all electrons reside in a single layer and $Q=0$ if they are equally distributed between the two layers,
one obtains an energy cost 
$$\frac{E_C}{N_{el}}\sim \nu\frac{e^2}{\epsilon l_B}\frac{d}{l_B},
$$
per particle in the charge-imbalenced state,
in agreement with a more sophisticated microscopic calculation \cite{moon}. In terms of the pseudo-spin magnetisation, this means
that in the ground-state configuration, with a homogeneous charge distribution over both layers, all pseudo-spins 
are oriented in the $xy$-plane. Remember that a pseudo-spin $\ua$ corresponds to an electron in the upper layer and 
$\da$ to one in the lower layer, and a configuration as the one described in Eq. (\ref{SpinWF}) is therefore excluded, whereas
the symmetric and anti-symmetric combinations
$$\chi_{+} = \bigotimes_{j=1}^N|+_j\rangle \qquad {\rm and} \qquad \chi_{-} = \bigotimes_{j=1}^N|-_j\rangle, $$
with $|\pm_j\rangle = (|\ua_j\rangle \pm |\da_j\rangle)/\sqrt{2}$ is not.
These two states, which correspond to a ferromagnet in the $x$- and the $y$-direction, respectively, may be generalised by choosing
any other direction described by the angle $\phi$ in the $xy$-plane,
\beq\label{Superfl}
\chi_{\phi} = \bigotimes_{j=1}^N|\phi_j\rangle,
\eeq
where $|\phi_j\rangle \equiv [|\ua\rangle + \exp(i\phi)|\da\rangle]/\sqrt{2}$. The states $\chi_{+}$ and $\chi_-$ are obtained
for $\phi=0$ and $\phi=\pi$ (modulo $2\pi$), respectively.

\begin{figure}
\begin{center}
\epsfig{figure=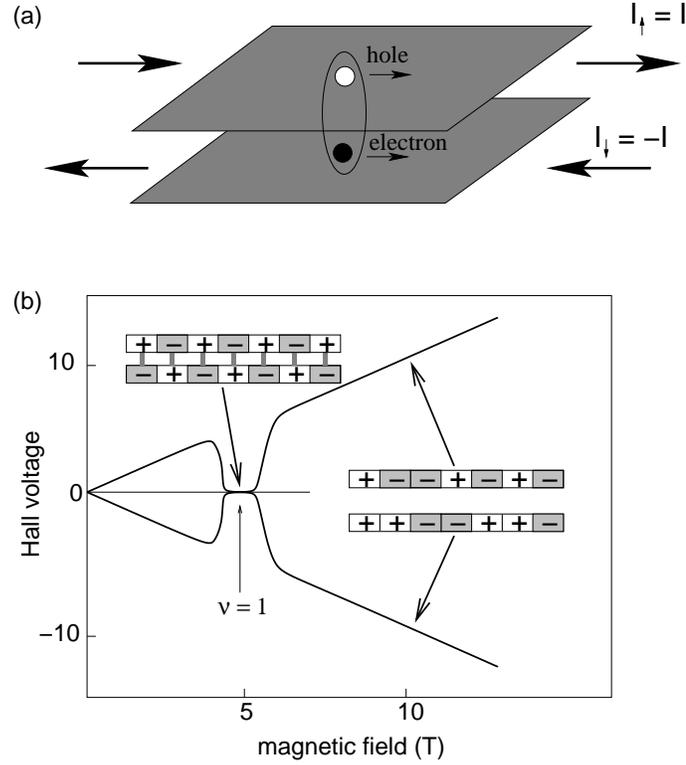,width=9cm,clip}
\end{center}
\caption{ Hall resistance measurement used to detect excitonic condensation, adopted from (Eisenstein and MacDonald, 2004).
{\sl (a)} Counterflow configuration, in which one drives a current $I_{\ua}=I$ through the upper layer that is
flowing in the opposite direction as that, $I_{\da}=-I$ in the lower layer. The hole component of the excitonic quantum
state in one layer thus moves in the same direction as the electron component in the other one.
{\sl (b)} The two curves schematically represent, when taking into account only excitonic superfluidity, 
the Hall resistance in both layers within the counterflow configuration. Because of the relative sign between 
the currents in the two layers, the measured Hall resistances are of opposite sign. Electrons with no interlayer correlations
yield the usual linear $B$-field dependence of the Hall resistance in order to compensated the Lorentz force
acting on them individually. In the case of exciton condensation (around
$B=5$ T), charge tranport is due to a uniform current of charge-neutral excitons, which are not affected by the Lorentz force,
and the Hall resistance vanishes, as it has been observed in the experiments (Kellogg {\sl et al.}, 2004; Tutuc {\sl et al.}, 2004).}
\label{fig25}
\end{figure}

Contrary to the case of the spin ferromagnet with full SU(2) symmetry, where a general state would be described in terms
of two angles $\theta$ and $\phi$, the different possible {\sl easy-plane} pseudo-spin ferromagnetic are characterised 
by the angle $\phi$ which may vary between $0$ and $2\pi$. The low-energy degrees of freedom are therefore described by
a different {\sl universality class} that turns out to be the same as the one that describes superfluidity or
superconductivity. The relation between superfluidity and the easy-plane pseudo-spin ferromagnet in bilayer systems at
$\nu=1$ may indeed be understood in the following manner: on the average, the average filling factor per layer 
is $\nu_{\ua}=\nu_{\da}=1/2$ in order to minimise the charging energy due to the capacitive term, i.e. there are as many
electrons as holes in the LLL of each layer. Naturally, because of the Coulomb interaction between the particles in the
two different layers, an electron in one layer wants to be bound to a hole in the other one. Since the number of
electrons in each layer equals, on the average, that of holes in the other one, all particles find their appropriate 
partner in the opposite layer. The electron-hole pair in the two layers may be viewed as a charge-neutral {\sl interlayer exciton} that
satisfies bosonic statistics [Fig. \ref{fig25}(a)]. Below a certain temperature, these bosons condense into a collective state that is
nothing other than the {\sl exciton superfluid} \cite{fertig,WZ,EI,moon}. 
The phase coherence between the different excitons is precisely described by the angle $\phi$.

The first experimental indication of excitonic superfluidity in bilayer quantum Hall systems was a zero-bias 
anomaly in tunneling experiments \cite{spielman}. Indeed, if one injects a charge in a tunneling experiment into one of
the layers and collects it in a contact at the other layer, the tunneling conductance $dI_z/dV$ is expected to be weak in the case
of uncorrelated electrons because of the Coulomb repulsion between electrons in the opposite layers.
However, below a critical value of $d/l_B$, where one
expects the interlayer correlations to be sufficiently strong to form a phase-coherent excitonic condensate, 
the injected electron systematically finds a hole in the other layer, 
such that tunneling between the layers is strongly enhanced. This strong enhancement, which due
to its reminiscence with the Josephson effect in superconductors \cite{tinkham} is also called 
{\sl quasi-Josephson effect},\footnote{Contrary to the Josephson effect, only the tunneling conductance $dI_z/dV$ is strongly
enhanced whereas the tunneling {\sl current} remains zero in the quasi-Josephson effect in bilayer systems.}
has indeed been observed experimentally \cite{spielman}.

Another strong indication for excitons in bilayer quantum Hall systems stems from transport measurements in the
counterflow configuration, where the current in the upper layer $I_{\ua}=I$ flows in the opposite direction as compared to
that in the lower layer $I_{\da}=-I$ [see Fig. \ref{fig25}(a)]. From a technical point of view, it is indeed
possible to contact the two layers separately such that one may measure the Hall resistance (and also the longitudinal
resistance) in both layers independently. In the case of exciton condensation, the charges involved in transport are 
{\sl zero} because the excitons are charge-neutral objects, which are not coupled to the magnetic field and thus not
affected by the Lorentz force. In addition to a vanishing longitudinal resistance, one would therefore expect a vanishing
Hall resistance because no density gradient between opposite edges is built up to compensate the Lorentz force 
\cite{WZ,EI}. This is schematically shown in Fig. \ref{fig25}(b). The simultaneous vanishing of the Hall and longitudinal
resistances was indeed observed in 2004 by two different experimental groups \cite{kellogg,tutuc}.

\subsection{SU(4) ferromagnetism in graphene}
\label{GraphFM}

The arguments in favour of a quantum Hall ferromagnetism may easily be generalised to the case of graphene, where
the Coulomb interaction respects to great accuracy the four-fold spin-valley degeneracy, as we have described above.
In order to avoid confusion about the filling factor, one first needs to remember that the filling factor 
$\nu_G$ in graphene
is defined with respect to the charge-neutral point, which happens to be in the centre of the central $n=0$ LL
(see Sec. \ref{relQHE}). Two of the four (degenerate) spin-valley branches are therefore completely filled at $\nu_G=0$,
which in non-relativistic quantum Hall systems would correspond rather to a filling factor $\nu=2$. Similarly the
filling factor $\nu=1$ would correspond to a graphene filling factor $\nu_G=-1$, whereas $\nu_G=1$ implies three
completely filled spin-valley branches ($\nu=3$). 

Let us first consider the filling factor $\nu_G=-1$ and see
how the above considerations apply to graphene with its SU(4) symmetry.\footnote{The filling factor $\nu_G=1$ is related
to $\nu_G=-1$ by particle-hole symmetry and therefore does not require a separate discussion.}
In the same manner as for the spin quantum Hall ferromagnet at $\nu=1$, the short-range component $v_0$ of the Coulomb potential
is screened in the completely anti-symmetric orbital wave function (\ref{eqVandermonde}), and the spin part of the wave
function must therefore be completely symmetric. Notice, however, that one may now distribute the electron over the 
four internal states ${|m;K,\ua\rangle}$, $|m;K,\da\rangle$, $|m;K',\ua\rangle$ and $|m;K',\da\rangle$. 
The general spin wave function is therefore a superposition of all these states
\beq\label{SU4spinWF}
\chi_{\rm SU(4)}=\bigotimes_{m=1}^{N}\left(u_{m,1}|m;K,\ua\rangle + u_{m,2}|m;K,\da\rangle + u_{m,3}|m;K',\ua\rangle + u_{m,4}|m;K',\da\rangle \right),
\eeq
where the complex coefficients $u_{m,i}$ satisfy the normalisation condition $\sum_{i=1}^4|u_{m,i}|^2=1$. In the case
of global coherence, all coefficients are independent of the guiding-centre quantum number $m$, $u_{m,i}=u_i$,
and one thus obtains the spin wave function of an {\sl SU(4) ferromagnetism} \cite{nomura,GMD,alicea,yang}. 
These arguments may also be generalised
to the case of $\nu_G=0$, where two branches are completely filled \cite{yang}, but the ground state does not reveal the same
degeneracy as the SU(4) ferromagnet at $\nu_{G}=\pm 1$.
Indeed, a general argument on $K$-component quantum Hall system
shows that one has generalised ferromagnetic states at all integer values of the filling factor $\nu=1, ...,K-1$
\cite{arovas}.

As a consequence of the SU(4) quantum Hall ferromagnet, one may expect a quantum Hall effect in graphene at the unusual filling factors
$\nu_G=0,\pm 1$. Remember that these states do not belong to the series (\ref{RelLLseries}), 
$\nu_G=\pm 2,\pm 4, ...$ of the RQHE which may be explained by LL quantisation within the picture of non-interacting 
relativistic particles. 
In the same manner as for the spin quantum Hall ferromagnet, the gapless spin-density-wave modes, which reveal a higher 
degeneracy due to the larger SU(4) symmetry, do not imply that the charged modes are also gapless. Indeed, the 
elementary charged excitations of the SU(4) quantum Hall ferromagnet are generalised skyrmions \cite{yang,DGLM}
which are separated by a gap from the ground state, which therefore describes an incompressible quantum liquid that
displays the quantum Hall effect. A quantum Hall effect has indeed been observed at these unusual filling factors \cite{zhang},
in agreement with the formation of an SU(4) quantum Hall ferromagnet. However, there exist alternative scenarios
to describe the appearance of a quantum Hall effect at these filling factors \cite{gusynin,fuchs,herbut} and a
clear indication of SU(4) quantum Hall ferromagnetism is yet lacking.

We finally emphasise that an SU(4) description is not restricted to graphene. Indeed, if one takes into account the
electron spin, the bilayer quantum Hall system and its excitations
may also be treated within the SU(4) framework \cite{arovas,ezawaSU4,ezawa,DGLM} although the interaction does not respect
the full SU(4) symmetry because of the asymmetry in the layer pseudo-spin described above.




\section{Multi-Component Wave Functions}
\label{MCWF}

\markboth{Brief Overview of Multicomponent Quantum-Hall Systems}{Multi-Component Wave Functions}

Until now, we have considered a multi-component quantum Hall effect at the integer filling factor $\nu=1$ (or other integer
fillings in the case of graphene) that is described in terms of the Vandermonde determinant (\ref{eqVandermonde}) 
$\prod_{k<l}(z_k - z_l)$ regardless of whether the particle at the position $z_k$ is in a state $\ua$ or $\da$. The
spin orientation has only been taken into account within a spin wave function that is multiplied to the Vandermonde 
determinant. One may naturally ask the question whether one may also describe other filling factors than $\nu=1$. 

A simple generalisation of the quantum Hall ferromagnetism to other filling factors consists of replacing the
Vandermonde determinant by, e.g., the Laughlin (\ref{LaughlinWF}) at $\nu=1/(2s+1)$ or the Jain wave function (\ref{JainWF})
at $\nu=p/(2sp+1)$ and to multiply it again with a spin wave function that is naturally ferromagnetic because the 
orbital wave function remains anti-symmetric. There are, however, more general states for which the orbital wave function
is not fully anti-symmetric, but only in the intra-component parts as it is required by the Pauli principle. These
states are described in terms of wave functions proposed by Halperin in 1983 \cite{halperin83} that we present in 
this section, as well as a natural generalisation to systems with more components than $K=2$.

\subsection{Halperin's wave function}

Halperin's wave function for spin-1/2 electrons
is a straight-forward generalisation of Laughlin's proposal (\ref{LaughlinWF}). We consider
the particle positions to be separated into two sets $\{z_1^{\ua},z_2^{\ua},\, ..., z_{N_{\ua}}^{\ua}\}$ for 
spin-$\ua$ particles and $\{z_1^{\da},z_2^{\da},\, ..., z_{N_{\da}}^{\da}\}$ for 
spin-$\da$ particles. If the particles with different spin orientation could be treated as independent from one another,
i.e. in the absence of an interaction between spin-$\ua$ and spin-$\da$ particles, one would simply write down
a product ansatz
\beq\label{LaughProd}
\psi_{\ua,m_1}^{L}(\{z_j^{\ua}\})\times \psi_{\da,m_2}^{L}(\{z_j^{\da}\}) 
= \prod_{k<l}^{N_{\ua}}\left(z_k^{\ua} - z_l^{\ua}\right)^{m_1}
\prod_{k<l}^{N_{\da}}\left(z_k^{\da} - z_l^{\da}\right)^{m_2}
\eeq
of two independent Laughlin wave functions that need not necessarily be described by the same exponent $m$. The total
filling factor would then be simply the sum $\nu=\nu_{\ua}+\nu_{\da}$ of the filling factors $\nu_{\ua}=1/m_1$ and
$\nu_{\da}=1/m_2$ for spin-$\ua$ and spin-$\da$ particles, respectively. 

Apart from the fact that this situation
is not particularly interesting, it is also unphysical because the Coulomb interaction does not depend on the spin orientation
of the particle pairs. In the wave function (\ref{LaughProd}), two particles of opposite spin orientation
may be at the same position, i.e. the wave function does not vanish in general for $z_k^{\ua}=z_l^{\da}$. Remember that such a 
double occupancy of the same position would be penalised by an energy cost on the order of the short-range component $v_0$
in a pseudopotential expansion. 

In order to account for these inter-component correlations, Halperin proposed to
add a factor $\prod_{k=1}^{N_{\ua}}\prod_{l=1}^{N_{\da}}(z_k^{\ua} - z_l^{\da})^n$ to the wave function (\ref{LaughProd})
the exponent of which does not necessarily need to be odd because particles of opposite spin orientation are not
constrained by the Pauli principle. Halperin's wave function
\beq\label{HalperinWF}
\psi_{m_1,m_2,n}^H (\{z_j^{\ua},z_j^{\da}\}) = \prod_{k<l}^{N_{\ua}}\left(z_k^{\ua} - z_l^{\ua}\right)^{m_1}
\prod_{k<l}^{N_{\da}}\left(z_k^{\da} - z_l^{\da}\right)^{m_2}\prod_{k=1}^{N_{\ua}}\prod_{l=1}^{N_{\da}}\left(z_k^{\ua} - z_l^{\da}\right)^n
\eeq
is therefore characterised by the set $(m_1,m_2,n)$ of three exponents.

In analogy with Laughlin's wave function, for which we have $\nu=1/m$,
the exponents fix the (component) filling factors, as one may see from the power-counting argument (see Sec. \ref{FQHE1}).
According to this argument,
the maximal exponent for a particular particle position cannot exceed the number of flux quanta $N_B$ threading the area
$\Amath$ of the 2D electron system. Apart from the shift that vanishes anyway in the thermodynamic limit, one obtains the
two equations
\beq\label{zerosHWF}
N_B=m_1 N_{\ua} + n N_{\da} \qquad {\rm and} \qquad N_B = m_2 N_{\da} + n N_{\ua}.
\eeq
This means that, contrary to the simpler case of Laughlin's wave function, the number of zeros in one component is not
simply given by the corresponding exponent times the number of particles in this component (first term in the above
expressions). Instead, it is also affected by the particles in the other component that contribute each a zero of order $n$
(second term) due to the mixed term in Halperin's wave function (\ref{HalperinWF}). In terms of the component filling factors,
\beq\label{CompFill}
\nu_{\sigma} = \frac{N_{\sigma}}{N_B}\, ,
\eeq
Eq. (\ref{zerosHWF}) may be rewritten in matrix form
\beq\label{Mmatrix}
\left(\begin{array}{c} 1 \\ 1\end{array}\right) = \left(\begin{array}{cc} m_1 & n \\ n & m_2\end{array}\right)
\left(\begin{array}{c} \nu_{\ua} \\ \nu_{\da}\end{array}\right) ,
\eeq
from which one obtains the component filling factors by matrix inversion
\beq\label{MmatrixInv}
\left(\begin{array}{c} \nu_{\ua} \\ \nu_{\da}\end{array}\right) = \frac{1}{m_1m_2-n^2}\left(\begin{array}{cc} m_2 & -n \\ -n & m_1\end{array}\right)
\left(\begin{array}{c} 1 \\ 1\end{array}\right),
\eeq
and one finds
\beq\label{TotFill}
\nu=\nu_{\ua}+\nu_{\da}=\frac{m_1 + m_2 - 2n}{m_1m_2-n^2}\ 
\eeq
for the total filling factor.

One first notices that, in Eq. (\ref{MmatrixInv}), not only the filling factors are fixed by the exponents but also,
for a given magnetic field (i.e. a given number of flux quanta), the number of particles per component. Contrary to what
one could have expected from the expression of Halperin's wave function (\ref{HalperinWF}), the numbers $N_{\sigma}$,
namely the ratio between them, cannot be chosen arbitrarily.

Furthermore, the above expressions (\ref{MmatrixInv}) and (\ref{TotFill}) for the filling factors
are ill-defined if the exponent matrix in Eq. (\ref{Mmatrix}) is not
invertible, i.e. when its determinant is zero, $m_1m_2-n^2=0$. The only physically relevant situation arises when 
all exponents are equal odd integers $m_1=m_2=n$. However, this result should not surprise us: we are then confronted again
with a completely anti-symmetric wave function, actually a Laughlin wave function, which requires a ferromagnetic spin
wave function. As we have seen above, in the discussion of the quantum Hall ferromagnetism, the ground-state manifold 
comprises states with different polarisation along the $z$-axis: the state with $N_{\ua}=N$ and $N_{\da}=0$ is an 
equally valid ground state as a state with $N_{\ua}=N_{\da}=N/2$ or $N_{\ua}=0$ and $N_{\da}=N$, where
$N=N_{\ua}+N_{\da}$ is the total number of particles. The component
filling factor is therefore not well-defined and depends on the polarisation 
\beq\label{Polarisation}
p_z=\frac{N_{\ua}-N_{\da}}{N}=\frac{\nu_{\ua}-\nu_{\da}}{\nu},
\eeq
whereas the total filling factor is simply given by $\nu=1/m$, in terms of the common odd exponent $m$. Notice that
contrary to the quantum Hall ferromagnet, a state with an invertible exponent matrix has a polarisation that is completely fixed, 
\beq\label{Polarisation2}
p_z=\frac{m_2-m_1}{m_1m_2-n^2}\ .
\eeq

We finally mention that not all states that can be written down in terms of Halperin's wave function are good candidates for the
description of the ground state chosen by the system. One may show, e.g. within a generalisation of Laughlin's plasma analogy
(presented in Sec. \ref{PlasmaLaugh}) to two or more components, that several of Halperin's wave functions do not describe
a homogeneous liquid but a liquid in which the different components phase-separate \cite{dGRG}. For two components, the condition
for a homogeneous state is simply that both the exponents $m_1$ and $m_2$, which describe the intra-component correlations, must be larger
than $n$ for the inter-component correlations. As an example, we may study the states $(3,3,1)$ and $(1,1,3)$, which would both
be candidates for a possible two-component FQHE at $\nu=1/2$ and which have indeed been investigated in the literature
\cite{MDYG}. However, only the first one describes a homogeneous liquid, such that the second one may be discarded right from
the beginning.

Furthermore, some of Halperin's wave functions, even if they satisfy the above-mentioned condition, turn out to be problematic
if the interaction is SU(2) symmetric, such as for the true electron spin. In this case, one may show that $(m,m,n)$ states
are only eigenstates of the total-spin operator, which commutes with the interaction Hamiltonian, if $n=m$ (i.e. in the
ferromagnetic state) or if $n=m-1$ \cite{PG}. This restriction may be omitted though in bilayer quantum Hall systems 
or in wide quantum wells where the interaction Hamiltonian is not pseudo-spin SU(2)-symmetric. 

\subsubsection{Physical relevance of Halperin states}

A physically relevant Halperin state is e.g. the unpolarised $(3,3,2)$ state which would occur at a filling factor $\nu=2/5$. 
Remember from the discussion of CF theory in Sec. \ref{FQHE3} that there is also a (naturally polarised) CF candidate, with
$p=2$ completely filled CF LLs, to describe the ground state at this filling factor. Which of them is now the better one? This question
could be answered within exact-diagonalisation calculations, which showed that, in the absence of a Zeeman effect, the
true ground state is described in terms of the unpolarised Halperin wave function $(3,3,2)$ \cite{CZ}. Notice, 
however, that the energy difference between the two states is extremely small, as may be seen from variational calculations
\cite{JainBook}, such that the polarised CF state becomes the ground state above a critical value of the energy $\Delta_Z$
associated with the Zeeman effect. This critical value would therefore describe a phase transition between an unpolarised 
and a fully polarised FQHE state. Such transitions have indeed been observed in polarisation experiments, where the strength
of the Zeeman effect was varied by a simultaneous change in the magnetic field and in the electronic density \cite{kang,kukush}.


\subsection{Generalised Halperin wave functions}

We would finally mention that Halperin's wave function may easily be generalised to describe possible FQHE states in systems with
a larger number of components, such as the four spin-valley components in graphene.
This generalised wave function for $K$-component quantum Hall systems may be written as a product
\beq\label{HalperinGen}
\psi_{m_1,...,m_K;n_{ij}}^{SU(K)}\left(\left\{z_{j_1}^{(1)},z_{j_2}^{(2)}, ...,z_{j_K}^{(K)}
\right\}\right)
=\psi_{m_1,...,m_K}^L \,\times\,\psi_{n_{ij}}^{inter} 
\eeq
of a product of Laughlin wave functions
$$
\psi_{m_1,...,m_K}^L=\prod_{j=1}^K\prod_{k_j<l_j}^{N_j}\left(
z_{k_j}^{(j)}-z_{l_j}^{(j)}\right)^{m_j}
$$
for each of the components and a term
$$
\psi_{n_{ij}}^{inter}=\prod_{i<j}^{K}\prod_{k_i}^{N_i}\prod_{k_j}^{N_j}
\left(z_{k_i}^{(i)}-z_{k_j}^{(j)}\right)^{n_{ij}}
$$
that takes into account the correlations between particles in different components \cite{GR}. Here, 
the indices $i$ and $j$ denote the component, $i,j=1,\, ...,K$, and $z_{k_i}^{(i)}$ is the complex
position of the $k_i$-th particle in the component $i$. 

Although the wave function (\ref{HalperinGen})
may seem scary at the first sight, it is as easily manipulated as Halperin's original wave function
(\ref{HalperinWF}). The component filling factors $\nu_j=N_j/N_B$ may be determined, in the same manner
as in the two-component case (\ref{Mmatrix}), with the help of the ``exponent matrix'' 
$\Mmath$ the off-diagonal terms of which are the exponents $(\Mmath)_{ij}=n_{ij}$ (for $i\neq j$), whereas 
the diagonal terms are simply the exponents corresponding to the intra-component correlations,
$(\Mmath)_{ii}=m_i$. The zero-counting argument yields the matrix equation
\beq\label{MmatrixK}
\left(\begin{array}{c} 1 \\ \vdots \\ 1\end{array}\right) = \Mmath
\left(\begin{array}{c} \nu_1 \\ \vdots \\ \nu_K\end{array}\right) 
\eeq
relating the component filling factors to the exponents, and if $\Mmath$ is invertible, all component
filling factors are fixed by the inverse equation
\beq\label{MmatrixKinv}
\left(\begin{array}{c} \nu_1 \\ \vdots \\ \nu_K \end{array}\right) = \Mmath^{-1}
\left(\begin{array}{c} 1 \\ \vdots \\ 1 \end{array}\right) .
\eeq
If the determinant $\det(\Mmath)$ is zero and the matrix thus not invertible, not all component filling factors can
be determined. In analogy with the two-component case this hints at underlying ferromagnetic states.
A perfect SU($K$) ferromagnetic state is obtained when all components are equal odd integers, $m_i=n_{ij}=m$, in
which case one obtains again a simple (fully anti-symmetric)
Laughlin wave function for all particles regardless of to which component they belong. For $K=4$ and $m=1$, this is just the
SU(4) ferromagnetic state at $\nu=1$ which we have already discussed in the context of the quantum Hall effect at $\nu_G=\pm 1$
in graphene (Sec. \ref{GraphFM}). 

Notice, however, that contrary to a two-component system, where one only needs to distinguish between an invertible and a 
non-invertible matrix, the situation is much richer for $K>2$. One may indeed have different ``degrees'' of invertibility 
that are described by the {\sl rank} of the matrix. Consider, e.g., the fully anti-symmetric wave function with
$m_i=n_{ij}=m$. In this case, Eq. (\ref{MmatrixK}) actually consists only of one single equation relating the component
filling factors, i.e. $1=m(\nu_1 + ... + \nu_K)=m\nu$, and all other lines of the matrix equation are simply copies of the
first one. The rank of this matrix is 1, i.e. only the total filling factor is fixed, $\nu=1/m$ [SU($K$) ferromagnet]
whereas in the case of an invertible matrix the rank is $K$ and the $K$ lines
in the matrix equation (\ref{MmatrixK}) represent (linearly) independent equations. If the rank of an exponent matrix
is smaller than $K$ but larger than 1, the resulting state is neither a full SU($K$) ferromagnet nor a state with
completely fixed component filling factors (or polarisations) -- it is rather a state with some intermediate ferromagnetic 
properties. 

As for two-component Halperin wave functions (\ref{HalperinWF}), a generalisation of Laughlin's plasma analogy allows one 
to distinguish between physical (i.e. homogeneous) and unphysical states (which show a phase separation of at least some of the
components). Indeed, the exponent matrix $\Mmath$ must have only positive eigenvalues in order to describe a homogeneous 
state \cite{dGRG}. We finally mention that $\Mmath$ encodes not only information concerning the filling factors (\ref{MmatrixKinv}),
but fully describes the quantum Hall state (\ref{HalperinGen}), such as its topological degeneracy,
the charges of its quasi-particle excitations as well as the statistical properties of the latter \cite{WenZee}.

\appendix

\chapter{Electronic Band Structure of Graphene}
\label{TBgraphene}

\markboth{Appendix}{Electronic Band Structure of Graphene}

In this appendix, we calculate the band structure of graphene in the tight-binding model
\cite{wallace}, the results of which we have summarised in Sec. \ref{SecGraph}. Because graphene's
honeycomb lattice consists of two distinct sublattices A and B, the electronic wave function 
\beq\label{eq2:06}
\psi_{\bk}(\br)=a_{\bk}\psi^{(A)}_{\bk}(\br)+b_{\bk}\psi^{(B)}_{\bk}(\br),
\eeq
is a superposition of two wave functions, for the A and B sublattice, respectively,
where $a_{\bk}$ and $b_{\bk}$ are complex functions of the quasi-momentum
$\bk$. Both $\psi^{(A)}_{\bk}(\br)$ and $\psi^{(B)}_{\bk}(\br)$ 
are Bloch functions with
\beq\label{eq2:07}
\psi_{\bk}^{(j)}(\br)=\sum_{\bR_l}e^{i\bk\cdot\bR_l}\phi^{(j)}(\br+\deltab_j-\bR_l),
\eeq
in terms of the atomic wave functions $phi^{(j)}(\br+\deltab_j-\bR_l)$ centred around the 
position $\bR_l-\deltab_j$, where $\deltab_j$ is
the vector which connects the sites $\bR_l$ of the underlying Bravais lattice with the
site of the $j$ atom within the unit cell. Typically one chooses the sites of one
of the sublattices, e.g. the A sublattice, to coincide with the sites of the 
Bravais lattice such that $\deltab_A=0$.

With the help of these wavefunctions, we may now search the solutions of the 
Schr\"odinger equation
$$H\psi_{\bk}=\epsilon_{\bk}\psi_{\bk},
$$
where $H$ is the full Hamiltonian for electrons on a lattice, which 
is of the type (\ref{LatticeHam}) mentioned in Sec. \ref{zeroB}.
Here, we have chosen an arbitrary representation, which is not necessarily 
that in real space.\footnote{The wavefunction $\psi_{\bk}(\br)$ is, thus, the
real space representation of the Hilbert vector $\psi_{\bk}$.}
Multiplication of the Schr\"odinger equation by $\psi_{\bk}^*$ 
from the left yields the equation $\psi_{\bk}^*H\psi_{\bk}=\epsilon_{\bk}
\psi_{\bk}^*\psi_{\bk}$, which may be rewritten in matrix form 
with the help of Eqs. (\ref{eq2:06}) and (\ref{eq2:07})
\beq\label{eq2:08}
\left(a_{\bk}^*,b_{\bk}^*\right)\Hmath_{\bk}
\left(\begin{array}{c} a_{\bk} \\ b_{\bk}\end{array}\right)=
\epsilon_{\bk}\left(a_{\bk}^*,b_{\bk}^*\right)\Smath_{\bk}
\left(\begin{array}{c} a_{\bk} \\ b_{\bk}\end{array}\right).
\eeq
Here, the Hamiltonian matrix is defined as
\beq\label{eq2:09}
\Hmath_{\bk}\equiv \left(
\begin{array}{cc}
 \psi_{\bk}^{(A)*}H\psi_{\bk}^{(A)} & \psi_{\bk}^{(A)*}H\psi_{\bk}^{(B)} \\
\psi_{\bk}^{(B)*}H\psi_{\bk}^{(A)} & \psi_{\bk}^{(B)*}H\psi_{\bk}^{(B)} 
\end{array}\right)
=\Hmath_{\bk}^{\dagger},
\eeq
and the overlap matrix
\beq\label{eq2:10}
\Smath_{\bk}\equiv \left(
\begin{array}{cc}
 \psi_{\bk}^{(A)*}\psi_{\bk}^{(A)} & \psi_{\bk}^{(A)*}\psi_{\bk}^{(B)} \\
\psi_{\bk}^{(B)*}\psi_{\bk}^{(A)} & \psi_{\bk}^{(B)*}\psi_{\bk}^{(B)} 
\end{array}\right)
=\Smath_{\bk}^{\dagger}
\eeq
accounts for the non-orthogonality of the trial wavefunctions.
The eigenvalues $\epsilon_{\bk}$ of the Schr\"odinger equation yield the 
energy bands, and they may be obtained from the secular equation
\beq\label{eq2:11}
\det\left[\Hmath_{\bk}-\epsilon_{\bk}^{\lambda}\Smath_{\bk}\right]=0,
\eeq
which needs to be satisfied for a non-zero solution of the wavefunctions,
i.e. for $a_{\bk}\neq 0$ and $b_{\bk}\neq 0$. The label $\lambda$ denotes 
the energy bands, and it is clear that there are as many energy bands as 
solutions of the secular equation (\ref{eq2:11}), i.e. two bands for the 
case of two atoms per unit cell.

\begin{figure}
\centering
\includegraphics[width=6.5cm,angle=0]{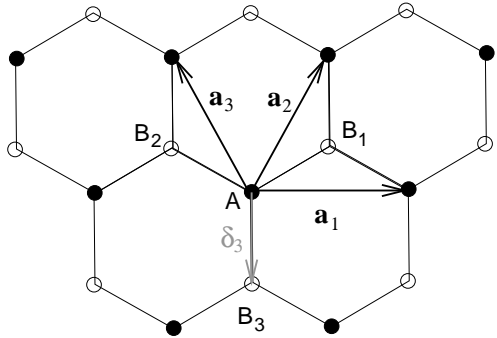}
\caption{ {Tight-binding model for the honeycomb lattice.}}
\label{fig2:01}
\end{figure}

From now on, we neglect the overlap of wave functions on neighbouring sites, 
such that the overlap matrix (\ref{eq2:10}) simply becomes the one matrix $\bone$ times 
the number of particles $N$ due to the normalisation of the wave functions.
The secular equation then tells us that the energy bands are just the 
eigenvalues of the Hamiltonian matrix (\ref{eq2:09}). Furthermore, one notices
that because the two sublattices are equivalent from a chemical point of
view, we have $\psi_{\bk}^{(A)*}H\psi_{\bk}^{(A)}=\psi_{\bk}^{(B)*}H\psi_{\bk}^{(B)}$,
and the diagonal terms therefore contribute just a constant shift to 
the band energies that we may set to zero. The only relevant terms are then
the off-diagonal terms in Eq. (\ref{eq2:09}),
$\Hmath_{\bk}^{AB}\equiv\psi_{\bk}^{(A)*}H\psi_{\bk}^{(B)} =Nt_{\bk}^{AB}$, with
the {\sl hopping term}
\beq\label{eq2:13}
t_{\bk}^{AB}\equiv \sum_{\bR_l} e^{i\bk\cdot\bR_l}\int d^2r\, 
\phi^{(A)*}(\br-\bR_k) H \phi^{B)}(\br+\deltab_{AB}-\bR_m)\ ,
\eeq
where $\deltab_{AB}$ is a vector that connects an A site to a B site.

In order to obtain the basic band structure of graphene, it is sufficient to consider 
a hopping only between nearest-neighbouring sites described by the {\sl hopping amplitude}
\beq\label{eq2:15}
t\equiv \int d^2r\, \phi^{A*}(\br) H \phi^{B}(\br+\deltab_3),
\eeq
where we have chosen $\deltab_{AB}=\deltab_3$ (see Fig. \ref{fig2:01}). Notice that one may
also take into account hopping to sites that are further away such as the next-nearest neighbours
which turn out to be on the same sublattice and which would thus yield diagonal terms to the 
Hamiltonian matrix. However, whereas we have $t\sim 3$ eV, the hopping amplitude for next-nearest-neighbour
hopping is roughly 10 times smaller \cite{antonioRev} and only marginally affects the low-energy
properties of electrons in graphene.

If we now consider an arbitrary site $A$ on the A sublattice (Fig. \ref{fig2:01}),
we may see that the hopping term (\ref{eq2:13}) consist
of three terms corresponding to the nearest neighbours $B_1$, $B_2$, and $B_3$, all of which
have the same hopping amplitude $t$. However, only the site $B_3$ is described by the
same lattice vector (shifted by $\deltab_3$) as the site $A$ and thus yields a zero
phase to the hopping matrix. The sites $B_1$ and $B_2$ correspond to lattice vectors
shifted by 
$$\ba_2=\frac{\sqrt{3}a}{2}(\be_x + \sqrt{3}\be_y) \qquad{\rm and} \qquad 
\ba_3\equiv\ba_2-\ba_1=\frac{\sqrt{3}a}{2}(-\be_x + \sqrt{3}\be_y),$$ 
respectively, where $a=|\deltab_3|=0.142$ nm is the distance between nearest-neighbour carbon atoms. Therefore, they
contribute a phase factor $\exp(i\bk\cdot\ba_2)$ and $\exp(i\bk\cdot\ba_3)$, 
respectively. The hopping term (\ref{eq2:13}) may therefore be written as
$$t_{\bk}^{AB}=t\gamma_{\bk}^*=\left(t_{\bk}^{BA}\right)^*,
$$
where we have defined the sum of the nearest-neighbour phase factors
\beq\label{eq2:18}
\gamma_{\bk}\equiv 1+e^{i\bk\cdot\ba_2}+e^{i\bk\cdot\ba_3}.
\eeq

The band dispersion may now easily be obtained by solving the secular equation (\ref{eq2:11}),
\beq\label{BandDisp}
\epsilon_{\lambda}(\bk) = \lambda\left|t_{\bk}^{AB}\right| = \lambda t\left|\gamma_{\bk}\right|,
\eeq
and is plotted in Fig. \ref{fig08}. The band dispersion is obviously particle-hole symmetric, and the valence
band ($\lambda=-$) touches the conduction band ($\lambda$) in the inequivalent points 
$$\pm \bK= \pm \frac{4\pi}{3\sqrt{3}a}\be_x\ ,$$ 
which
one determines by setting $\gamma_{\pm \bK}=0$
and which coincide with the two inequivalent BZ corners $K$ and $K'$. 
Because the whole band structure is half-filled in undoped graphene,
as we have mentioned in Sec. \ref{zeroB}, the Fermi energy lies exactly in these points $K$ and $K'$.

\section*{Continuum Limit}

The low-energy electronic properties may be obtained by expanding the band structure in the vicinity
of these points, and the low-energy Hamiltonian is obtained simply by expanding the sum of the phase factors
(\ref{eq2:18}) around $K$ and $K'$,
\beqn\label{eq2:30}
\nn
\gamma_{\bp}^{\pm}\equiv \gamma_{\bk=\pm \bK+\bp} &=& 1 + e^{\pm i\bK\cdot\ba_2}
e^{i\bp\cdot\ba_2} + e^{\pm i\bK\cdot\ba_3}e^{i\bp\cdot\ba_3}
\\
\nn
&\simeq& 1 + e^{\pm i 2\pi/3}\left[1 + i\bp\cdot\ba_2\right] + e^{\mp i 2\pi/3}\left[1 + i\bp\cdot\ba_3 \right] \\
\nn
&=& \gamma_{\bp}^{\pm (0)}+\gamma_{\bp}^{\pm (1)}
\eeqn
By definition of the Dirac points and their
position at the BZ corners $K$ and $K'$, we have 
$\gamma_{\bp}^{\pm (0)}=\gamma_{\pm\bK}=0$. We limit the expansion to first order in $|\bp|a$. Notice that, in order
to simplify the notations, we have used a system of units with $\hbar=1$, i.e. where the momentum has the same units as
the wave vector.

The first order term is given by 
\beqn\label{eq2:31}
\nn
\gamma_{\bp}^{\pm (1)} &=& i\frac{\sqrt{3}a}{2}\left[(p_x+\sqrt{3}p_y)e^{\pm i2\pi/3}
+ (-p_x+\sqrt{3}p_y)e^{\mp i2\pi/3} \right]\\
&=& \mp \frac{3a}{2}(p_x \pm i p_y),
\eeqn
which is obtained with the help of $\sin(\pm 2\pi/3)=\pm \sqrt{3}/2$ and
$\cos(\pm 2\pi/3)=-1/2$. This yields the effective low-energy Hamiltonian
\beq\label{eq2:32}
H_{\bp}^{\xi}=\xi v(p_x\sigma^x + \xi p_y\sigma^y),
\eeq
in terms of the Fermi velocity
\beq\label{eq2:32b}
v \equiv \frac{3ta}{2\hbar}\ .
\eeq
The index $\xi=\pm $ denotes the valleys $K$ and $K'$, and one obtains at the $K$ point the Dirac Hamiltonian
mentioned in (\ref{0BHams})
\beq\label{DirHamK}
H_D=v\bp\cdot\sigmab\, ,
\eeq
whereas the low-energy Hamiltonian at the $K'$ point reads
\beq\label{DirHamKp}
H_D^{\prime}=-v\bp\cdot\sigmab^*\, ,
\eeq
with $\sigmab^*=(\sigma^x,-\sigma^y)$. Both Hamiltonians yield the same energy spectrum which is therefore
{\sl two-fold valley-degenerate}.

Notice that if one prefers to avoid the complex conjugation in the Hamiltonian (\ref{DirHamKp}), one simply
changes the representation by interchanging the A and B sublattices, in which case one may write the 
Hamiltonians for the two valleys $K$ ($\xi=+$) and $K'$ ($\xi=-$) in a compact form,
\beq\label{DirHamBis}
H_D^{\xi}=\xi H_D=\xi v\bp\cdot\sigmab\, .
\eeq

\chapter{Landau Levels of Massive Dirac Particles}
\label{MassLL}

\markboth{Appendix}{Landau Levels of Massive Dirac Particles}

\section*{Mass Confinement of Dirac Fermions at $B=0$}

\begin{figure}
\centering
\includegraphics[width=12cm,angle=0]{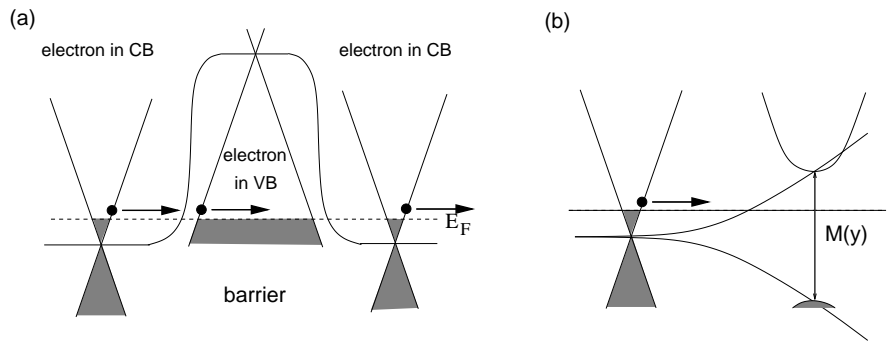}
\caption{ {\sl (a)} Klein tunneling through a barrier. An incident electron in the conduction band (CB) above the Fermi energy, 
which is at the Dirac point before the barrier, transverses the barrier as en electron above the Fermi energy in the valence band (VB).
The valence band is partially emptied because the Dirac point has shifted to a higher energy corresponding to the barrier height.
{\sl (b)} Mass confinement. A gap opens when the particle approaches the edge, which becomes a forbidden region
where no quantum state can be found at the energy corresponding to that of the incident electron.}
\label{fig2:02}
\end{figure}

Even in the absence of a magnetic field, electronic confinement in graphene turns out to be quite tricky because 
a simple-minded approach in terms of a potential $V_{\rm conf}=V(y)\bone$ cannot confine Dirac electrons. This 
fact is due to an intrinsically relativistic effect that is called the {\sl Klein paradox}, according to which
a (massless) relativistic particle may transverse a potential barrier without being backscattered \cite{klein}.
This effect may be understood in the following manner: consider an incident electron in the region with $V=0$ the
energy of which is slightly above the Fermi energy. In the potential barrier, the Dirac point is shifted to a higher
energy that corresponds to the barrier height and the Fermi energy lies now in the valence band, where the electron
may still find a quantum state (with the same wave-vector direction and the same velocity $v$) -- instead of 
moving as an electron in the conduction band, it thus simply moves in the same direction as an electron in the valence
band [Fig. \ref{fig2:02}(a)]. This is in stark contrast with quantum mechanical tunneling of a non-relativistic particle,
for which the transmission probability through a potential barrier is exponentially suppressed because of a lacking
quantum state at the same energy as that of the incident electron.

The problem is circumvented by a so-called {\sl mass confinement}
\beq\label{MassConfA}
V_{\rm conf}=V(y)\,\sigma^z=\left(\begin{array}{cc} V(y) & 0 \\ 0 & -V(y) \end{array}\right),
\eeq
and we discuss first the simpler case of a constant mass term $M\sigma^z$ that needs to be added to the Dirac 
Hamiltonian. That this term yields indeed a mass may be seen from the Dirac Hamiltonian at $B=0$ 
\beq\label{MassDir}
H_D^m=v\bp\cdot\sigmab + M\sigma^z=\left(
\begin{array}{cc}
 M & v(p_x - ip_y) \\ v(p_x + ip_y) & -M
\end{array}\right),
\eeq 
the diagonalisation of which yields the energy spectrum
$$\epsilon_{\lambda}(\bp)= \lambda\sqrt{v^2|\bp|^2 + M^2},$$
which is gapped at zero momentum. This is nothing other than the dispersion relation of a relativistic particle\footnote{
The sign $\lambda=-$ corresponds to the anti-particle.}
with mass $m$ such that $M=mv^2$. Qualitatively one may see from Fig. \ref{fig2:02}(b) why a mass confinement is 
more efficient than a potential barrier. Indeed, when the particle approaches the edge with $M(y)\neq 0$ a gap opens. 
An electron slightly above the Dirac point may then only propagate in the region with $M=0$, whereas at the edge its
energy lies in the gap which is a forbidden region, and the electron is thus confined. 

Similarly to the $B=0$ case, one may find the energy spectrum of the massive Dirac Hamiltonian (\ref{MassDir}) in 
a perpendicular magnetic field, which reads, in terms of the ladder operators $a$ and $a^{\dagger}$,
\beq\label{MassDirB}
H_D^B=  \left(\begin{array}{cc}
M & v(\Pi_x- i \Pi_y) \\ v(\Pi_x+ i \Pi_y) & -M
        \end{array}\right) = \left(\begin{array}{cc}
M & \sqrt{2}\frac{\hbar v}{l_B} a \\ \sqrt{2}\frac{\hbar v}{l_B} a^{\dagger} & -M
        \end{array}\right) .
\eeq
Its eigenvalues may be obtained in the same manner as in the $M=0$ case (c.f. Sec. \ref{RelLLsec}), and one obtains
\beq\label{MassiveLL}
\epsilon_{\lambda n} = \lambda \sqrt{M^2 + 2\frac{\hbar^2 v^2}{l_B^2} n} 
\eeq
for the massive relativistic LLs, $n\neq 0$. 

Special care needs to be taken in the discussion of the central LL $n=0$, which necessarily shifts away from zero energy.
The associated quantum state (\ref{spinN0}) is zero in the first component $u_0$, whereas the second component is 
given by $v_0=|0\rangle$. In order to satisfy the second line in the eigenvalue equation 
$$
H_D^B \psi_0 = \epsilon_0 \psi_0 \qquad {\Leftrightarrow} \qquad
\left(\begin{array}{cc}
M & \sqrt{2}\frac{\hbar v}{l_B} a \\ \sqrt{2}\frac{\hbar v}{l_B} a^{\dagger} & -M
        \end{array}\right)\left(\begin{array}{c} 0 \\ |0\rangle \end{array}\right) 
= \epsilon _0 \left(\begin{array}{c} 0 \\ |0\rangle \end{array}\right) ,
$$
one needs to fulfil 
\beq\label{CondMass}
\sqrt{2}\frac{\hbar v}{l_B}\, a^{\dagger}\, u_0 = (\epsilon_0 + M) v_0 \qquad \Leftrightarrow \qquad
0=(\epsilon_0 + M)|0\rangle,
\eeq
such that the only solution is $\epsilon_0 = -M$. The relativistic $n=0$ LL is therefore shifted to negative energies
and does no longer satisfy particle-hole symmetry. This effect is called {\sl parity anomaly} and depends
on the {\sl sign of the mass}. 

In the case of graphene, we need to remember that there are two copies of the energy spectrum, one at the $K$ point
and one at the $K'$ point. As we have disussed in Appendix \ref{TBgraphene}, the Hamiltonian (\ref{MassDirB}) describes
the low-energy properties at the $K$ point whereas we need to interchange the A and B sublattices at the $K'$ point and add
a global sign in front of the off-diagonal terms [see Eq. (\ref{DirHamBis})],
\beq\label{MassDirBp}
H_D^{B\prime}=   \left(\begin{array}{cc}
- M & - \sqrt{2}\frac{\hbar v}{l_B} a \\ -\sqrt{2}\frac{\hbar v}{l_B} a^{\dagger} & M
        \end{array}\right) = -H_D^{B\prime}.
\eeq
Naturally, the eigenstates of this Hamiltonian are the same as those of the Hamiltonian (\ref{MassDirB}) at the 
$K$ point, but the eigenvalues change their sign. Due to the particle-hole symmetry of the levels (\ref{MassiveLL}),
the global sign does not affect the energy spectrum for $n\neq 0$. However, the $n=0$ LL, which
does not respect particle-hole symmetry, must again be treated apart, and one finds in the same manner as for the $K$ point
the condition corresponding to Eq. (\ref{CondMass}),
\beq\label{CondMassbis}
-\sqrt{2}\frac{\hbar v}{l_B}\, a^{\dagger}\, u_0 = (\epsilon_0 - M) v_0 \qquad \Leftrightarrow \qquad
0=(\epsilon_0 - M)|0\rangle.
\eeq
One notices that the $n=0$ LL level at the $K'$ point shifts to positive energies as a function of the mass, such that 
the overall level spectrum for graphene, when one takes into account {\sl both} valleys, 
is again particle-hole symmetric, but the valley degeneracy is lifted for
$n=0$.

The case of a mass term that varies in the $y$-direction, such as for the mass confinement potential, may finally
be treated in the same manner as we have discussed in Sec. \ref{secInv}: the system remains translation-invariant in
the $x$-direction, such that the Landau gauge is the appropriate gauge and the wave vector $k$ in this direction
is a good quantum number. Because this wave vector determines the position of the eigenstate in the $y$-direction,
$y_0=kl_B^2$, the energy spectrum is given by the expression (\ref{MassSpec}),
\beq\label{MassSpecA}
\epsilon_{\lambda n,y_0;\xi} = \lambda \sqrt{M^2(y_0) + 2\frac{\hbar^2 v^2}{l_B^2} n}, 
\eeq
for $n\neq 0$ and both valleys $\xi=\pm$, whereas the $n=0$ LL is found at 
\beq\label{MassSpec0}
\epsilon_{n=0,y_0;\xi} = -\xi M(y_0).
\eeq

\clearpage

\markboth{Bibliography}{Bibliography}

\bibliographystyle{cj} 
\bibliography{refsQHE} 

\begin{thebibliography}{100}

\bibitem{PG}
Prange, R. and Girvin, E., S.~M. (1990) {\em The Quantum Hall Effect\/}.
  Springer.

\bibitem{yoshioka}
Yoshioka, D. (2002) {\em The Quantum Hall Effect\/}. Springer.

\bibitem{ezawa}
Ezawa, Z.~F. (2000) {\em Quantum Hall Effects -- Field Theoretical Approach and
  Related Topics\/}. World Scientific.

\bibitem{GirvinLH}
Girvin, S.~M. (1999) {\em The Quantum Hall Effect: Novel Excitations and Broken
  Symmetries, {\rm in A. Comptet, T. Jolicoeur, S. Ouvry and F. David (Eds.)}
  Topological Aspects of Low-Dimensional Systems -- \'Ecole d'\'Ete de Physique
  Th\'eorique LXIX\/}. Springer.

\bibitem{MS}
Murthy, G. and Shankar, R. (2003) {\em Rev. Mod. Phys.\/}, {\bf 75}, 1101.

\bibitem{graph1}
Novoselov, K.~S., Geim, A.~K., Morosov, S.~V., Jiang, D., Katsnelson, M.~I.,
  Grigorieva, I.~V., Dubonos, S.~V., and Firsov, A.~A. (2005) {\em Nature\/},
  {\bf 438}, 197.

\bibitem{graph2}
Zhang, Y., Tan, Y.-W., Stormer, H.~L., and Kim, P. (2005) {\em Nature\/}, {\bf
  438}, 201.

\bibitem{montam}
Akkermans, E. and Montambaux, G. (2008) {\em Mesoscopic Physics of Electrons
  and Photons\/}. Cambridge UP.

\bibitem{SdH}
Shubnikov, L.~W. and {de Haas}, W.~J. (1930) {\em Proceedings of the Royal
  Netherlands Society of Arts and Science\/}, {\bf 33}, 130 and 163.

\bibitem{KDP}
v.~Klitzing, K., Dorda, G., and Pepper, M. (1980) {\em Phys. Rev. Lett.\/},
  {\bf 45}, 494.

\bibitem{metrology1}
Poirier, W. and Schopfer, F. (2009) {\em Eur. Phys. J. Special Topics\/}, {\bf
  172}, 207.

\bibitem{metrology2}
Poirier, W. and Schopfer, F. (2009) {\em Int. J. Mod. Phys. B\/}, {\bf 23},
  2779.

\bibitem{TSG}
Tsui, D.~C., St\"ormer, H., and Gossard, A.~C. (1983) {\em Phys. Rev. Lett.\/},
  {\bf 48}, 1559.

\bibitem{laughlin}
Laughlin, R.~B. (1983) {\em Phys. Rev. Lett.\/}, {\bf 50}, 1395.

\bibitem{Jain1}
Jain, J.~K. (1989) {\em Phys. Rev. Lett.\/}, {\bf 63}, 199.

\bibitem{Jain2}
Jain, J.~K. (1990) {\em Phys. Rev. B\/}, {\bf 41}, 7653.

\bibitem{willett}
Willett, R.~L., Eisenstein, J.~P., Stormer, H.~L., Tsui, D.~C., Gossard, A.~C.,
  and English, J.~H. (1987) {\em Phys. Rev. Lett.\/}, {\bf 59}, 1776.

\bibitem{MR}
Moore, G. and Read, N. (1991) {\em Nucl. Phys. B\/}, {\bf 360}, 362.

\bibitem{GWW}
Greiter, M., Wen, X.-G., and Wilczek, F. (1991) {\em Phys. Rev. Lett.\/}, {\bf
  66}, 3205.

\bibitem{Pan}
Pan, W., Stormer, H.~L., Tsui, D.~C., Pfeiffer, L.~N., Baldwin, K.~W., and
  West, K.~W. (2003) {\em Phys. Rev. Lett.\/}, {\bf 90}, 016801.

\bibitem{antonioRev}
{Castro Neto}, A.~H., Guinea, F., Peres, N. M.~R., Novoselov, K.~S., and Geim,
  A.~K. (2009) {\em Rev. Mod. Phys.\/}, {\bf 81}, 109.

\bibitem{zhang}
Zhang, Y., Jiang, Z., Small, J.~P., Purewal, M.~S., Tan, Y.-W., Fazlollahi, M.,
  Chudow, J.~D., Jaszczak, J.~A., Stormer, H.~L., and Kim, P. (2006) {\em Phys.
  Rev. Lett.\/}, {\bf 98}, 197403.

\bibitem{grapheneFQHE1}
Du, X., Skachko, I., Duerr, F., Luican, A., and Andrei, E.~Y. (2009) {\em
  Nature\/}, p. doi:10.1038/nature08522.

\bibitem{grapheneFQHE2}
Bolotin, K.~I., Ghahari, F., Shulman, M.~D., Stormer, H.~L., and Kim, P. (2009)
  {\em preprint\/}, p. arXiv:0910.2763.

\bibitem{AM}
Ashcroft, N.~W. and Mermin, N.~D. (1976) {\em Solid State Physics\/}. Harcourt.

\bibitem{kittel}
Kittel, C. (2005) {\em Introduction to Solid State Physics\/}. Wiley, 8th Ed.

\bibitem{jackson}
Jackson, J.~D. (1999) {\em Classical Electrondynamics\/}. Wiley, 3rd ed.

\bibitem{CT}
Cohen-Tannoudji, C., Diu, B., and Lalo\"e, F. (1973) {\em Quantum Mechanics\/}.
  Hermann.

\bibitem{mcclure}
McClure, J.~W. (1956) {\em Phys. Rev.\/}, {\bf 104}, 666.

\bibitem{berger}
Berger, C., Song, Z., Li, T., Ogbazghi, A.~Y., Feng, R., Dai, Z., Marchenkov,
  A.~N., Conrad, E.~H., First, P.~N., and {de Heer}, W.~A. (2004) {\em J. Phys.
  Chem.\/}, {\bf 108}, 19912.

\bibitem{sadowski}
Sadowski, M.~L., Martinez, G., Potemski, M., Berger, C., and {de Heer}, W.~A.
  (2006) {\em Phys. Rev. Lett.\/}, {\bf 97}, 266405.

\bibitem{jiang}
Jiang, Z., Henriksen, E.~A., L.~C.~Tung, Y.-J.~W., Schwartz, M.~E., Han, M.~Y.,
  Kim, P., and Stormer, H.~L. (2007) {\em Phys. Rev. Lett.\/}, {\bf 98},
  197403.

\bibitem{champel}
Champel, T. and Florens, S. (2007) {\em Phys. Rev. B\/}, {\bf 75}, 245326.

\bibitem{AALR}
Abrahams, E., Anderson, P.~W., Licciardello, D.~C., and Ramakrishnan, T.~V.
  (1979) {\em Phys. Rev. Lett.\/}, {\bf 42}, 673.

\bibitem{butt}
B\"uttiker, M. (1992) {\em The Quantum Hall Effect in Open Conductors, {\rm in
  M. Reed (Ed.)} Nanostructured Systems (Semiconductors and Semimetals, {\rm
  {\bf 35}, 191})\/}. Academic Press.

\bibitem{BILP}
B\"uttiker, M., Imry, Y., Landauer, R., and Pinhas, S. (1985) {\em Phys. Rev.
  B\/}, {\bf 31}, 6207.

\bibitem{datta}
Datta, S. (1995) {\em Electronic Transport in Mesoscopic Systems\/}. Cambridge
  UP.

\bibitem{klass}
Kla\ss, U., Dietsche, W., v.~Klitzing, K., and Ploog, K. (1991) {\em Z. Phys.
  B:Cond. Matt.\/}, {\bf 82}, 351.

\bibitem{buett}
B\"uttiker, M. (1988) {\em Phys. Rev. B\/}, {\bf 38}, 9375.

\bibitem{hashimoto}
Hashimoto, K., Sohrmann, C., Wiebe, J., Inaoka, T., Meier, F., Hirayama, Y.,
  R\"omer, R.~A., Wiesendanger, R., and Morgenstern, M. (2008) {\em Phys. Rev.
  Lett.\/}, {\bf 101}, 256802.

\bibitem{sondhiRev}
Sondhi, S.~L., Girvin, S.~M., Carini, J.~P., and Shahar, D. (1997) {\em Rev.
  Mod. Phys.\/}, {\bf 69}, 315.

\bibitem{sachdev}
Sachdev, S. (1999) {\em Quantum Phase Transitions\/}. Cambridge UP.

\bibitem{wei}
Wei, H.~P., Tsui, D.~C., Paalanen, M.~A., and Pruisken, A. M.~M. (1988) {\em
  Phys. Rev. Lett.\/}, {\bf 61}, 1294.

\bibitem{wei2}
Wei, H.~P., Engel, L.~W., and Tsui, D.~C. (1994) {\em Phys. Rev. B\/}, {\bf
  50}, 14609.

\bibitem{li1}
Li, W., Csathy, A., Tsui, D.~C., Pfeiffer, L.~N., and West, K.~W. (2005) {\em
  Phys. Rev. Lett.\/}, {\bf 94}, 206807.

\bibitem{li2}
Li, W., Vicente, C.~L., Xia, J.~S., Pan, W., Tsui, D.~C., Pfeiffer, L.~N., and
  West, K.~W. (2009) {\em Phys. Rev. Lett.\/}, {\bf 102}, 216801.

\bibitem{CC}
Chalker, J.~T. and Coddington, P.~D. (1988) {\em J. Phys. C\/}, {\bf 21}, 2665.

\bibitem{huckestein}
Huckestein, B. (1995) {\em Rev. Mod. Phys.\/}, {\bf 67}, 357.

\bibitem{Huck}
Huckestein, B. and Backhaus, M. (1999) {\em Phys. Rev. Lett.\/}, {\bf 82},
  5100.

\bibitem{slevin}
Slevin, K. and Ohtsuki, T. (2009) {\em Phys. Rev. B\/}, {\bf 80}, 041304.

\bibitem{BreyFertig}
Brey, L. and Fertig, H. (2006) {\em Phys. Rev. B\/}, {\bf 73}, 195408.

\bibitem{mahan}
Mahan, G.~D. (1993) {\em Many-Particle Physics\/}. Plenum Press, 2nd Ed.

\bibitem{GV}
Giuliani, G.~F. and Vignale, G. (2005) {\em Quantum Theory of Electron
  Liquids\/}. Cambridge UP.

\bibitem{KH}
Kallin, C. and Halperin, B.~I. (1984) {\em Phys. Rev. B\/}, {\bf 30}, 5655.

\bibitem{iyengar}
Iyengar, A., Wang, J., Fertig, H.~A., and Brey, L. (2007) {\em Phys. Rev. B\/},
  {\bf 75}, 125430.

\bibitem{RFG}
Rold\'an, R., Fuchs, J.-N., and Goerbig, M.~O. (2009) {\em Phys. Rev. B\/},
  {\bf 80}, 085408.

\bibitem{wigner}
Wigner, E. (1934) {\em Phys. Rev.\/}, {\bf 102}, 46.

\bibitem{FPA}
Fukuyama, H., Platzman, P.~M., and Anderson, P.~W. (1979) {\em Phys. Rev. B\/},
  {\bf 19}, 5211.

\bibitem{glatt}
Andrei, E.~Y., Deville, G., Glattli, D.~C., Williams, F. I.~B., Paris, E., and
  Etienne, B. (1988) {\em Phys. Rev. Lett.\/}, {\bf 60}, 2765.

\bibitem{gervais}
Gervais, G., Engel, L.~W., Stormer, H.~L., Tsui, D.~C., Baldwin, K.~W., West,
  K.~W., and Pfeiffer, L.~N. (2004) {\em Phys. Rev. Lett.\/}, {\bf 93}, 266804.

\bibitem{cooper}
Cooper, N.~R. (2008) {\em Advances in Physics\/}, {\bf 57}, 539.

\bibitem{haldane}
Haldane, F. D.~M. (1983) {\em Phys. Rev. Lett.\/}, {\bf 51}, 605.

\bibitem{HaldRez}
Haldane, F. D.~M. and Rezayi, E.~H. (1985) {\em Phys. Rev. Lett.\/}, {\bf 54},
  237.

\bibitem{FOC}
Fano, G., Ortolani, F., and Colombo, E. (1986) {\em Phys. Rev. B\/}, {\bf 34},
  2670.

\bibitem{GMP}
Girvin, S.~M., MacDonald, A.~H., and Platzman, P.~M. (1986) {\em Phys. Rev.
  B\/}, {\bf 33}, 2481.

\bibitem{tinkham}
Tinkham, M. (2004) {\em Introduction to Superconductivity\/}. Dover
  Publications, 2nd Ed.

\bibitem{jach}
Girvin, S.~M. and Jach, T. (1984) {\em Phys. Rev. B\/}, {\bf 29}, 5617.

\bibitem{SN1}
de~Picciotto, R., Reznikov, M., Heidblum, M., Umansky, V., Bunin, G., and
  Mahalu, D. (1997) {\em Nature\/}, {\bf 389}, 162.

\bibitem{SN2}
Saminadayar, L., Glattli, D.~C., Jin, Y., and Etienne, B. (1997) {\em Phys.
  Rev. Lett.\/}, {\bf 79}, 2526.

\bibitem{nayak}
Nayak, C., Simon, S.~H., Stern, A., Friedman, M., and {Das Sarma}, S. (2008)
  {\em Rev. Mod. Phys.\/}, {\bf 80}, 1083.

\bibitem{mermin}
Mermin, N.~D. (1979) {\em Rev. Mod. Phys.\/}, {\bf 51}, 591.

\bibitem{haldane2}
Haldane, F. D.~M. (1991) {\em Phys. Rev. Lett.\/}, {\bf 67}, 937.

\bibitem{luhmann}
Luhman, D.~R., Pan, W., Tsui, D.~C., Pfeiffer, L.~N., Baldwin, K.~W., and West,
  K.~W. (2008) {\em Phys. Rev. Lett.\/}, {\bf 101}, 266804.

\bibitem{shabani}
Shabani, J., Gokmen, T., and Shayegan, M. (2009) {\em Phys. Rev. Lett.\/}, {\bf
  103}, 046805.

\bibitem{halperin}
Halperin, B.~I. (1984) {\em Phys. Rev. Lett.\/}, {\bf 52}, 1583.

\bibitem{JainBook}
Jain, J.~K. (2007) {\em Composite Fermions\/}. Cambridge UP.

\bibitem{LF}
Lopez, A. and Fradkin, E. (1991) {\em Phys. Rev. B\/}, {\bf 44}, 5246.

\bibitem{HLR}
Halperin, B.~I., Lee, P.~A., and Read, N. (1993) {\em Phys. Rev. B\/}, {\bf
  47}, 7312.

\bibitem{Heinonen}
Heinonen, E., O. (1998) {\em Composite Fermions\/}. World Scientific.

\bibitem{RR}
Rezayi, E.~H. and Read, N. (1994) {\em Phys. Rev. Lett.\/}, {\bf 72}, 100.

\bibitem{papic}
Papi\'c, Z., M\"oller, G., Milovanovi\'c, M., Regnault, N., and Goerbig, M.~O.
  (2009) {\em Phys. Rev. B\/}, {\bf 79}, 245327.

\bibitem{Wojs}
W\'ojs, A. and Quinn, J.~J. (2000) {\em Philos. Mag. B\/}, {\bf 80}, 1405.

\bibitem{kitaev}
Kitaev, A.~Y. (2003) {\em Ann. Phys. (N.Y.)\/}, {\bf 303}, 2.

\bibitem{halperin83}
Halperin, B.~I. (1983) {\em Helv. Phys. Acta\/}, {\bf 56}, 75.

\bibitem{albofath}
Abolfath, M., Belkhir, L., and Nafari, N. (1997) {\em Phys. Rev. B\/}, {\bf
  55}, 10643.

\bibitem{sondhi}
Sondhi, S.~L., Karlhede, A., Kivelson, S.~A., and Rezayi, E.~H. (1993) {\em
  Phys. Rev. B\/}, {\bf 47}, 16419.

\bibitem{moon}
Moon, K., Mori, H., Yang, K., Girvin, S.~M., MacDonald, A.~H., Zheng, I.,
  Yoshioka, D., and Zhang, S.-C. (1995) {\em Phys. Rev. B\/}, {\bf 51}, 5143.

\bibitem{fertig}
Fertig, H.~A. (1989) {\em Phys. Rev. B\/}, {\bf 40}, 1087.

\bibitem{WZ}
Wen, X.-G. and Zee, A. (1992) {\em Phys. Rev. Lett\/}, {\bf 69}, 1811.

\bibitem{EI}
Ezawa, Z.~F. and Iwazaki, A. (1993) {\em Phys. Rev. B\/}, {\bf 47}, 7295.

\bibitem{spielman}
Spielman, I.~B., Eisenstein, J.~P., Pfeiffer, L.~N., and West, K.~W. (2000)
  {\em Phys. Rev. Lett.\/}, {\bf 84}, 5808.

\bibitem{kellogg}
Kellogg, M., Eisenstein, J.~P., and an~K.~W.~West, L. N.~P. (2004) {\em Phys.
  Rev. Lett.\/}, {\bf 036801}.

\bibitem{tutuc}
Tutuc, E., Shayegan, M., and Huse, D.~A. (2004) {\em Phys. Rev. Lett\/}, {\bf
  93}, 036802.

\bibitem{nomura}
Nomura, K. and MacDonald, A.~H. (2006) {\em Phys. Rev. Lett.\/}, {\bf 96},
  256602.

\bibitem{GMD}
Goerbig, M.~O., Dou\c{c}ot, B., and Moessner, R. (2006) {\em Phys. Rev. B\/},
  {\bf 74}, 161407.

\bibitem{alicea}
Alicea, J. and Fisher, M. P.~A. (2006) {\em Phys. Rev. B\/}, {\bf 74}, 075422.

\bibitem{yang}
Yang, K., {Das Sarma}, S., and MacDonald, A.~H. (2006) {\em Phys. Rev. B\/},
  {\bf 74}, 075423.

\bibitem{arovas}
Arovas, D.~P., Karlhede, A., and Lillieh\"o\"ok, D. (1999) {\em Phys. Rev.
  B\/}, {\bf 59}, 13147.

\bibitem{DGLM}
Dou\c{c}ot, B., Goerbig, M.~O., Lederer, P., and Moessner, R. (2008) {\em Phys.
  Rev. B\/}, {\bf 78}, 195327.

\bibitem{gusynin}
Gusynin, V.~P., Miransky, V.~A., Sharapov, S.~G., and Shovkovy, I.~A. (2006)
  {\em Phys. Rev. B\/}, {\bf 74}, 195429.

\bibitem{fuchs}
Fuchs, J.-N. and Lederer, P. (2007) {\em Phys. Rev. Lett.\/}, {\bf 98}, 016803.

\bibitem{herbut}
Herbut, I.~F. (2007) {\em Phys. Rev. B\/}, {\bf 75}, 165411.

\bibitem{ezawaSU4}
Ezawa, Z.~F. (1999) {\em Phys. Rev. Lett.\/}, {\bf 82}, 3512.

\bibitem{dGRG}
{de Gail}, R., Regnault, N., and Goerbig, M.~O. (2008) {\em Phys. Rev. B\/},
  {\bf 77}, 165310.

\bibitem{MDYG}
MacDonald, A.~H., Yoshioka, D., and Girvin, S.~M. (1989) {\em Phys. Rev. B\/},
  {\bf 39}, 8044.

\bibitem{CZ}
Chakraborty, T. and Zhang, F.~C. (1984) {\em Phys. Rev. B\/}, {\bf 29}, 7032.

\bibitem{kang}
Kang, W., Young, J.~B., Hannahs, S.~T., Palm, E., Campman, K.~L., and Gossard,
  A.~C. (1997) {\em Phys. Rev. B\/}, {\bf 56}, R12776.

\bibitem{kukush}
Kukushkin, I.~K., v.~Klitzing, K., and Eberl, K. (1999) {\em Phys. Rev.
  Lett.\/}, {\bf 82}, 3665.

\bibitem{GR}
Goerbig, M.~O. and Regnault, N. (2007) {\em Phys. Rev. B\/}, {\bf 75}, 241405.

\bibitem{WenZee}
Wen, X.-G. and Zee, A. (1992) {\em Phys. Rev. B\/}, {\bf 46}, 2290.

\bibitem{wallace}
Wallace, P.~R. (1947) {\em Phys. Rev.\/}, {\bf 71}, 622.

\bibitem{klein}
Klein, O. (1929) {\em Z. Phys.\/}, {\bf 53}, 157.

\end{thebibliography}


\end{document}